\documentclass{aa}  

\usepackage{graphicx}
\usepackage{multirow}
\usepackage{txfonts}
\usepackage[]{hyperref}
\usepackage{ulem}
\usepackage{xcolor}
\newcommand{\kms}{km\,s$^{-1}$}
\newcommand{\dego}{$^{\circ}$}

\begin{document}

\title{Millimetre observations of the S-type AGB star $\chi$ Cygni: 
variability of the emission of the inner envelope}


   \author{D.T. Hoai\inst{1}
          \and
          J.M. Winters\inst{2}
          \and
          P.T. Nhung\inst{1}
          \and
          P. Darriulat\inst{1}
          \and
          T. Le Bertre\inst{3}
          }

   \institute{Department of Astrophysics, Vietnam National Space Center (VNSC), Vietnam Academy of Science and Technology (VAST), 18 Hoang Quoc Viet, Cau Giay, Ha Noi, Vietnam
   \email{dthoai@vnsc.org.vn}\\
         \and
          Institut de Radioastronomie Millim\'{e}trique (IRAM), 300 rue de la Piscine, Domaine Universitaire, 38406 St. Martin d’Hères, France\\
         \and  
          LERMA, UMR 8112, CNRS and Observatoire de Paris, PSL Research University, 61 av. de l’Observatoire, 75014 Paris, France \\
             }

   \date{Received xx; accepted xx}

 
  \abstract
      {New observations are presented of millimetre line emissions of the circumstellar envelope (CSE) of  Asymptotic Giant Branch (AGB) star $\chi$ Cygni using the recently upgraded NOEMA array. $\chi$ Cygni is an S-type Mira variable, at the border between oxygen-rich and carbon-rich, that has been observed for over 40 years to display features giving evidence for the strong role played by pulsation-associated shock waves in the generation of its wind. The new observations give evidence for a bright H$^{12}$CN(3-2) line emission confined to the very close neighbourhood of the star, however significantly more extended in 2024 than in 2023. Interpretation of such variability in terms of maser emission is considered but found to raise significant unanswered questions. Moreover, other unexpected features are observed in the very close neighbourhood of the star, including low Si$^{16}$O(6-5)/Si$^{17}$O(6-5), $^{28}$SiO(5-4)/$^{29}$SiO(5-4) and $^{12}$CO(2-1)/$^{13}$CO(2-1) line emission ratios. Possibly confirming the important role played by shocks, we discuss a measurement of the SiO(5-4)/SiO(6-5) emission ratio, evidence for a recent mass ejection particularly enhanced in the north-western red-shifted octant of the circumstellar envelope, leaving a depression of emission in its wake, and patterns of enhanced CO(2-1) line-emission suggesting an interpretation in terms of episodic outflows enhanced over solid angles associated with the surface of convective cells and on a time scale of a few decades. Unravelling the mechanisms underlying such newly observed features is very challenging and confirmation of the reported observations with improved sensitivity and angular resolution would be highly welcome. Observation of SiO maser emission in the ($\nu$=1,$J$=6-5) transition is reported for the first time.}


   \maketitle

\section{Introduction}
Recent high-resolution imaging of nearby Asymptotic Giant Branch (AGB) stars at visible and infrared wavelengths has revealed inhomogeneous distributions of gas and dust in the inner circumstellar environment, with changes in morphology and grain sizes occurring over the course of weeks or months, giving support to 3-D hydro-dynamical models, to which we refer as ``standard models'' \citep[][and references therein]{Hofner2019, Freytag2023}. Such models describe large-scale convective flows below the photosphere, which, together with stellar pulsations, produce shock waves that give rise to ballistic gas motions, typically peaking around 2 stellar radii, which lift it in regions where dust can form and start the acceleration process \citep{Liljegren2018}. Recent high angular resolution observations in the visible/infrared and at millimetre wavelengths have given support to this picture \citep[e.g.][]{Paladini2018, Paladini2019, Lagadec2019}.

The important role played by shocks in levitating the inner atmosphere of AGB stars to distances where dust can form had already been recognised over forty years ago.  \citet{Hinkle1982}, using observations of a time series of infrared spectra (1.6-2.5 $\mu$m) spanning more than three pulsation cycles of $\chi$ Cygni, had given evidence for the photometric variability to be associated with an outwardly propagating wave, which traverses the photosphere during the pre-maximum and maximum phases of the light curve. They established that a shock front forms between gas in-falling from the previous cycle and material already accelerated by the emerging wave, and that, at this interface, the gas is heated from $\sim$2000 to over 4000 K, molecules being dissociated, and hydrogen being probably ionised. Soon after, \citet{Fox1984}, \citet{Fox1985} and \citet{Bertschinger1985} used these observations, together with a catalogue of Balmer line emission profiles of eight other Mira variables, to propose periodic shock wave models of Mira variable atmospheres. Their models show good fits to the observations, a typical shock being expected to decelerate from $\sim$30 km\,s$^{-1}$ at maximum light to $\sim$20 km\,s$^{-1}$ at a stellar phase of 0.25.

At the end of the nineties, the availability of millimetre observations of molecular line emissions, \citep{Bujarrabal1994a, Bujarrabal1994b, Olofsson1998} including in particular the oxygen-rich IK Tau \citep{Duari1999, Willacy1997}, the carbon-rich IRC+10216 \citep{Willacy1998} and a sample of 30 AGB stars of M-, S-, and C-types \citep{Bieging2000} allowed for significant progress in the understanding of the associated shock chemistry. The latter authors concluded that pulsation-driven shocks result in the formation of organic molecules like HCN in the circumstellar envelopes (CSEs) of M stars \citep[see also][]{Charnley1995}, the HCN/SiO intensity ratios of lines with similar excitation energies separating clearly the carbon stars from the M and S stars, and the latter showing a trend of increasing HCN/SiO intensity ratios with increasing mass-loss rate. Over the past decade, further progress has been achieved on the understanding of the mechanism governing the formation of HCN in the inner CSE of thermally pulsing AGB stars \citep{Marigo2016, Millar2016, Gobrecht2016}, in particular giving evidence for the first shock to emerge very close to the photosphere and for the gas temperature to cool primarily by emission of radiation in the post-shock relaxation phase. \citet{Schoier2013}, from the study of a large sample of AGB stars of different chemical types, have evaluated the median values of the HCN abundances with respect to H$_2$ as 3$\times$10$^{-5}$, 7$\times$10$^{-7}$ and 10$^{-7}$ for C-, S- and M-type AGB stars, respectively.

At the same time, 3-D hydro-dynamical models \citep{Hofner2019, Freytag2023} have been developed, including the role played by the partition of the photosphere in the form of giant convective cells covering solid angles at steradian scale. Indeed, direct evidence for their presence has been obtained by \citet{Paladini2018} using the Precision Integrated-Optics Near-infrared Imaging ExpeRiment (PIONIER) at the Interferometer of the Very Large Telescope (VLTI) to observe the surface of $\pi^1$ Gru. Using the same instrument, giant convective cells have been revealed on the surface of Betelgeuse \citep{Montarges2016} and their presence has been confirmed by spectro-polarimeter observations \citep{LopezAriste2018} using NARVAL at Pic du Midi; the latter give evidence for slow evolution of their pattern on a 2-3-year timescale, accompanied by a much faster and smaller amplitude variability on a week to month time scale. NARVAL observations have also given evidence for asymmetric shocks in $\chi$ Cygni, induced by the interplay between pulsation and convection, suggesting the presence of a highly anisotropic and variable velocity field near the surface of the star \citep{LopezAriste2019}.  Yet, while providing a very useful guide to current research, having received general support and suffered no contradiction, standard models still require many additional observations before being reliably validated \citep{Darriulat2024}. In particular, observations of the highest possible angular resolution at both millimetre and visible/infrared wavelengths, performed in conjunction with measurements of the light curve, are necessary to better specify the respective roles played by convection and stellar pulsations.

Recently, millimetre observations of the continuum and molecular line emissions of the CSE of M- and S-types AGB stars have concentrated on three fronts: 1) study of the continuum emission of the stellar photosphere and surrounding hot dust, giving evidence for occasional hot spots and variability; 2) measurements of the radial dependence of the abundance of molecular species giving evidence, in particular, for the early formation of large aluminium-rich transparent dust grains; 3) measurements of the morpho-kinematics of the inner CSE layer using the emission of molecular lines. In this context, the main results are \citep{Darriulat2024}: presence of a gravitationally bound dense reservoir of gas within $\sim$2 stellar radii from the centre of the star \citep[e.g.][]{Wong2016}; evidence for the early formation of large transparent dust grains within the same radial range \citep[e.g.][]{Takigawa2017}; evidence for out-flowing and in-falling gas within a very few stellar radii  from the photosphere, possibly reaching velocities much larger than the terminal velocities \citep[e.g.][]{Nhung2022, Vlemmings2017}; occasional presence of rotation confined to the same radial range \citep[e.g.][]{Vlemmings2018}; outflows covering limited solid angles, at steradian scale, having velocities of the order of 10 km s$^{-1}$ or less, and lasting for a period at century scale \citep[e.g.][]{Hoai2022a}; mass ejections of similar morphology but lasting for only a few years \citep[e.g.][]{Hoai2022b, Nhung2021}; evidence for episodes of enhanced mass loss \citep[e.g.][]{Nhung2019}.

In the present article, we analyse millimetre observations of $\chi$ Cygni, the AGB star that had been found over forty years ago by \citet{Hinkle1982} to give evidence for pulsation related shock waves triggering the nascent wind. It is a very bright red giant, showing one of the largest variations in apparent magnitude of any pulsating variable star. It is a Mira variable with a period of 408 days \citep{Kholopov1999, Samus2009} and a spectral type S6+/1e \citep{Keenan1980}. Its light curve is regularly monitored and obeys the period-luminosity relation characteristic of the fundamental-mode pulsation of Mira variables \citep{Wood2015}. With a C/O ratio close to unity \citep{Duari2000, Justtanont2010} it is at the border between oxygen-rich and carbon-rich AGB stars. The presence of technetium in its spectrum \citep{Vanture1991, Lebzelter2003, Uttenthaler2019} and a high $^{12}$C/$^{13}$C abundance ratio of 33 to 40 \citep{DeBeck2010, Milam2009, Ramstedt2014, Hinkle2016, Wallerstein2011} confirm such advanced stage of evolution. Its distance from Earth has been measured between 150 pc \citep{Knapp2003} and 180 pc \citep{vanLeeuwen2007}; in the present article we adopt an intermediate value of 165 pc. \citet{Lacour2009}, using optical interferometry IOTA observations, estimate its mass between 1.4 and 3.6 M$_\odot$, its radius between 348 and 480 solar radii (10-14 mas), its luminosity between 6000 and 9000 solar luminosities, its temperature between 2440 and 2740 K. They imaged the stellar disc at four different stellar phases revealing strong variability of its apparent diameter, with amplitude reaching 40\% of the stellar radius, and the presence of hot cells on the photosphere. \citet{DeBeck2010}, using single dish observations of the emission of nine different molecular lines, measured a mass loss rate of 2.4$\times$10$^{-7}$ M$_\odot$ yr$^{-1}$. Numerous observations of SiO masers have been published \citep{Olofsson1981, Schwartz1982, Bujarrabal1987, Alcolea1992, Rizzo2021, Boboltz2004, GomezGarrido2020}. In the millimetre and sub-millimetre range, single dish observations have used the Herschel Space Telescope \citep{Justtanont2010, Schoier2011}, the IRAM 30m telescope \citep{Ramstedt2009}, the 11m telescope of the National Radio Astronomy Observatory (NRAO) at Kitt Peak \citep{Lo1977} and the James Clerk Maxwell Telescope (JCMT) \citep{Duari2000}, the latter giving evidence for  HCN being formed within $<\sim$20 stellar radii, suggesting that shock chemistry plays an active role; interferometry observations have used the PdBI to detect SiO \citep{Lucas1992} and CO \citep{CastroCarrizo2010} line emissions. In the mid-infrared, \citet{Tevousjan2004} using ISI interferometry observations, measured an inner radius of the dust shell as large as 4.3-4.4 stellar radii and \citet{Nowotny2010} have suggested the presence of shocks near the photosphere from measurements of the dependence on stellar phase of the radial velocity of infrared 4 $\mu$m emission. Finally, \citet{Lebre2014, Lebre2015}, using NARVAL on the 2 m telescope at Pic-du-Midi, have given evidence for a weak surface magnetic field (a few Gauss), suggesting that shock waves may have an efficient compressive effect, and \citet{LopezAriste2019} interpret the azimuthal asymmetry of the degree of linear polarisation as evidence for the pulsation shock wave to produce an inhomogeneous radial velocity field. \citet{Herpin2006}, using the IRAM 30m telescope, have measured SiO maser polarisation.

Like RS Cnc, which has been recently studied in much detail using Northern Extended Millimeter Array (NOEMA) \citep{Winters2022}, $\chi$ Cygni is too boreal to be comfortably observed by Atacama Large Millimeter/submillimeter Array (ALMA). Its proximity to Earth, its location at the border between oxygen-rich and carbon-rich, the important role it has played over the past decades in giving evidence for pulsation-related shocks levitating the inner layer of the stellar atmosphere, make it an ideal target for millimetre observations using the recently upgraded NOEMA \citep{Neri2022}.

\section{Observations and data reduction}

\subsection{Observations and data calibration and reduction}
\begin{figure*}
  \centering
  \includegraphics[width=1\linewidth]{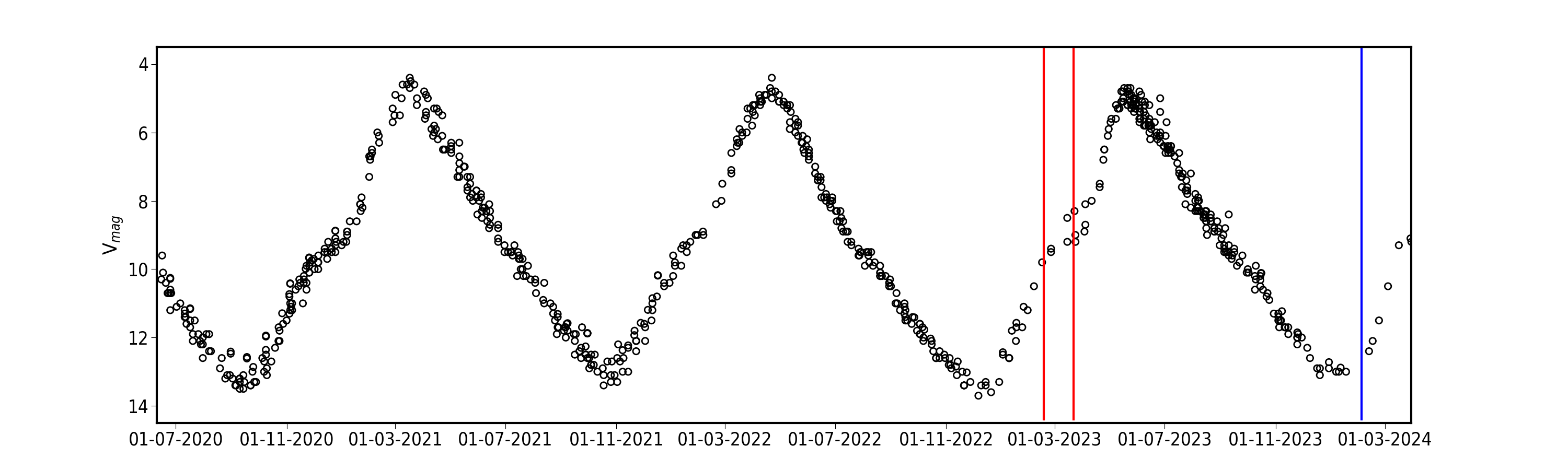}
  \caption{Light curve \citep[taken from][]{BAA} covering the time of the 2023 (red) and 2024 (blue) observations. The stellar phases are 0.75 and 0.85 in 2023 and 0.65 in 2024.}
  \label{fig1}
\end{figure*}

$\chi$ Cygni was observed with NOEMA using 12 antennas in the extended A- and B-configurations in February and March 2023, midway between minimum and maximum light (Figure~\ref{fig1}) and in February 2024, closer to minimum light. Observations were using two individual frequency setups covering a total frequency range of $\sim$32 GHz in the 1.3 mm atmospheric window. The low frequency setup covers frequency ranges of 213.8 GHz to 221.6 GHz in LSB and 229.4 GHz to 237.2 GHz in USB and the high-frequency setup covers 242.8 GHz to 250.6 GHz in LSB and 258.4 GHz to 266.2 GHz in USB. The high-frequency setup is identical to the one previously used to observe RS Cnc \citep{Winters2022} whereas the low-frequency setup was shifted upwards by 1 GHz. We used the new 250 kHz mode of the PolyFiX correlator to cover the full 32 GHz bandpass with a continuous spectral resolution of 250 kHz, corresponding to intrinsic velocity resolutions between 0.35 \kms\ and 0.28 \kms. All data were finally re-binned to a spectral resolution of 0.5 \kms. Phase and amplitude calibration was done on the two quasars J1948+359 and 2013+370, which were observed every $\sim$20 min. Pointing and focus of the telescopes was checked about every hour, and corrected when necessary. The bandpass was calibrated on different strong quasars (3C279, 2013+370, 2200+420). The absolute flux scale was fixed on MWC349, whose flux is regularly monitored and calibrated on planets whose fluxes are known. The accuracy of the absolute flux calibration at 1.3 mm is estimated to be better than 15\%. The data were calibrated and imaged within the GILDAS\footnote{https://www.iram.fr/IRAMFR/GILDAS} suite of software packages using CLIC for the NOEMA data calibration and $uv$ table creation and the MAPPING package for merging, subsequent self-calibration and imaging of the data sets. Continuum data were extracted for each sideband of the two frequency setups individually by filtering out spectral lines and phase self-calibration was performed on the corresponding frequency averaged data. The gain tables containing the self-calibration solutions were then applied to the spectral line $uv$ tables using the SELFCAL procedures provided in MAPPING. The resulting data sets were imaged applying robust weighting with a threshold of 0.1 to increase the spatial resolution by typically about a factor 2. The resulting dirty maps were then CLEANed using the H\"{o}gbom algorithm \citep{Hogbom1974}. The final beam characteristics and sensitivities of the individual and combined data sets from A- and B-configuration are listed in Table~\ref{tab2}.

\begin{figure*}
  \centering
  \includegraphics[width=0.3\linewidth]{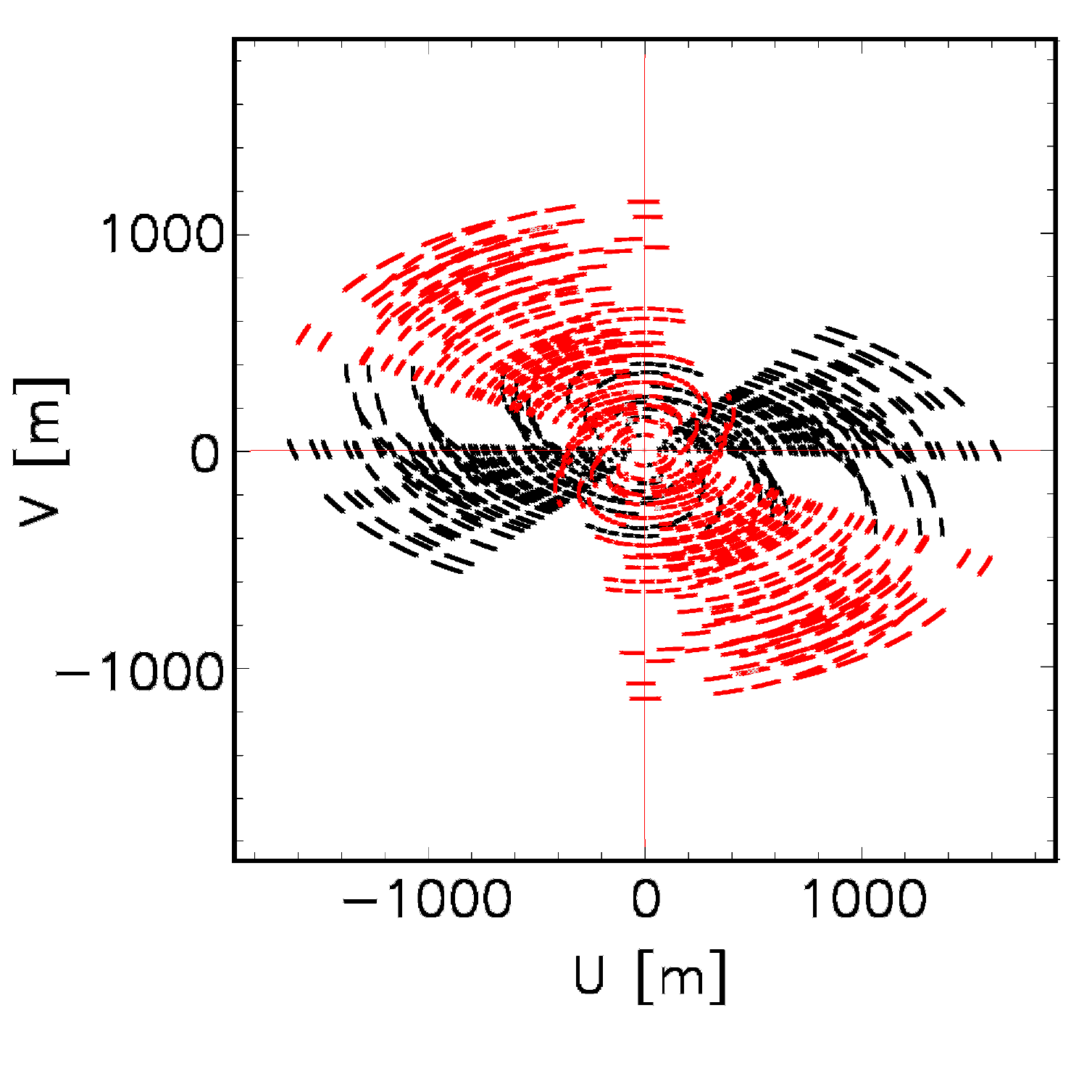}
  \includegraphics[width=0.3\linewidth]{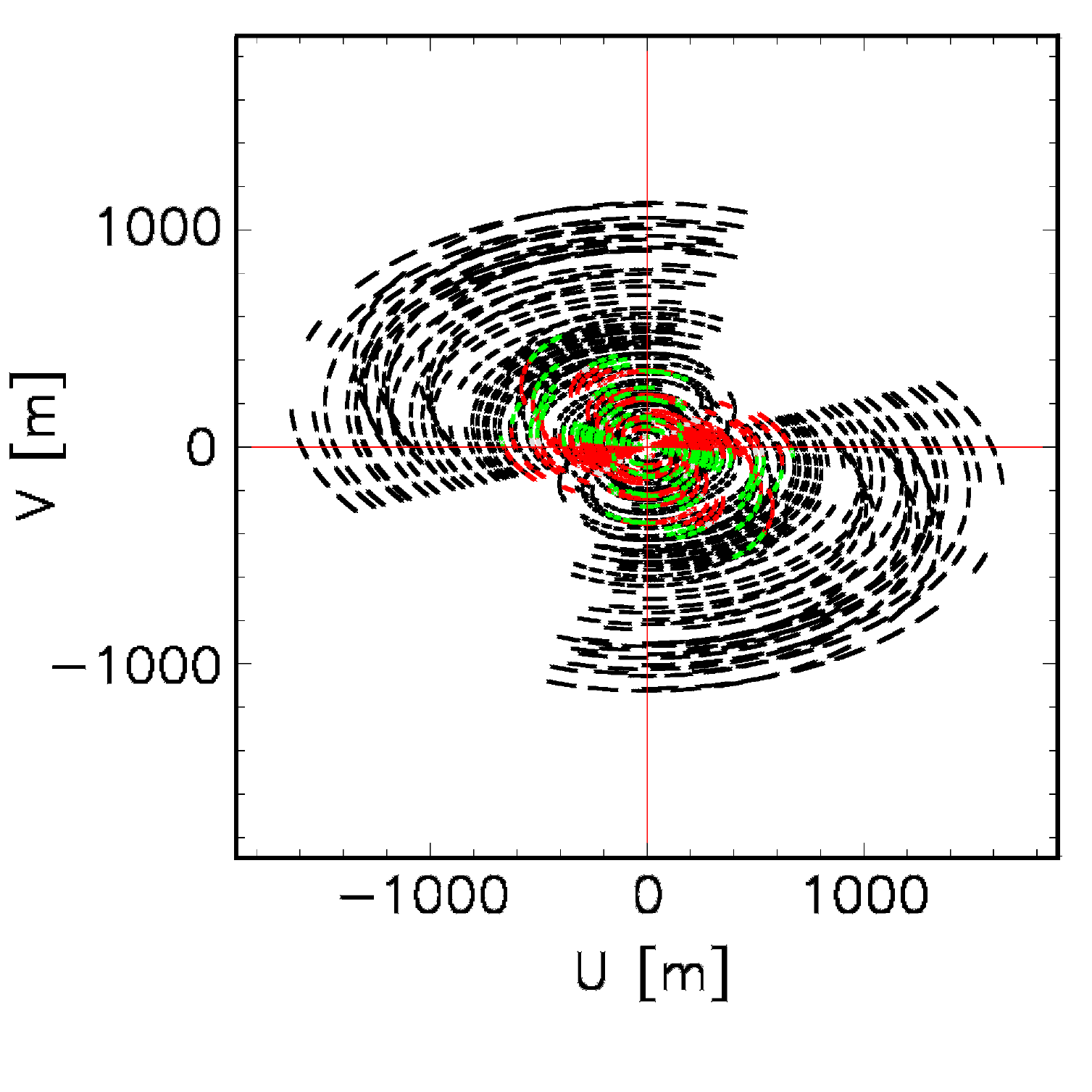}
  \includegraphics[width=0.3\linewidth]{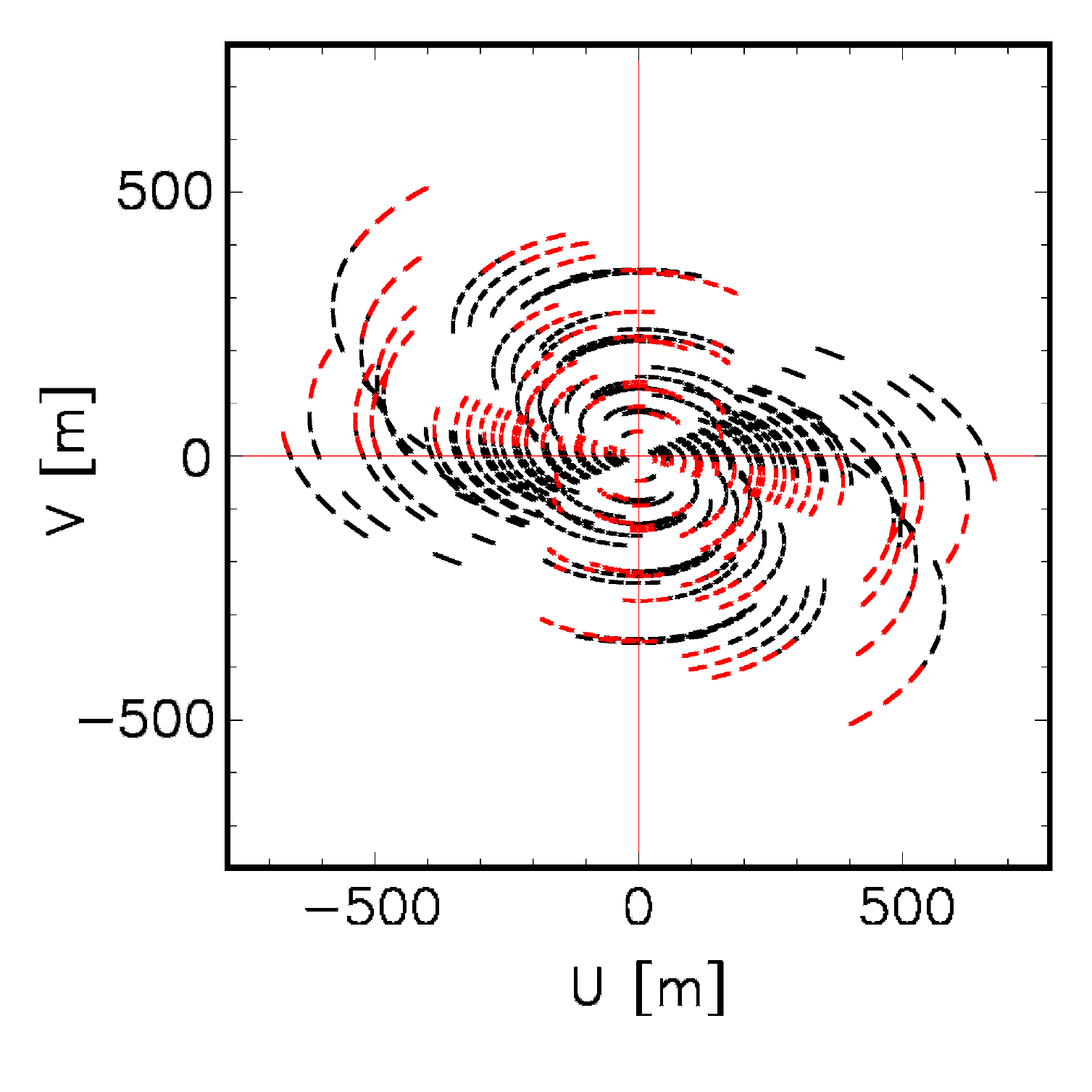}
  \caption{From left to right: $uv$ coverages for the high-frequency setup using A-configuration in 2023, the low-frequency setup using merged A+B-configuration in 2023 and the high-frequency setup using the B-configuration in 2024. For the high-frequency observations, red is for February 21 and black for February 18. For the low-frequency 2023 observations, black is for February 13, red for March 21 and green for March 22. For the high-frequency 2024 observations, black is for February 3 and red for February 5.}
  \label{fig2}
\end{figure*}

The low frequency setup was observed in A- and B-configurations in 2023, each during 4 hours. The corresponding data have been merged, resulting in a maximum recoverable scale of $\sim$2 arcsec. The baselines ranged from 23 m to 1649 m. The high frequency setup was observed in the A-configuration during 4 hours in 2023 and in B-configuration during 2.6 hours in 2024, with baselines ranging from 65 to 1612 m. The associated $uv$ coverages are illustrated in Figure~\ref{fig2} and details are given in Table~\ref{tab1}.

Evidence for major differences between the 2023 and 2024 observations of the emissions of the H$^{12}$CN(3-2) line is illustrated in Figure~\ref{fig3}. It prevented merging the associated A- and B-configuration data for this line. We comment on this in more detail in Sub-section 2.2 below.

We studied carefully the impact of the lack of short spacing by imaging disc sources of different sizes and different radial dependence of the brightness. We found that reliable imaging is obtained for angular distances $R$$<$1.5-2 arcsec for the merged A+B data, and $R$$<$0.6-0.8 arcsec for the A-configuration data.

In what follows, unless otherwise stated, we present line emission results obtained after subtraction of the contribution of continuum emission. Table~\ref{tab2} summarises some properties of strong line emissions. Some other clearly detected line emissions, not considered in the present article, include: SO(6$_5$-5$_4$), SO(7$_6$-6$_5$), H$_2$O at 232.68670 GHz, SiS$\nu$=0,12-11), SiS($\nu$=0,13-12), AlF($\nu$=0, $J$=7-6, $F$=15/2-17/2), TiO($\nu$=1,7-6), $^{29}$Si$^{17}$O($\nu$=0,6-5) and PN($N$=5-4,$J$=6-5). Information on these lines is given in Table \ref{tabb1} and their spectra are shown in Figure \ref{figb1}. 

We use coordinates centred on the continuum emission (Figure~\ref{fig4}), $x$ pointing east, $y$ pointing north and $z$ pointing away from Earth. Its position in 2023 is found to be the same for both low frequency and high frequency setups and each of the lower and upper sidebands: RA=19h 50min 33.87986s and Dec=32$^o$ 54' 49.6350''. The 2024 data have been re-centred to take into account the proper motion of $\chi$ Cygni between 2023 and 2024. The flux density of the continuum emission is measured to be 54 mJy, within $\pm$8 mJy depending on frequency. The angular resolution does not allow for resolving the stellar disc. We checked that the 2023 and 2024 measurements agree within 0.8 mJy, well within uncertainty.

The angular distance to the centre of the star is calculated as $R$$=$$\sqrt{x^2+y^2}$. Position angles, $\omega$, are measured counter-clockwise from north. The Doppler velocity, $V_z$, is relative to the adopted systemic velocity of 10 \kms\ in the local standard of rest (LSR) frame. 

\subsection{Variability of the H$^{12}$CN(3-2) and SiO($\nu$=1,$J$=6-5) line emissions}
\begin{figure*}
  \centering
  \includegraphics[width=0.7\linewidth]{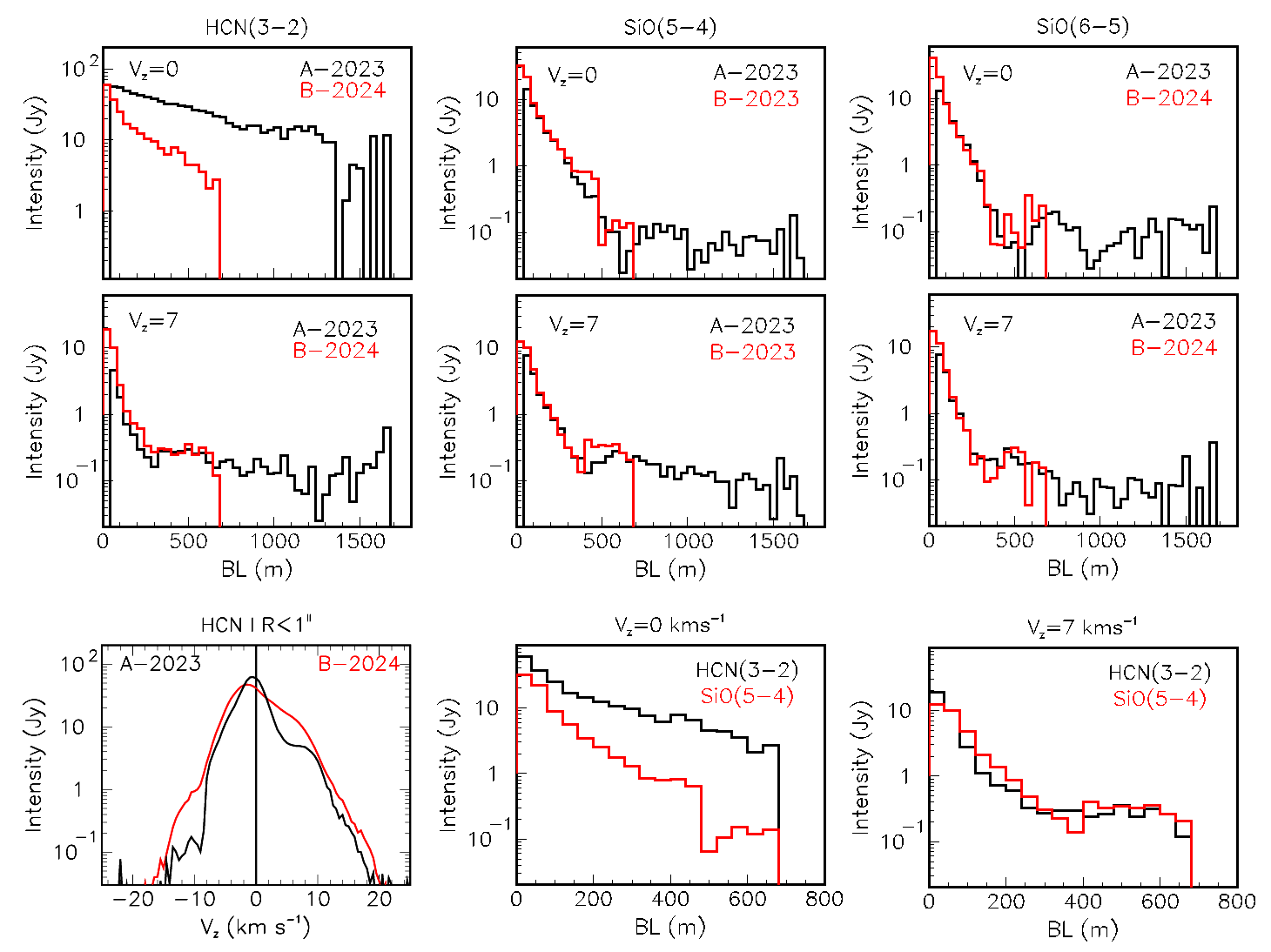}
  \caption{Distributions of the absolute values of the observed visibilities (Jy) over baseline lengths (m). In the upper and middle rows, we show separately the A-configuration (black) and B-configuration (red) for three different emission lines and two different Doppler velocity channels. The upper row corresponds to the frequency channel of the systemic star velocity ($V_z$=0). The middle row corresponds to $V_z$$=$7 \kms. In each triplet, from left to right, the lines are H$^{12}$CN(3-2), Si$^{16}$O($\nu$=0,$J$=5-4) and Si$^{16}$O($\nu$=0,$J$=6-5). A-configuration observations were made in 2023 while B-configuration observations were made in 2023 for the low frequency setup, here Si$^{16}$O(5-4), and in 2024 for the high frequency setup, here H$^{12}$CN(3-2) and Si$^{16}$O(6-5). Lower row: the left panel compares the H$^{12}$CN(3-2) Doppler velocity spectra integrated over $R$$<$1 arcsec observed in 2023 (A-configuration, black) and in 2024 (B-configuration, red). The rightmost panels show only data observed using the B-configuration. They correspond to Doppler velocity channels of 0 and 7 \kms\ respectively. Black is for H$^{12}$CN(3-2) and red for SiO(5-4).}
  \label{fig3}
\end{figure*}

\begin{figure}
  \centering
  \includegraphics[width=0.9\linewidth, trim= 0cm 1.3cm 0cm 2.8cm,clip]{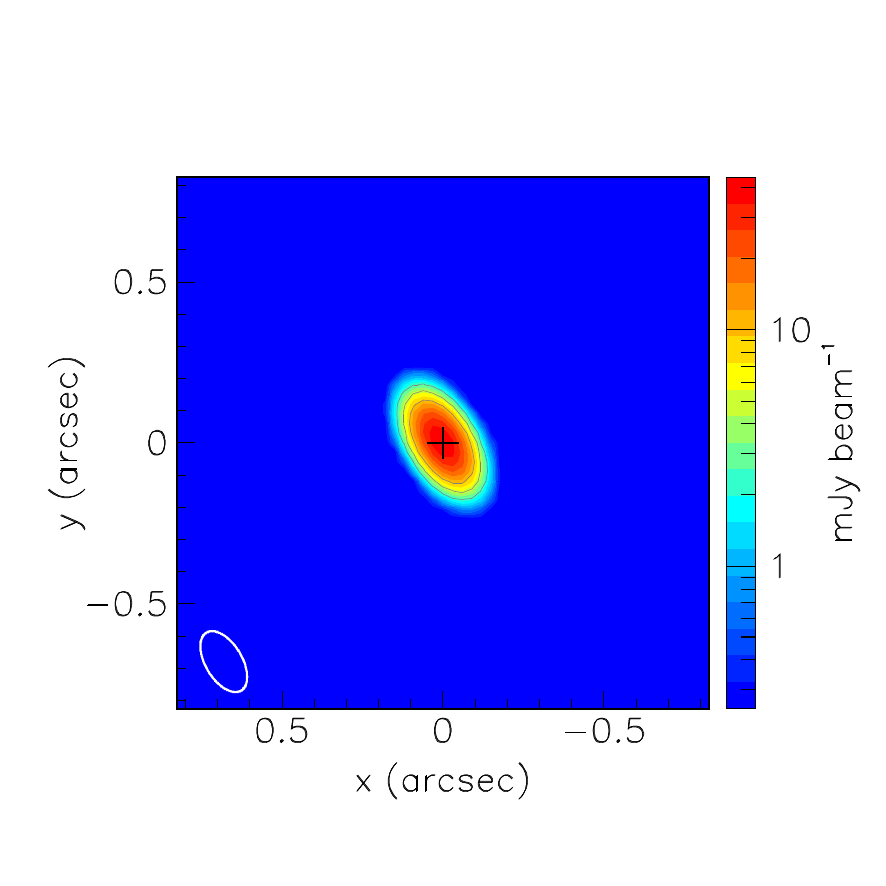}
  \caption{Continuum emission (A+B merged data, 2023). The beam (FWHM) is shown in the lower-left corner. The origin of coordinates is at RA=19h 50min 33.87986s and Dec=32$^o$ 54' 49.6350'' (see text). The colour scale is logarithmic. The noise level is 0.5 mJy\,beam$^{-1}$. The contours are at 5, 10, 20 times the noise level.}
  \label{fig4}
\end{figure}

Figure~\ref{fig3} compares the dependence on baseline length of the absolute values of the visibility for the emission of three molecular lines, Si$^{16}$O(5-4), Si$^{16}$O(6-5) and H$^{12}$CN(3-2) for A- and B-configurations separately and for two Doppler velocity channels. The justification for such a choice will become apparent later (see Section 3). For the time being, we simply note the spectacular singularity of the H$^{12}$CN(3-2) emission, giving evidence for a major variability between the 2023 and 2024 observations at Doppler velocities close to the systemic velocity of the star. This is in contrast to the Si$^{16}$O(5-4) and Si$^{16}$O(6-5) emissions, which show very similar distributions at short baselines, independently from the year of observation. However, we also note some smaller differences on these two lines, preventing us from ascertaining that their emission did not vary between 2023 and 2024. Similar small differences are also present for the other line emissions reported in the present article, however none being as spectacular as for H$^{12}$CN(3-2) at $V_z$=0. The variability of the SiO($\nu$=1,$J$=6-5) line emission is studied in Sub-section 3.8 and is reliably understood as caused by a maser: its interpretation does not pose a particular problem. In contrast, the interpretation of the variability of the H$^{12}$CN(3-2) line is challenging: while it is natural to think of masers as a possible cause, it is very unusual to observe such strong maser emission from a vibrational ground state, it only rarely occurred in the case of carbon-stars. We discuss this issue in Sub-sections 3.2 and 5.1. 

\begin{table*}
  \centering
  \caption{Properties of some line emissions and associated beam sizes and noise levels in 2023. }
  \label{tab2}
  \begin{tabular}{cccccc}
    \hline
    Line
    &Frequency (GHz)
    &$E_{up}$
    &$A_{ji}$
    &Beam size/PA
    &Noise\\
    &(GHz)&(K)&(Hz)&(mas$^2$/$^o$)&(mJy beam$^{-1}$)\\
    \hline
    \multicolumn{6}{c}{Low frequency setup, A+B configuration (2023)}\\
    \hline
    SiO($\nu$=0,5-4)&
    217.1049&
    31.26&
    5.20$\times$10$^{-4}$&
    244$\times$142/38&
    5.2\\
    $^{29}$SiO($\nu$=0,5-4)&
    214.3858&
    30.9&
    5.01$\times$10$^{-4}$&
    247$\times$144/38&
    5.2\\
    $^{12}$CO(2-1)&
    230.5380&
    16.6&
    6.91$\times$10$^{-7}$&
    228$\times$132/37&
    5.5\\
    $^{13}$CO(2-1)&
    220.3987&
    15.9&
    6.04$\times$10$^{-7}$&
    240$\times$137/39&
    4.9\\
    \hline
    \multicolumn{6}{c}{High frequency setup, A-configuration (2023)}\\
    \hline
    SiO($\nu$=0, 6-5)&
    260.5180&
    43.76&
    9.12$\times$10$^{-4}$&
    207$\times$112/27&
    7.3\\
    SiO($\nu$=1, 6-5)&
    258.7073&
    1812&
    9.90$\times$10$^{-4}$&
    207$\times$112/27&
    7.3\\
    H$^{12}$CN(3-2)&
    265.8864&
    25.5&
    8.4$\times$10$^{-4}$&
    204$\times$109/27&
    9.1\\
    H$^{13}$CN(3-2)&
    259.0118&
    24.9&
    7.7$\times$10$^{-4}$&
    211$\times$113/28&
    7.3\\
    Si$^{17}$O(6-5)&
    250.7447&
    42.1&
    8.13$\times$10$^{-4}$&
    213$\times$117/27&6.6\\
    CS(5-4)&
    244.9356&
    35.3&
    3.0$\times$10$^{-4}$&
    215$\times$120/27&6.5\\
    \hline    
  \end{tabular} 
 \tablefoot{In 2024, the beam is typically 467$\times$278 mas$^2$ FWHM at PA=43$^o$ and the noise level is $\sim$8 mJy beam$^{-1}$. Measured line emissions are listed in Table~\ref{tab3}. Line parameters are from the CDMS \citep{Muller2005}.}

\end{table*}

\begin{table*}
  \centering
  \caption{Observations: setups, antenna configurations and dates.}
  \begin{tabular}{ccccc}
    \hline
    Setup&Frequency&Configuration&Observation date&Time on source\\
    \hline
    \multirow{2}{*}{Low frequency}
    &214-222 GHz
    &A
    &13 Feb 2023
    &4 hours\\

    &229-237 GHz
    &B
    &21 \& 22 Mar 2023
    &4 hours\\
    \hline
    \multirow{2}{*}{High frequency}
    &243-251 GHz
    &A
    &18 \& 21 Feb 2023
    &4 hours\\
    &258-266 GHz
    &B
    &3 \& 5 Feb 2024
    &2.6 hours\\
    \hline
  \end{tabular}
  \label{tab1}
\end{table*}

\section{The inner CSE}
\subsection{General properties}
The observations reported below, using the upgraded NOEMA array, probe the inner CSE of $\chi$ Cygni with a higher angular resolution than previously possible, revealing new details of its morpho-kinematics. In the present section, we limit the study to the interior of a circle of 1 arcsec radius centred on the star, within which we have established that reliable imaging has been achieved. In addition to the dominant CO and SiO molecular line emissions, we report the emissions of the HCN and CS lines, which, as being emitted by carbon-bearing molecules, are of specific interest to the study of an S-type star. As a general rule, unless specified otherwise, we limit the study to 2023 data using the A-configuration: as B-configuration data were obtained from two different observing periods, a possible variability makes their inclusion less straightforward. However, we have repeated the analysis presented in the present section using the merged A+B-configuration data and we have made sure that all our conclusions are robust in this respect. Altogether, we cover the following molecular lines: $^{12}$CO(2-1), $^{13}$CO(2-1), Si$^{16}$O(5-4), Si$^{16}$O(6-5), Si$^{17}$O(6-5), H$^{12}$CN(3-2), H$^{13}$CN(3-2) and $^{12}$CS(5-4).

\begin{figure*}
  \centering
  \includegraphics[width=0.31\linewidth, trim= 0.5cm 1.5cm 2.5cm 1.8cm,clip]{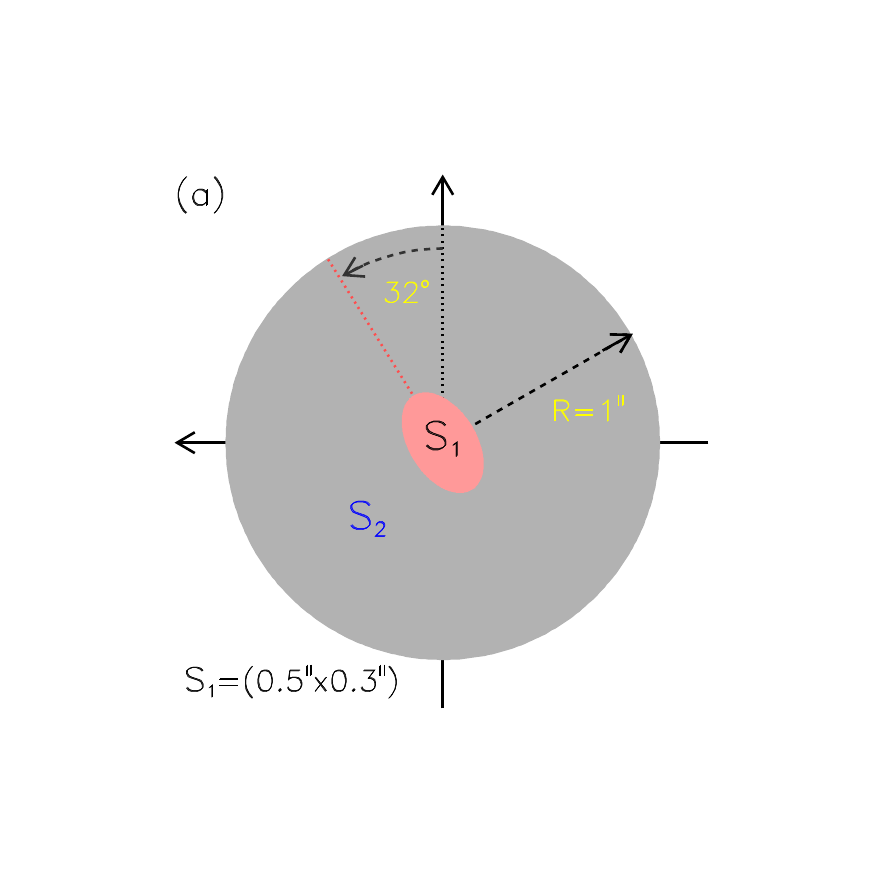}
  \includegraphics[width=0.27\linewidth, trim= 0.5cm 1.cm 2.5cm 2.1cm,clip]{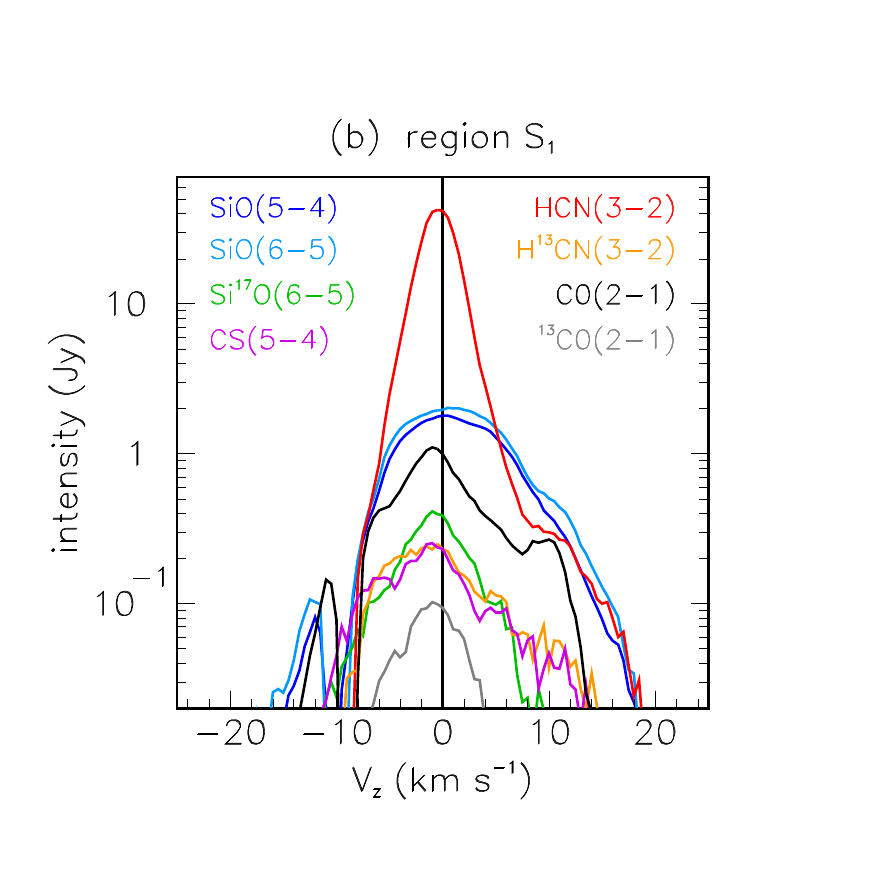}
  \includegraphics[width=0.27\linewidth, trim= 0.5cm 1.cm 2.5cm 2.1cm,clip]{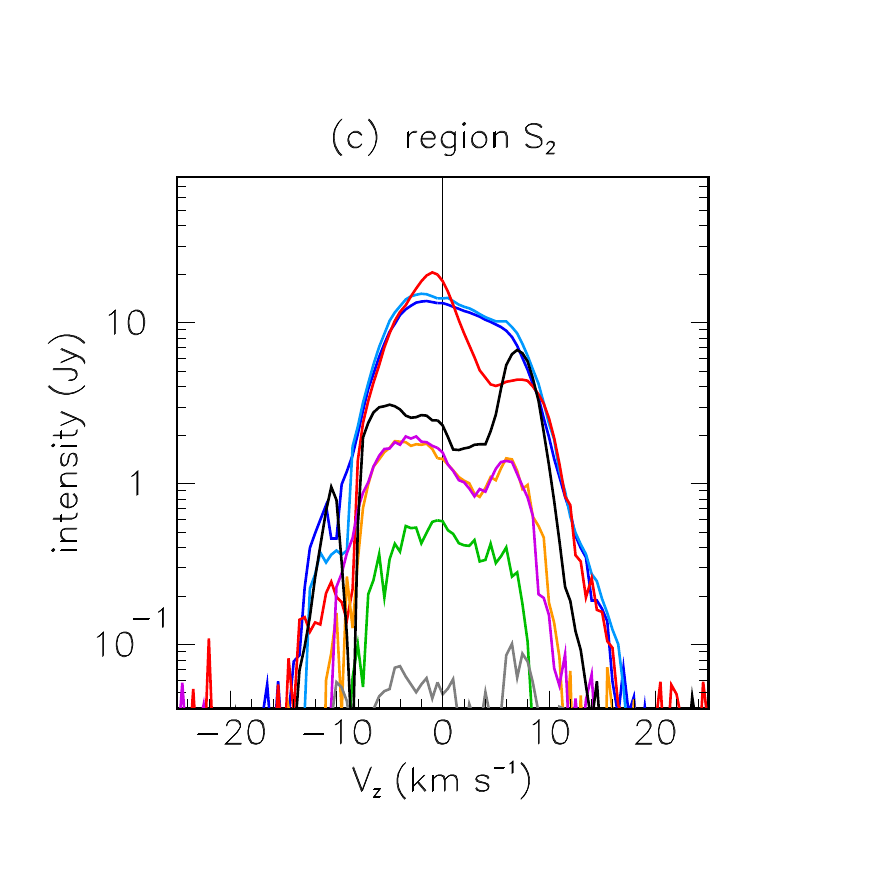}\\
  \includegraphics[width=0.27\linewidth, trim= 0.5cm 1.3cm 2.5cm 1.5cm,clip]{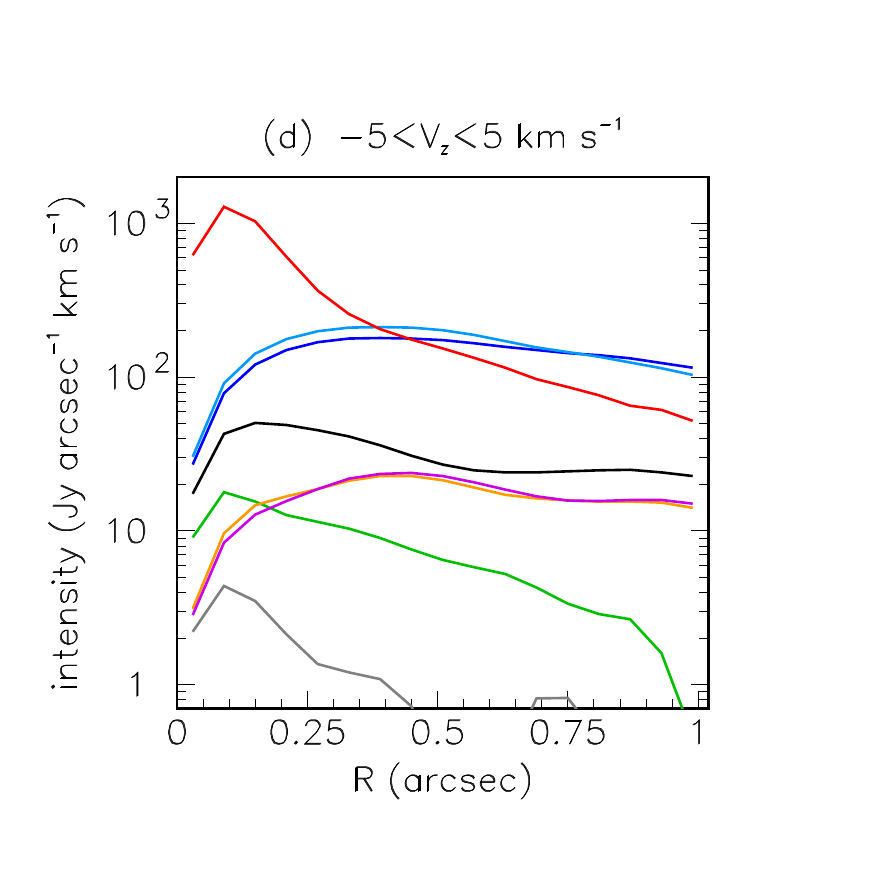}
  \includegraphics[width=0.27\linewidth, trim= 0.5cm 1.3cm 2.5cm 1.5cm,clip]{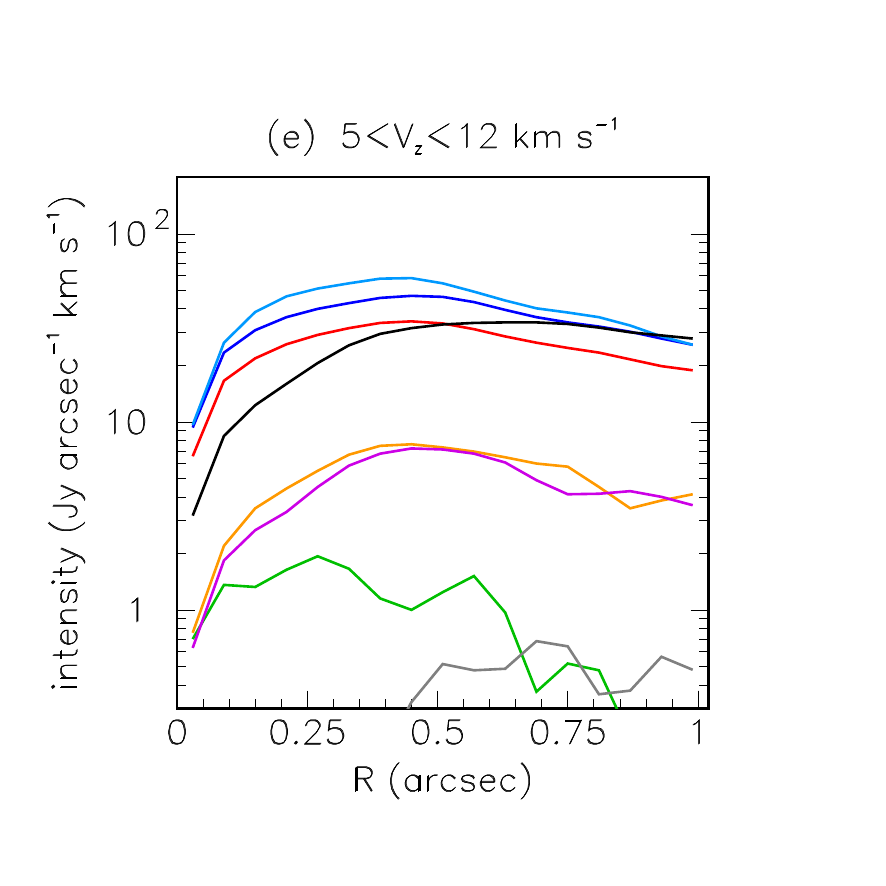}
  \caption{A-configuration data. a: Definition of regions S$_1$ and S$_2$; b and c: Doppler velocity spectra integrated over regions S$_1$ (b) and S$_2$ (c), respectively; d and e: $R$ distributions integrated over $|V_z|$$<$5 \kms\ (d) and 5$<$$V_z$$<$12 \kms\ (e), respectively. They are azimuthally integrated over the whole ring, such that the areas under the curves measure the total emission.}
  \label{fig5}
\end{figure*}

As will become apparent in what follows, the observations reveal different properties in the very close, unresolved neighbourhood of the star and farther away. We therefore distinguish between two complementary regions within the circle $R$$<$1 arcsec, S$_1$ and S$_2$ (Figure~\ref{fig5}a). S$_1$ corresponds approximately to twice the average beam size, namely 0.5$\times$0.3 arcsec$^2$ at a position angle of 32$^o$. S$_2$ is its complement in the circle $R$$<$1 arcsec. Moreover, we distinguish between the central Doppler velocity interval, $|V_z|$$<$5 \kms, and larger red-shifted Doppler velocities, 5$<$$V_z$$<$12 \kms, for which the effect of absorption is expected to be smaller than in the blue-shifted hemisphereAs will become apparent in Section 3.2, the observations reveal different properties in the very close, unresolved neighbourhood of the star and farther away.

Table~\ref{fig3} lists, for each line, the intensities measured in regions S$_1$ and S$_2$ in each Doppler velocity interval separately. Figure~\ref{fig5}b,c displays Doppler velocity spectra integrated over regions S$_1$ and S$_2$ separately and Figure~\ref{fig5}d,e displays radial distributions integrated over the low and high Doppler velocity intervals separately. The left panel shows a clear absorption peak on the blue-shifted side, revealing progressive acceleration up to a terminal velocity of $\sim$8-10 \kms. In S$_1$, all line emissions peak near systemic velocity, the H$^{12}$CN(3-2) line standing out as particularly narrow: its FWHM is only 4.0$\pm$0.2 \kms, meaning a $\sigma$ of $\sim$1.7$\pm$0.1 \kms\ for a bin size of 0.5 \kms; we study this feature in Sub-section 3.2. In S$_2$, the $^{12}$CO(2-1) spectrum shows, in addition to the central peak, enhancements near terminal velocity, affected by absorption on the blue-shifted side, typical of an extended spherical CSE having reached terminal velocity at short distances from the star. A similar red-shifted enhancement is also visible on other lines; as they cover much shorter radial distances, its relative intensity is significantly smaller.  In the higher Doppler velocity interval, Figure~\ref{fig5}e shows that all line emissions have similar radial dependences: we study this $V_z$ interval in Sub-section 3.3. In the low $|V_z|$ interval, the radial distributions displayed in Figure~\ref{fig5}d give again evidence for the H$^{12}$CN line standing out as particularly narrow, possibly revealing the dominance of a gravitationally bound static layer. It may also suggest a similar trend for the Si$^{17}$O(6-5) and the $^{13}$CO(2-1) lines, however with a much lower signal-to-noise ratio.

The observations reveal different properties in the very close, unresolved neighbourhood of the star (S$_1$) and farther away (S$_2$) The H$^{12}$CN(3-2) intensity is nearly an order of magnitude larger than the SiO intensities in S$_1$ and close to them in S$_2$; relative to $^{12}$CO(2-1) emission, H$^{13}$CN(3-2) emission is over three times larger in S$_2$ than in S$_1$. Comparing the relative values taken by the line emission intensities in the regions of the data-cube listed in Table \ref{tab3} provides useful information on the relative abundances of the associated molecules and on the mechanism of excitation of the upper states. However, a general concern, abundantly discussed in the published literature, is the impact of absorption on the line emission of densely populated molecules, such as $^{12}$C$^{16}$O and Si$^{16}$O. We compare in Sub-section 3.4 the Si$^{16}$O(5-4) and Si$^{16}$O(6-5) emissions and in Sub-section 3.5 the emissions of the $^{12}$C and $^{13}$C isotopologues of the CO(2-1) and HCN(3-2) molecular lines and the emissions of the $^{16}$O and $^{17}$O isotopologues of the SiO(6-5) molecular line. 

\begin{table*}
  \centering
  \caption{Intensities (Jy\,\kms) of line emissions integrated in two regions, S$_1$ and S$_2$, of the circle $R$$<$1 arcsec.}
    \label{tab3}
  \begin{tabular}{ccccccccc}
\hline
&$^{12}$CO(2-1)&
$^{13}$CO(2-1)&
Si$^{16}$O(5-4)&
Si$^{16}$O(6-5)&
Si$^{17}$O(6-5)&
H$^{12}$CN(3-2)&
H$^{13}$CN(3-2)&
CS(5-4)\\
\hline
$|V_z|$$<$5 \kms, S$_1$ &
7.2$\pm$0.1 &
0.6$\pm$0.1 &
15.8$\pm$0.1 &
18.3$\pm$0.2 &
2.6$\pm$0.2 &
183 &
1.9$\pm$0.2 &
1.7$\pm$0.2\\
$|V_z|$$<$5 \kms, S$_2$ &
23.9$\pm$0.5 &
0.5$\pm$0.5 &
124&
135&
4.7$\pm$0.7 &
124&
14.5$\pm$0.7 &
15.0$\pm$0.7\\
5$<$$V_z$$<$12 \kms, S$_1$ &
1.64$\pm$0.07 &
0.02$\pm$0.07 &
4.21$\pm$0.07 &
5.04$\pm$0.11 &
0.21$\pm$0.11 &
3.05$\pm$0.11 &
0.44$\pm$0.11 &
0.37$\pm$0.11\\
5$<$$V_z$$<$12 \kms, S$_2$ &
23.7$\pm$0.4 &
0.3$\pm$0.4 &
30.3$\pm$0.4 &
35.6$\pm$0.6 &
0.7$\pm$0.6 &
22.0$\pm$0.6 &
4.6$\pm$0.6 &
4.2$\pm$0.6\\
$|V_z|$$<$20 \kms &
66.6$\pm$0.9 &
1.8$\pm$0.9 &
195$\pm$1 &
215$\pm$2 &
9.2$\pm$1.4 &
348$\pm$2 &
26$\pm$2 &
26$\pm$2\\
\hline
  \end{tabular}

  \tablefoot{S$_1$ is defined as the interior of an ellipse having axes of 0.5 and 0.3 arcsec at a position angle of 32$^o$. S$_2$ is the complement of S$_1$ in the circle $R$$<$1 arcsec (Figure~\ref{fig5}a).  In each region, two Doppler velocity intervals are considered: $|V_z|$$<$5 \kms\ and 5$<$$V_z$$<$12 \kms. Uncertainties are at 3$\sigma$ of the noise level (Table~\ref{tab2}). All values are for 2023 A-configuration observations, with the exception of $^{12,13}$CO(2-1) data, which are merged 2023 A+B data.}
\end{table*}

\subsection{Low Doppler velocities: the HCN singularity}
\begin{figure*}
  \centering
  \includegraphics[width=0.23\linewidth, trim= 0.8cm 1.1cm 2.5cm 1.cm,clip]{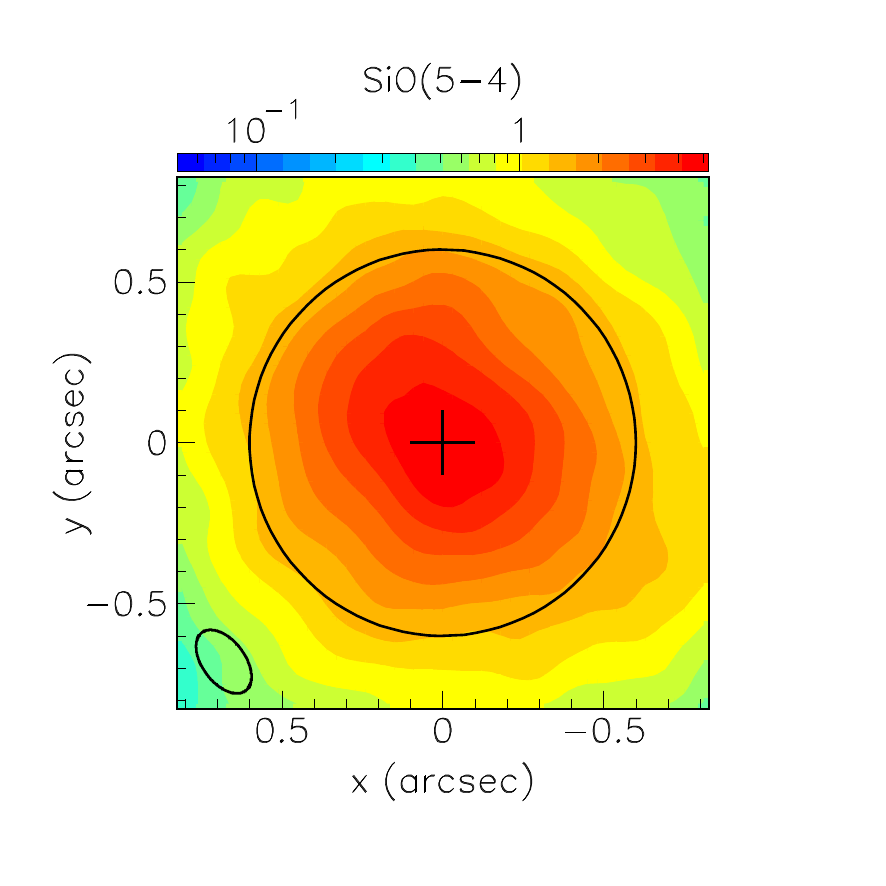}
  \includegraphics[width=0.23\linewidth, trim= 0.8cm 1.1cm 2.5cm 1.cm,clip]{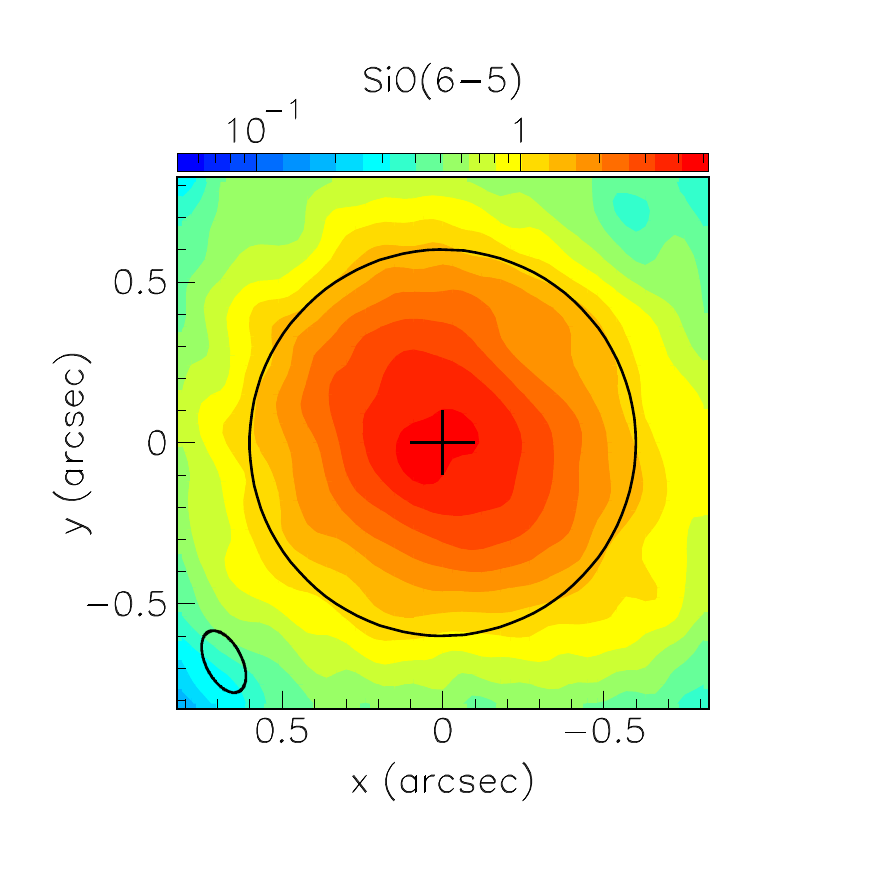}
  \includegraphics[width=0.23\linewidth, trim= 0.8cm 1.1cm 2.5cm 1.cm,clip]{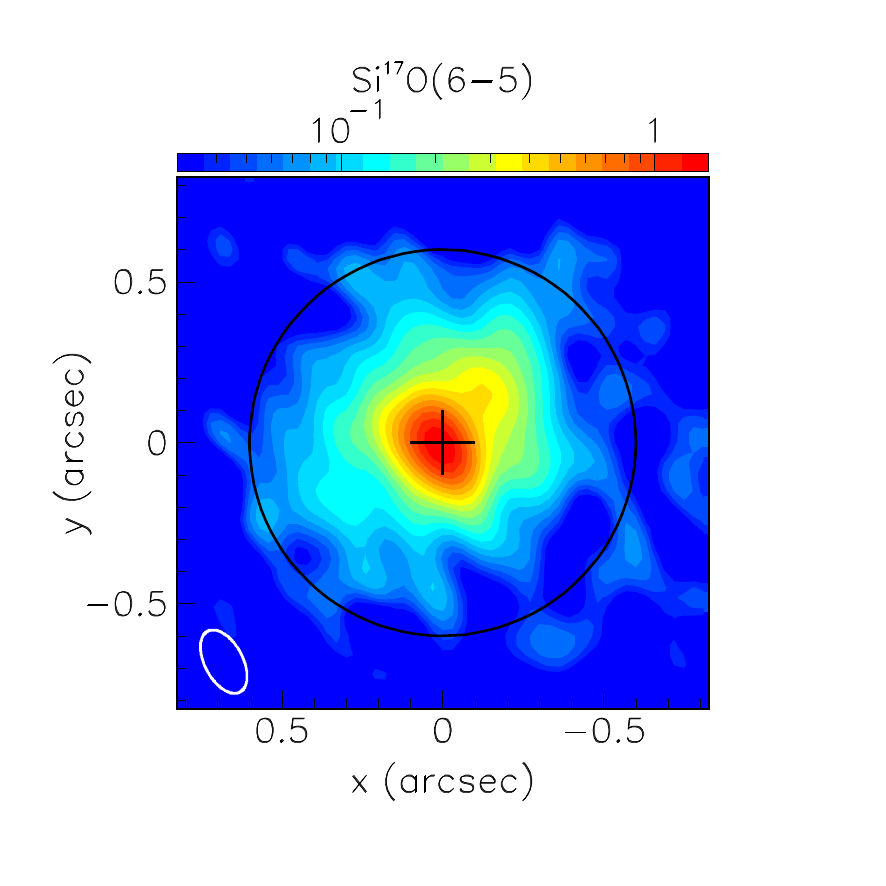}
  \includegraphics[width=0.23\linewidth, trim= 0.8cm 1.1cm 2.5cm 1.cm,clip]{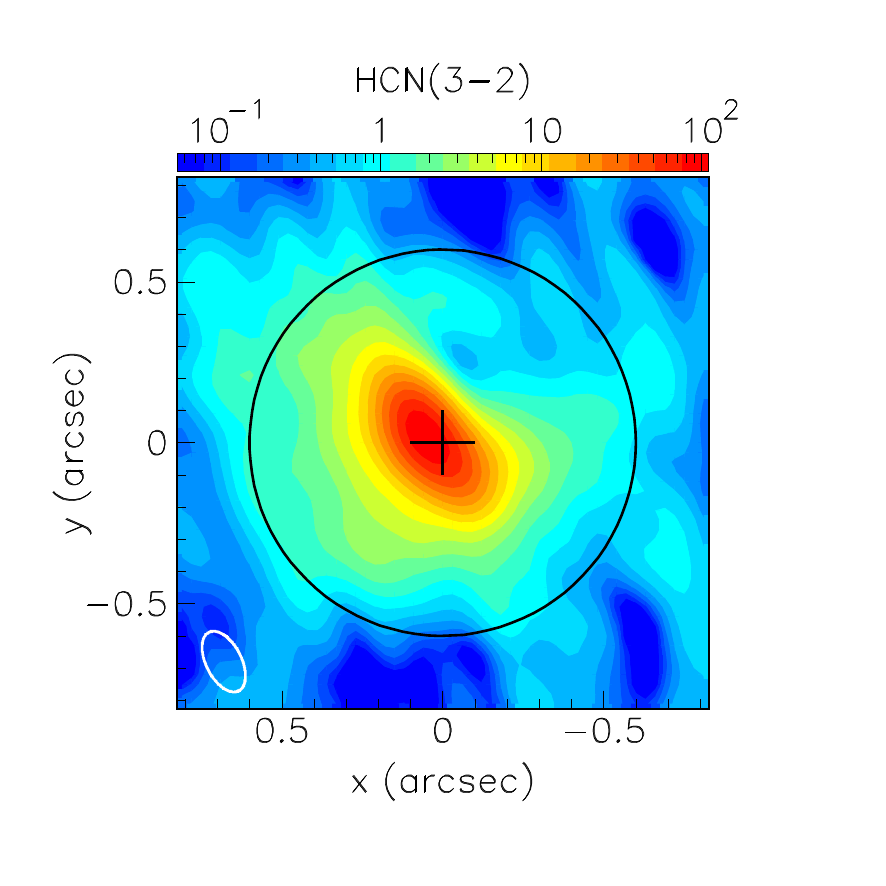}\\
  \includegraphics[width=0.23\linewidth, trim= 0.8cm 1.1cm 2.5cm 1.cm,clip]{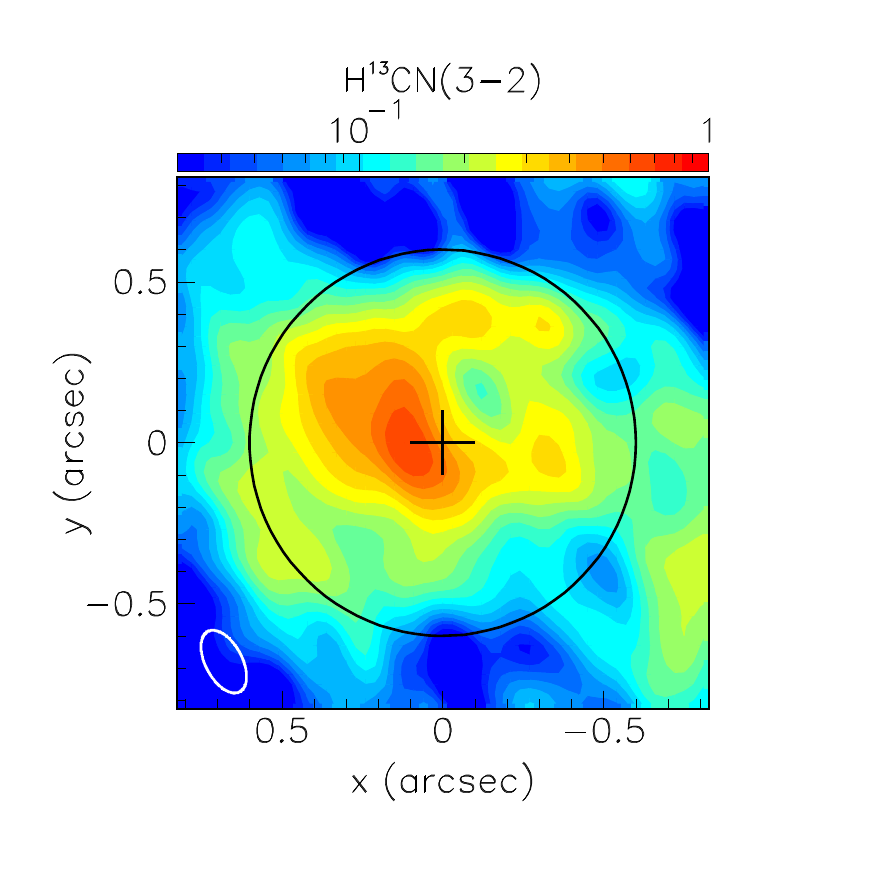}
  \includegraphics[width=0.23\linewidth, trim= 0.8cm 1.1cm 2.5cm 1.cm,clip]{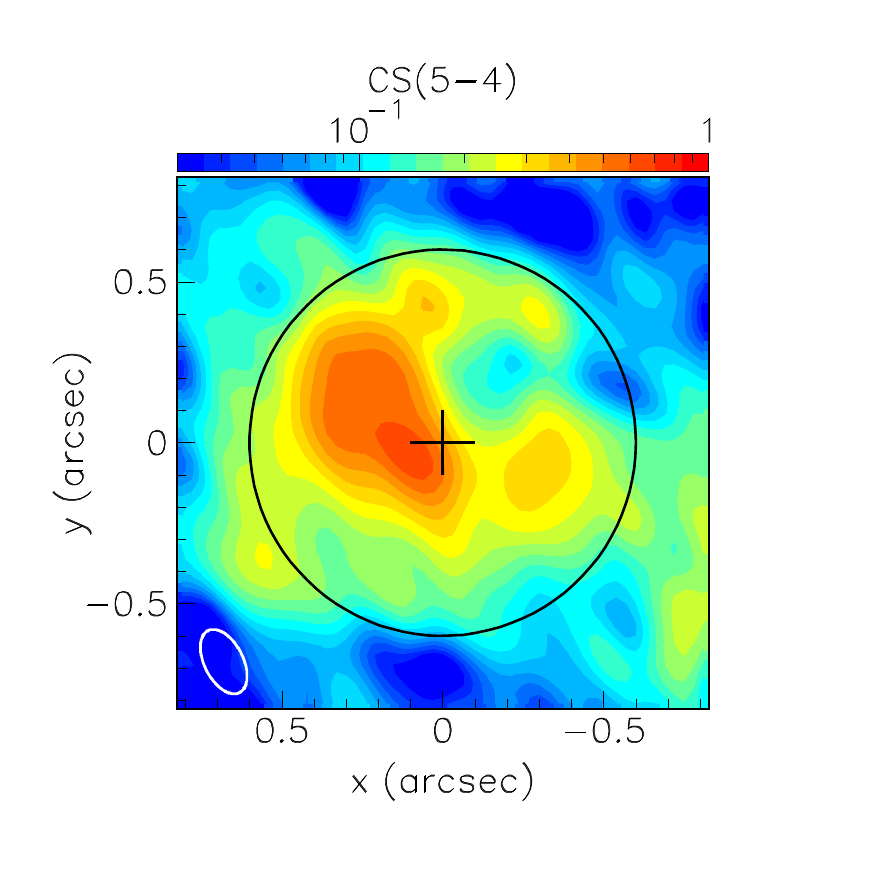}
  \includegraphics[width=0.23\linewidth, trim= 0.8cm 1.1cm 2.5cm 1.cm,clip]{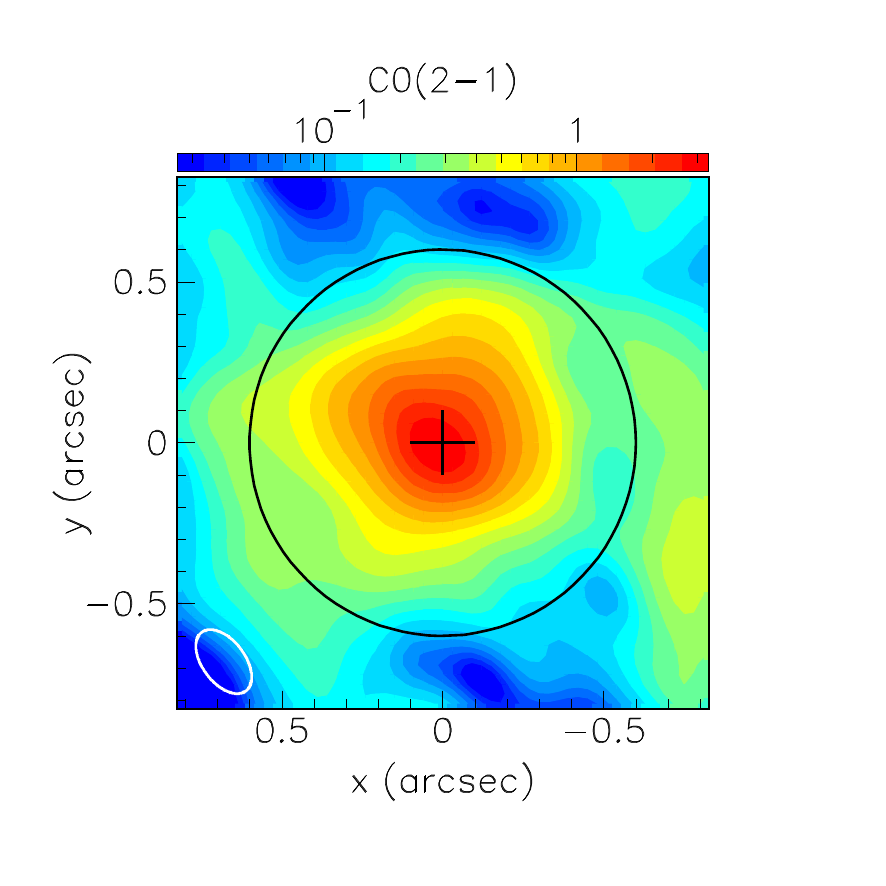}
  \includegraphics[width=0.23\linewidth, trim= 0.8cm 1.1cm 2.5cm 1.cm,clip]{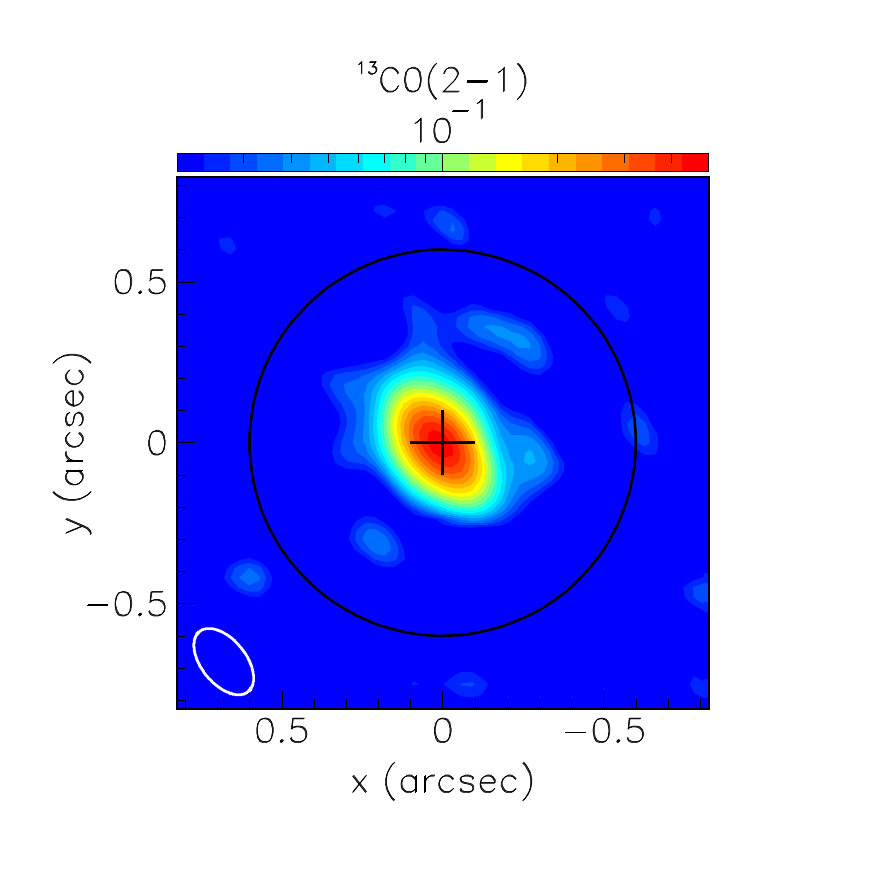}
  \caption{Intensity maps integrated over $|V_z|$$<$5 \kms. Lines are indicated on top of each panel. Circles of radius 0.6 arcsec show the region of reliable imaging. The beam (FWHM) is shown in the lower-left corner of each panel. The colour scale is in units of Jy\,beam$^{-1}$\kms. The mean noise level is 15 mJy\,beam$^{-1}$\kms. }
  \label{fig6}
\end{figure*}

Using A-configuration data, maps of the line intensities integrated over the $|V_z|$$<$5 \kms\ interval are displayed in Figure~\ref{fig6} for each line separately. In contrast to $^{12}$CO(2-1) emission, which was shown by \citet{CastroCarrizo2010} to extend up to $\sim$40 arcsec and which we study in Section 4, the Si$^{16}$O line emission was shown earlier by \citet{Lucas1992} and \citet{Ramstedt2009} to be confined to short distances from the star, the latter authors quoting a half-maximum radius of 1.6 arcsec. Similarly, \citet{Schoier2013} quote a half-maximum radius of 1.8$\pm$0.8 arcsec for H$^{12}$CN emission. The confinement of SiO molecules is a general feature of M- and S-types AGB stars, commonly interpreted as the result of their condensation on dust grains and of photo-dissociation. We see from the intensity maps displayed in Figure~\ref{fig6} and from the spectra displayed in Figure~\ref{fig5}b that all lines give evidence for the presence of a slowly expanding reservoir of gas surrounding the star and covering a radial range similar to, or smaller than that covered by Si$^{16}$O emission.

\begin{figure*}
  \centering
  \includegraphics[width=0.32\linewidth, trim= 0.8cm 1.1cm 0.5cm 1.cm,clip]{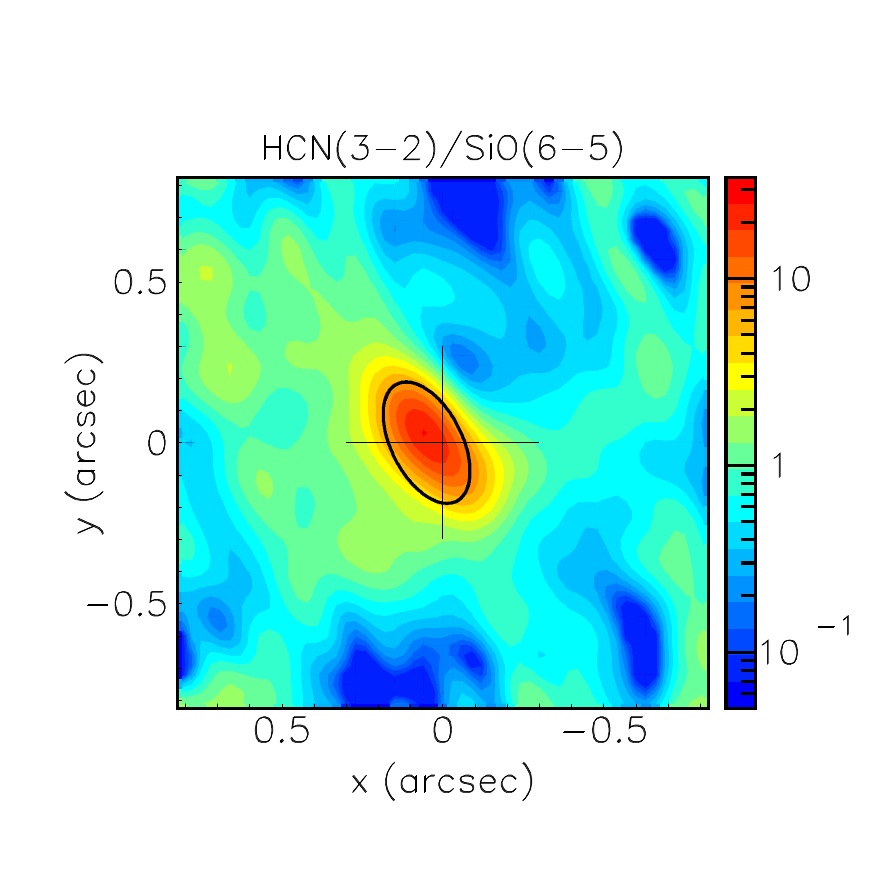}
  \includegraphics[width=0.32\linewidth, trim= 0.8cm 1.1cm 0.5cm 1.cm,clip]{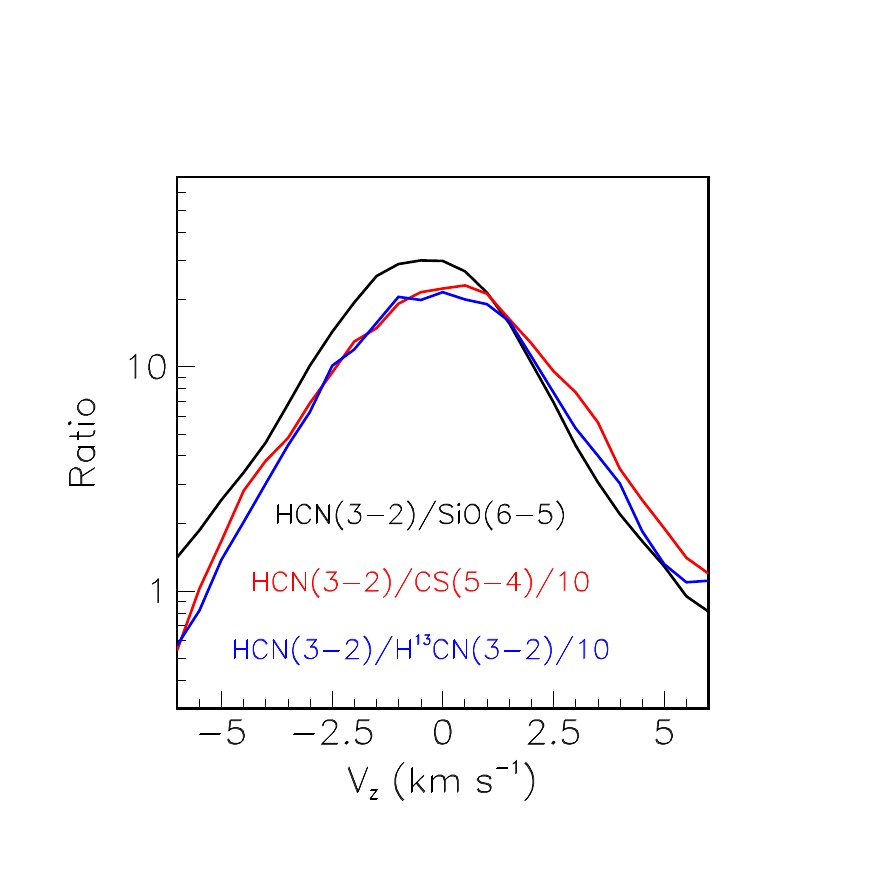}
  \caption{Left: ratio of H$^{12}$CN(3-2) over Si$^{16}$O(6-5) intensities integrated over $|V_z|$$<$5 \kms. The ellipse has a size of 2$\times$2=4 beam FWHM area, centred at 50 mas east. Right: ratio of the Doppler velocity spectra of H$^{12}$CN(3-2) to Si$^{16}$O(6-5) (black), H$^{13}$CN(3-2) (blue, divided by 10) and $^{12}$CS(5-4) (red, divided by 10) emissions integrated over the ellipse shown on the left panel.}
  \label{fig7}
\end{figure*}

As was shown in the preceding sub-section by the radial distributions displayed in Figure~\ref{fig5}d, we observe an outstanding confinement of H$^{12}$CN(3-2) emission in the immediate neighbourhood of the star. We further illustrate its singularity by displaying in Figure~\ref{fig7} a map and Doppler velocity distributions of the ratio of its emission to those of the Si$^{16}$O(6-5), H$^{13}$CN(3-2) and $^{12}$CS(5-4) lines, all using A-configuration data. \citet{Duari2000} had previously inferred from the intensity of emission of high excitation transitions that H$^{12}$CN molecules were formed within 200-300 mas from the centre of the star, but their single-dish observations could not reveal that the emission of low excitation lines would be confined to such short distances. The fact that all high-frequency emissions are simultaneously observed in the same sideband, implying in particular a common calibration, leaves little room for an explanation in terms of instrumental or data processing flaws. As neither H$^{13}$CN(3-2) nor $^{12}$CS(5-4) emission displays such a spectacular confinement near the photosphere as H$^{12}$CN(3-2) emission does, the singularity of the latter is not a straightforward result of HCN being a carbon-bearing molecule, enhanced by the large C/O ratio of the star. In summary, we can state with confidence that in the 2023 A-configuration data, the interval of low Doppler velocities, $|V_z|$$<$5 \kms, hosts an H$^{12}$CN(3-2) emission component that has no significant H$^{13}$CN(3-2) counterpart, has both a narrow $V_z$ width and a narrow spatial extension, and suggests the presence of an unresolved, gravitationally-bound, static layer.

In contrast, the H$^{12}$CN(3-2) line emission observed in 2024 using B-configuration, as was already apparent from Figure~\ref{fig3}, shows quite a different picture. Comparing the H$^{12}$CN(3-2) emissions observed in S$_1$ and S$_2$ in the $|V_z|$$<$5 \kms\ interval, we find that their ratio is $\sim$1.5 in 2023 and $\sim$0.2 in 2024 while their sum increases by only 7\%.  This suggests that the total emission has simply been smeared around the star between 2023 and 2024 but that the total emission has stayed approximately constant, as illustrated in the lower-left panel of Figure~\ref{fig3} by comparing Doppler velocity spectra observed in 2023 and 2024.   The beam associated with the B-configuration 2024 observations covers region S$_1$, implying that some emission having its source within this region is detected in S$_2$, causing a decrease of the detected S$_1$/S$_2$ emission ratio. To tell apart this instrumental effect from the actual expansion of the emission, we smeared the radial distribution observed in 2023 using the A-configuration to predict what it would produce using the B-configuration. Comparing the result with the actual 2024 observation, we estimate that the radial extension of the H$^{12}$CN(3-2) emission has increased by $\sim$0.1 arcsec between the two years. If such an increase were caused by an isotropic expansion of the gas volume, it would mean an expansion velocity at the scale of $\sim$80 \kms, conflicting with the observed narrow width of the line profile, $\sim$$\pm$2 \kms.

Such singular behaviour could not have been revealed by earlier single-dish observations. The ratio between the H$^{12}$CN(3-2) and Si$^{16}$O(5-4) line intensities measured by \citet{Bieging2000} in 1998 on $\chi$ Cygni using the Heinrich Hertz Submillimeter Telescope was 12.3/10.8=1.14, and these authors quote an average value of 1.2$\pm$0.5 for S stars. This result, being obtained with a single-dish telescope, covers a broad angular range, at variance with the present observations, making the comparison difficult. The H$^{12}$CN(3-2)/Si$^{16}$O(5-4) emission ratios that we measure over the whole Doppler velocity interval, $|V_z|$$<$20 \kms, vary between 1.78 and 1.55, the larger value corresponding to a circle $R$$<$1 arcsec in the A-configuration and the lower value to a circle $R$$<$2 arcsec in the B-configuration (namely in 2023 for SiO and in 2024 for HCN). These are, in all cases, well above the 1998 value of 1.14 measured by Bieging et al. Taken at face value, this suggests that the singular component of HCN emission that we observe is absent from the 1998 observation and changed in morphology, but not in total intensity, between 2023 and 2024. Indeed, the lower row of Figure~\ref{fig3} shows that in the high velocity channel, $V_z$$=$7 \kms, the H$^{12}$CN(3-2) and Si$^{16}$O(5-4) emissions are very close from each other, but in the low velocity channel, $V_z$$=$0, the two lines display very different distributions. This may suggest that H$ ^{12}$CN(3-2) and Si$^{16}$O(5-4) emissions would have a similar morphology if one could ignore the enhanced H$^{12}$CN(3-2) emission confined to low Doppler velocities. It raises the question of the evolution of the situation in the years to come: will the enhanced H$^{12}$CN(3-2) emission disappear, will it stay at its 2024 level, will it evolve to yet an other morphology? To answer such a question, new A-configuration observations need to be made in 2025; B-configuration observations would improve the $uv$ coverage but they should be made close in time to allow for reliable merging. In the present situation, we are left with a major question that we are unable to answer with sufficient confidence: does the confined H$^{12}$CN(3-2) enhanced emission observed in 2023-2024 reveal the presence close to the photosphere of an enhanced molecular abundance or simply an enhanced excitation of the line, due to masing and/or shock dynamics? 

\subsection{The red-shifted north-western octant: evidence for mass ejection}
\begin{figure*}
  \centering
\includegraphics[width=0.23\linewidth, trim= 0.9cm 1.1cm 2.3cm 1.cm,clip]{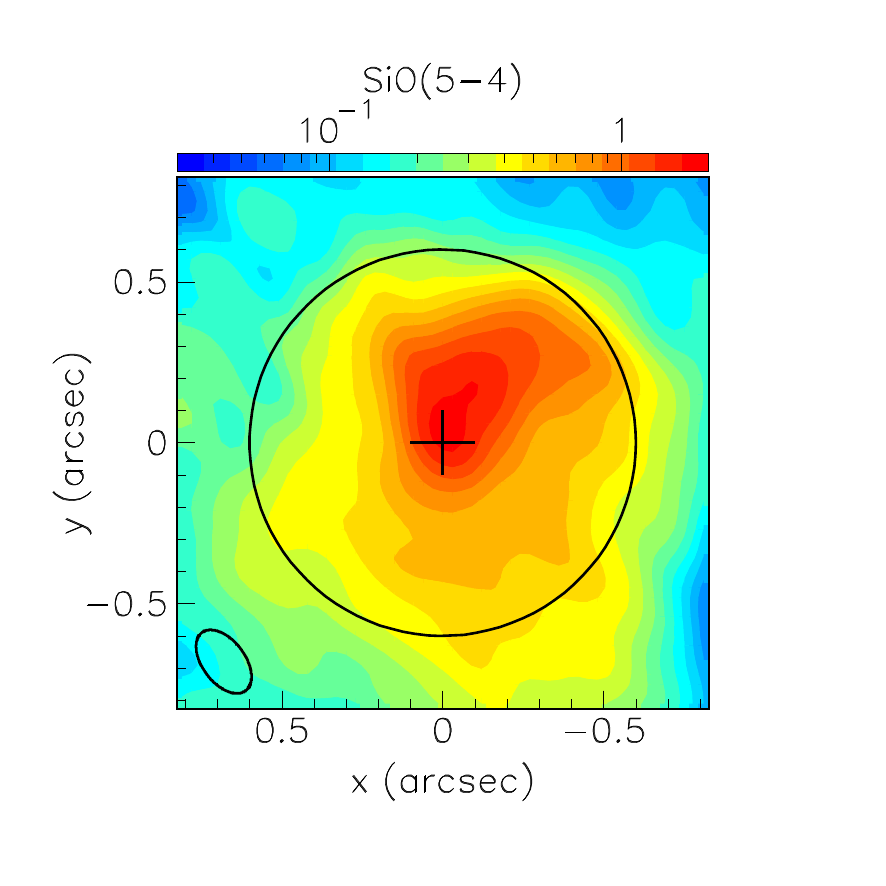}
\includegraphics[width=0.23\linewidth, trim= 0.9cm 1.1cm 2.3cm 1.cm,clip]{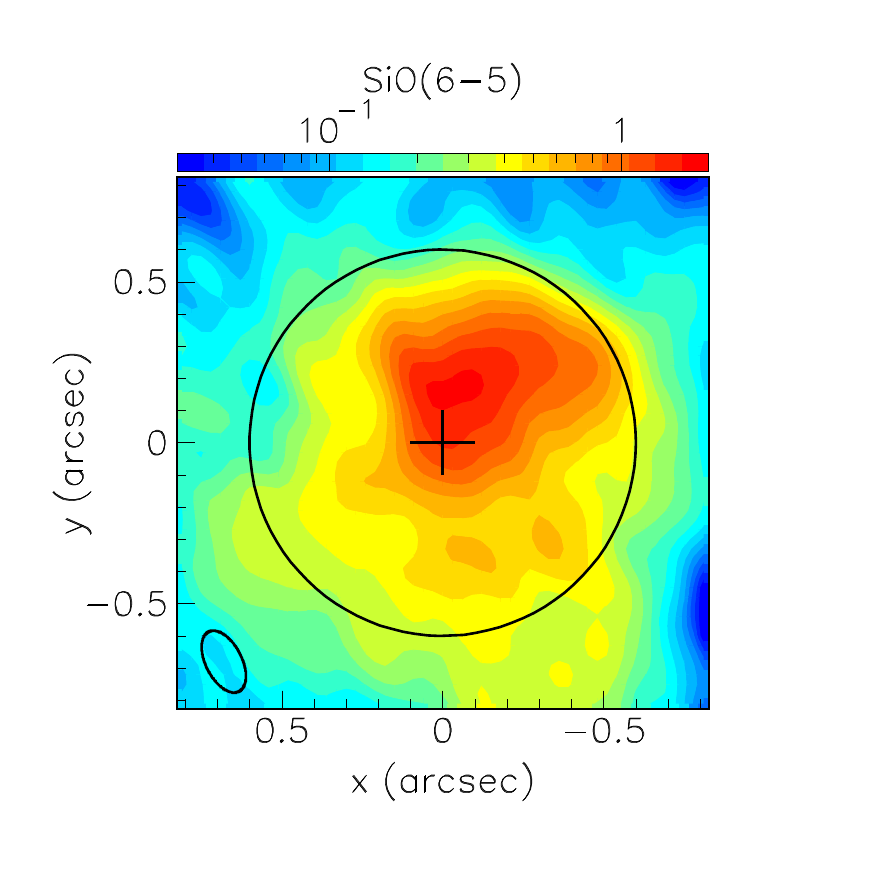}
\includegraphics[width=0.23\linewidth, trim= 0.9cm 1.1cm 2.3cm 1.cm,clip]{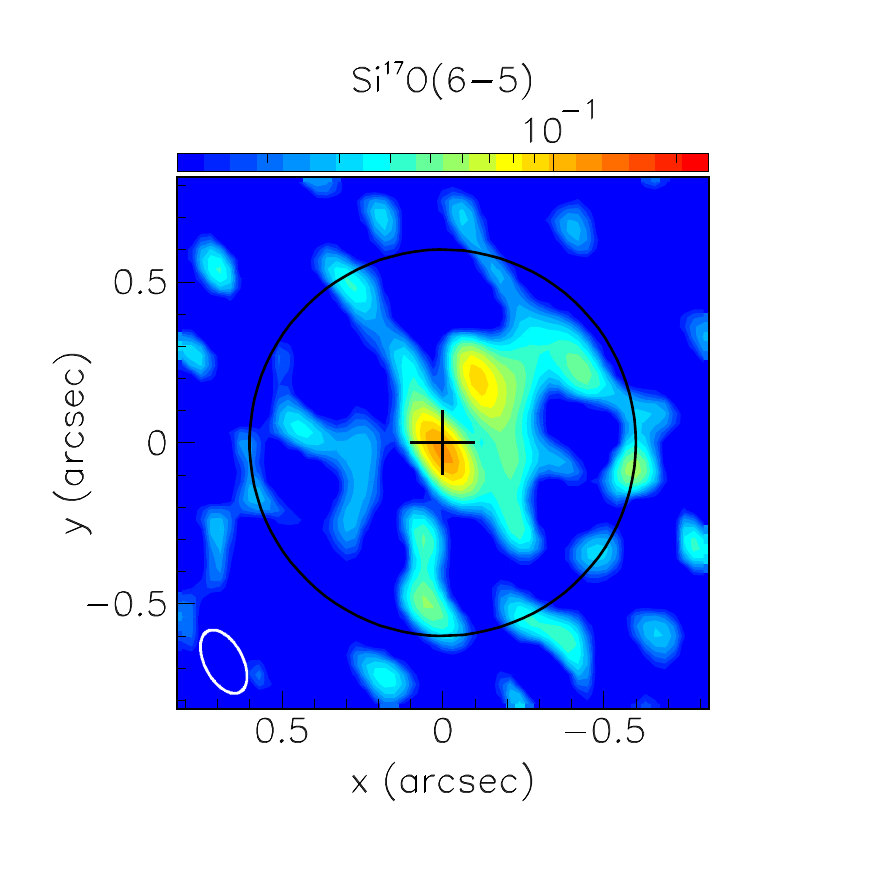}
\includegraphics[width=0.23\linewidth, trim= 0.9cm 1.1cm 2.3cm 1.cm,clip]{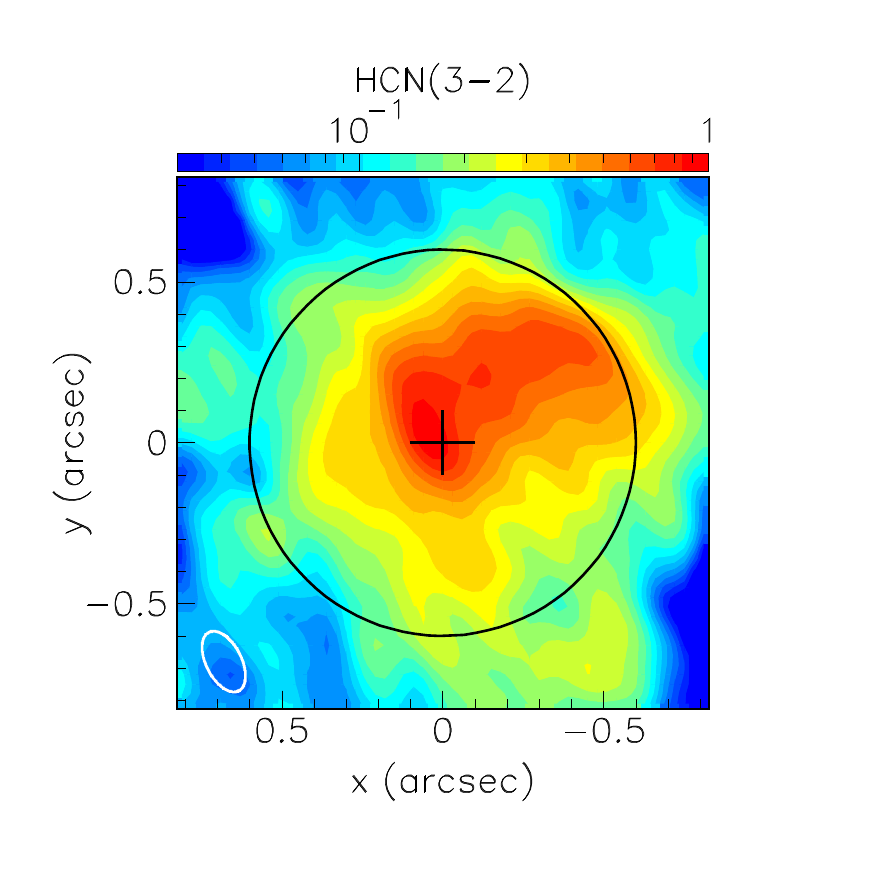}\\
\includegraphics[width=0.23\linewidth, trim= 0.9cm 1.1cm 2.3cm 1.cm,clip]{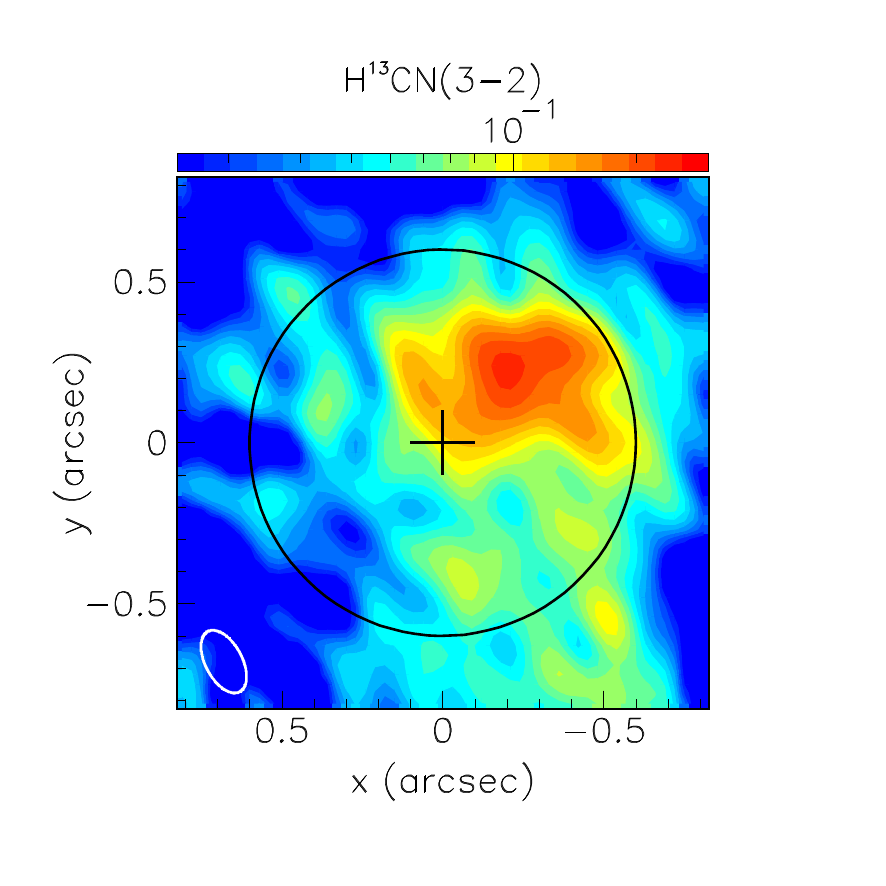}
\includegraphics[width=0.23\linewidth, trim= 0.9cm 1.1cm 2.3cm 1.cm,clip]{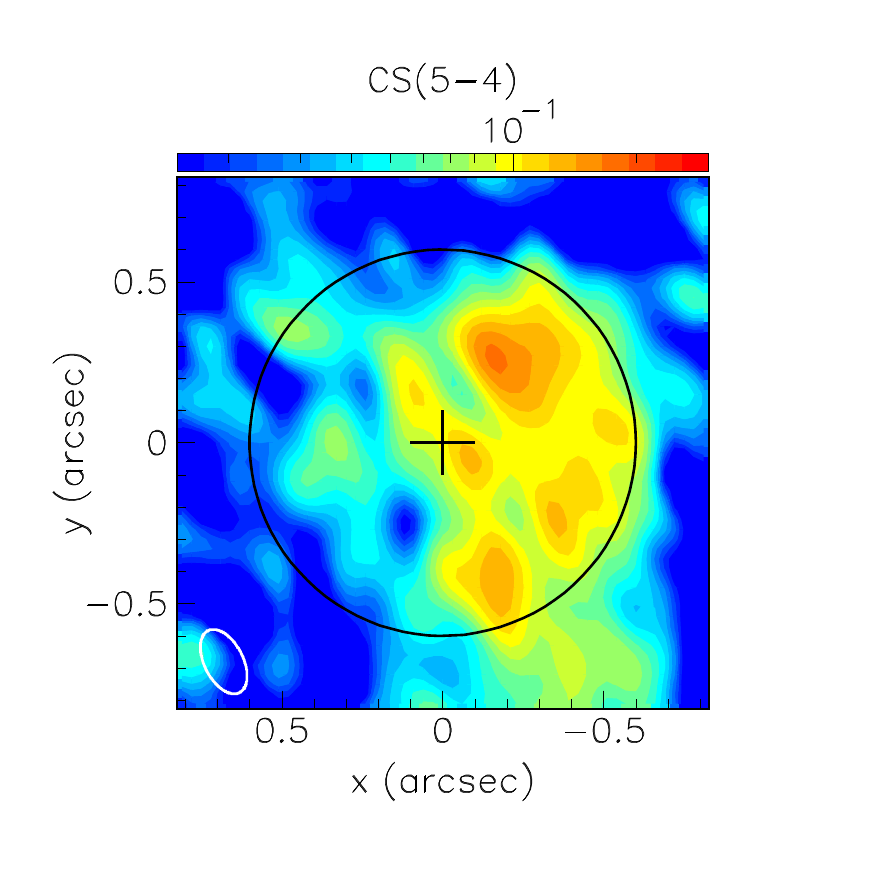}
\includegraphics[width=0.23\linewidth, trim= 0.9cm 1.1cm 2.3cm 1.cm,clip]{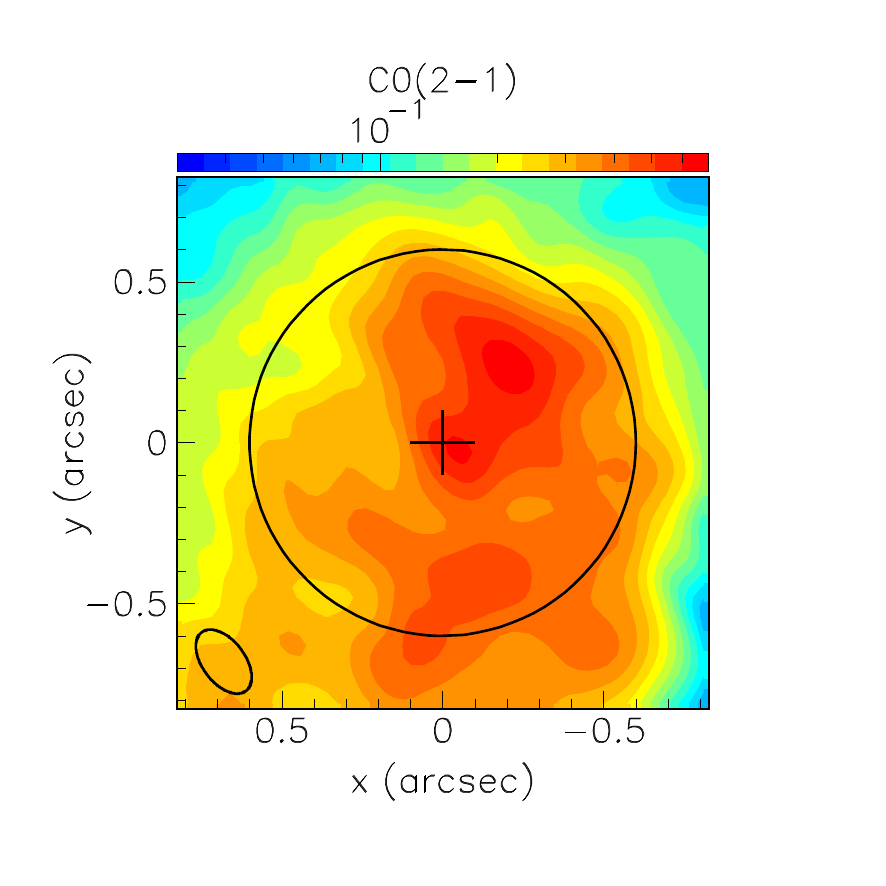}
\includegraphics[width=0.23\linewidth, trim= 0.9cm 1.1cm 2.3cm 1.cm,clip]{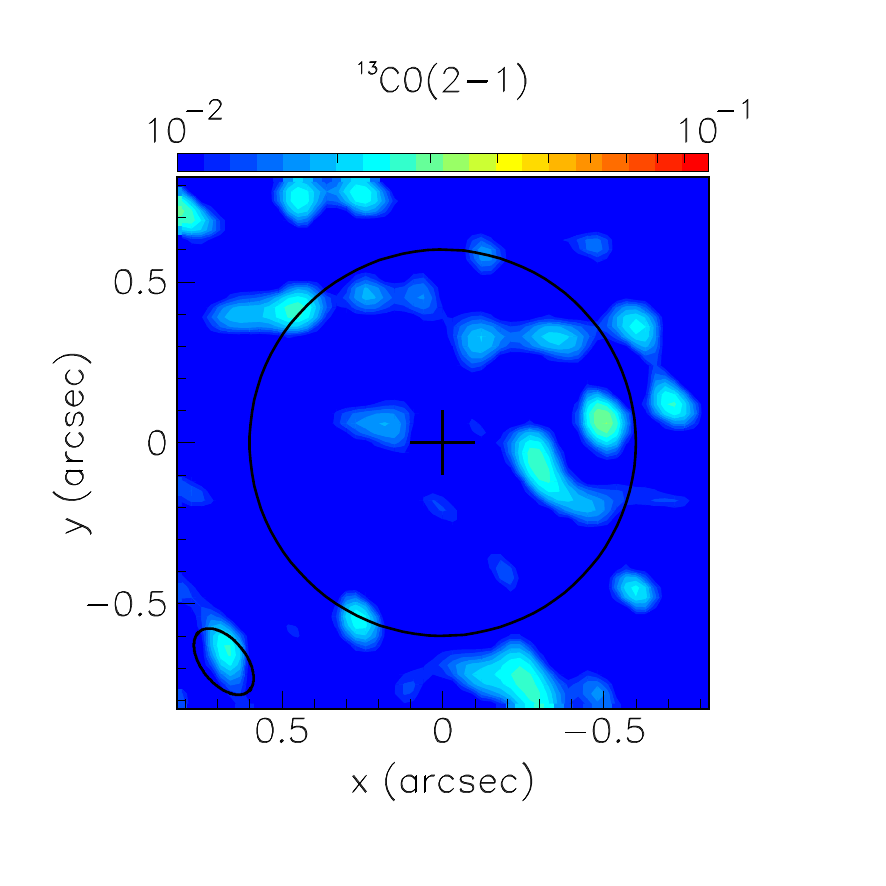}

  \caption{ Intensity maps integrated over 5$<$$V_z$$<$12 \kms\ (A-configuration 2023 observations). Lines are indicated on top of each panel. Circles of radius 0.6 arcsec show the region of reliable imaging. The beam (FWHM) is shown in the lower-left corner of each panel. The colour scale is in units of Jy\,beam$^{-1}$\kms.  The mean noise level is 12 mJy\,beam$^{-1}$\kms.}
  \label{fig8}
\end{figure*}

\begin{figure*}
  \centering
  
  \includegraphics[width=0.23\linewidth, trim= 0.9cm 1.1cm 2.3cm 1.cm,clip]{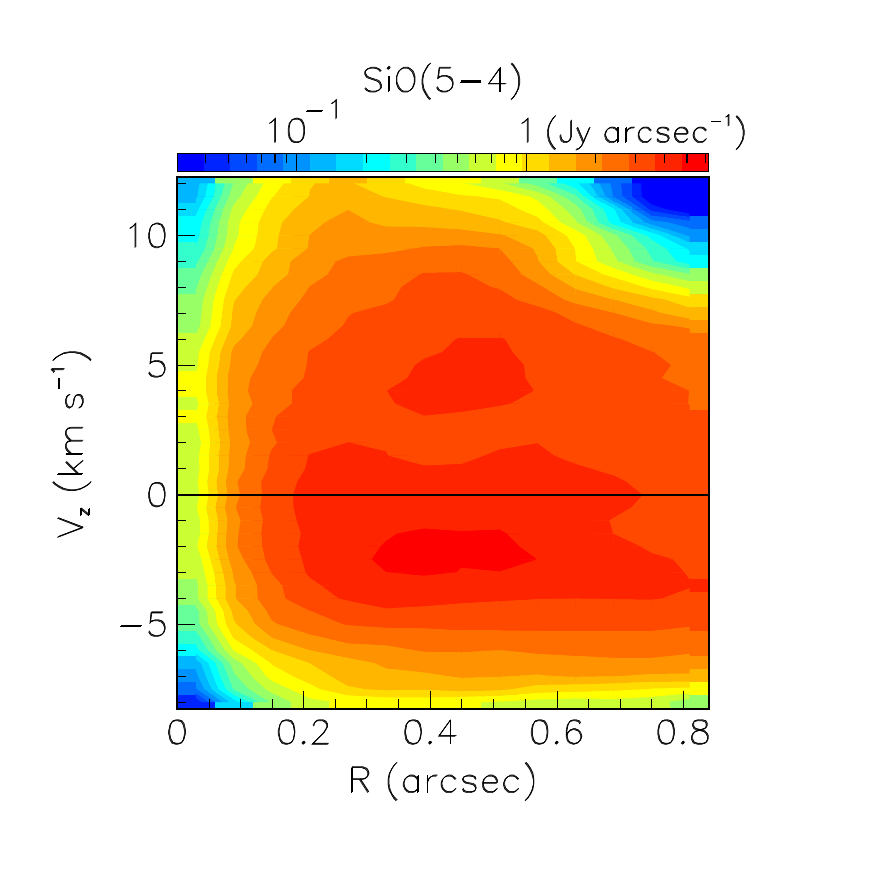}
  \includegraphics[width=0.23\linewidth, trim= 0.9cm 1.1cm 2.3cm 1.cm,clip]{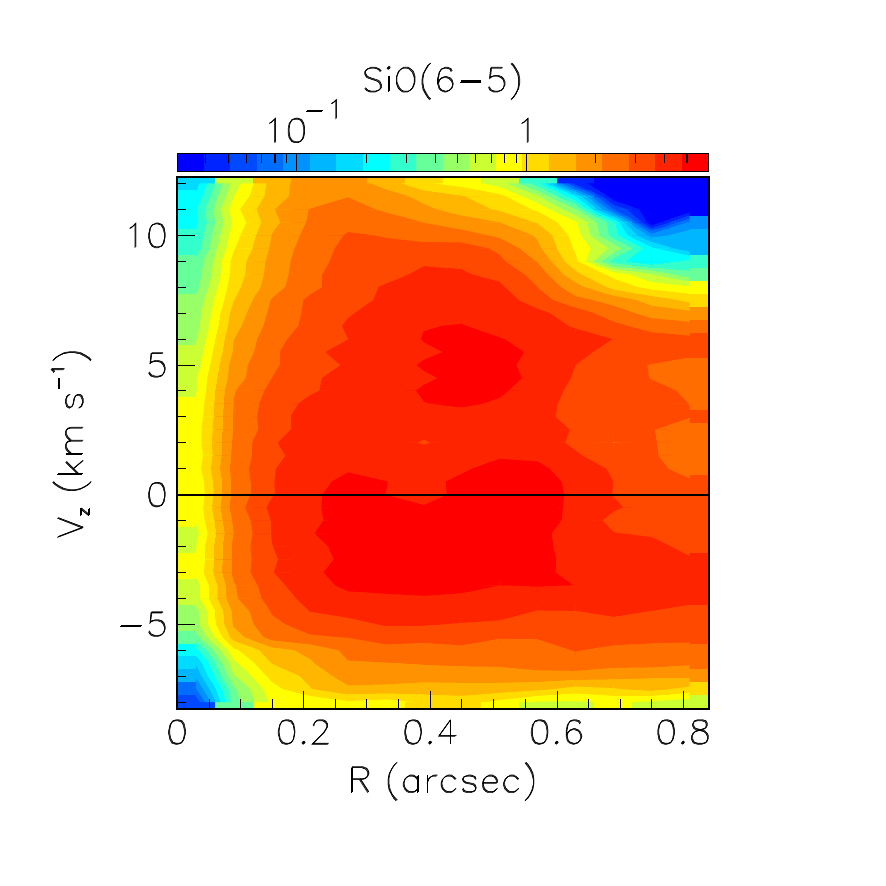}
  \includegraphics[width=0.23\linewidth, trim= 0.9cm 1.1cm 2.3cm 1.cm,clip]{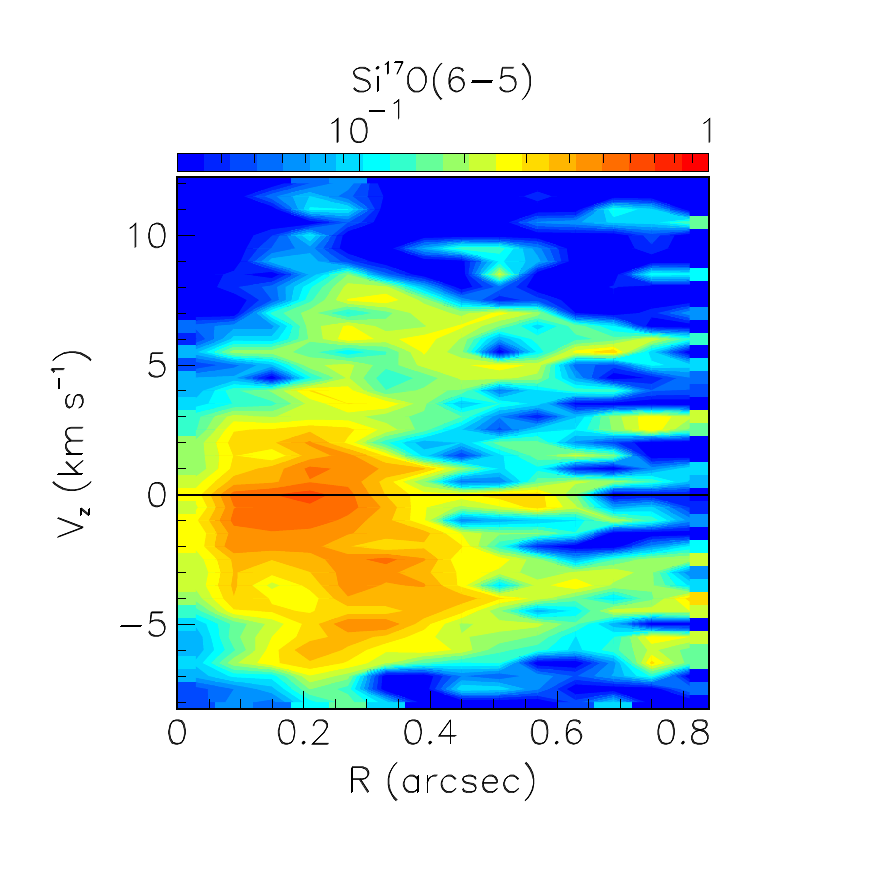}
  \includegraphics[width=0.23\linewidth, trim= 0.9cm 1.1cm 2.3cm 1.cm,clip]{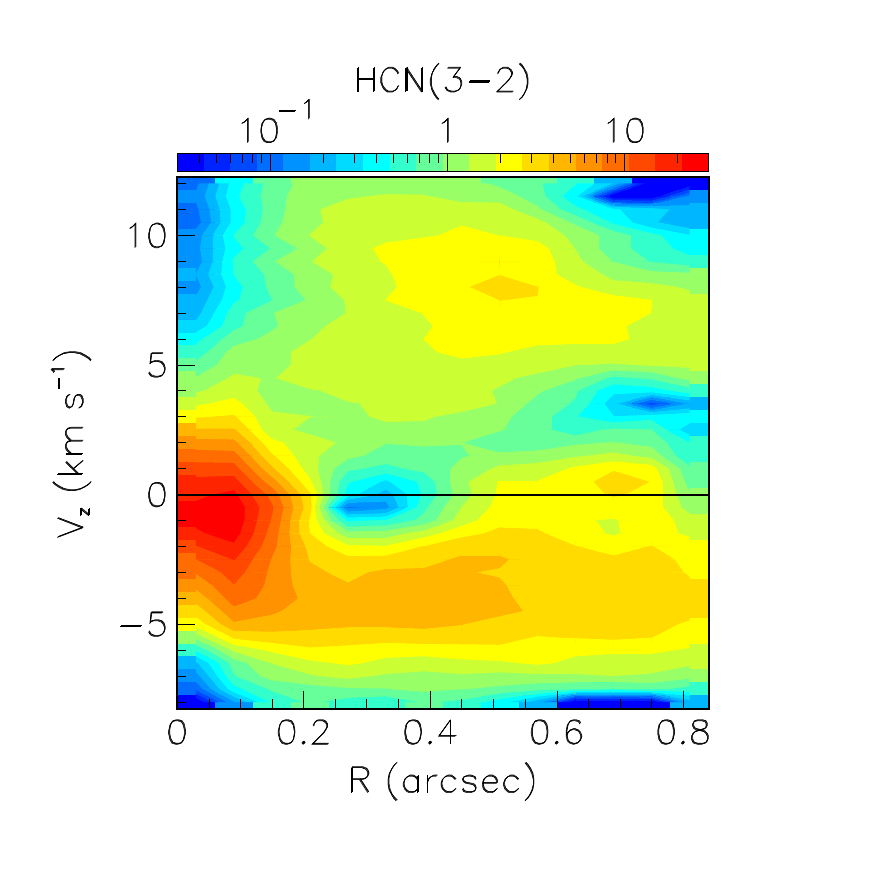}\\
  \includegraphics[width=0.23\linewidth, trim= 0.9cm 1.1cm 2.3cm 1.cm,clip]{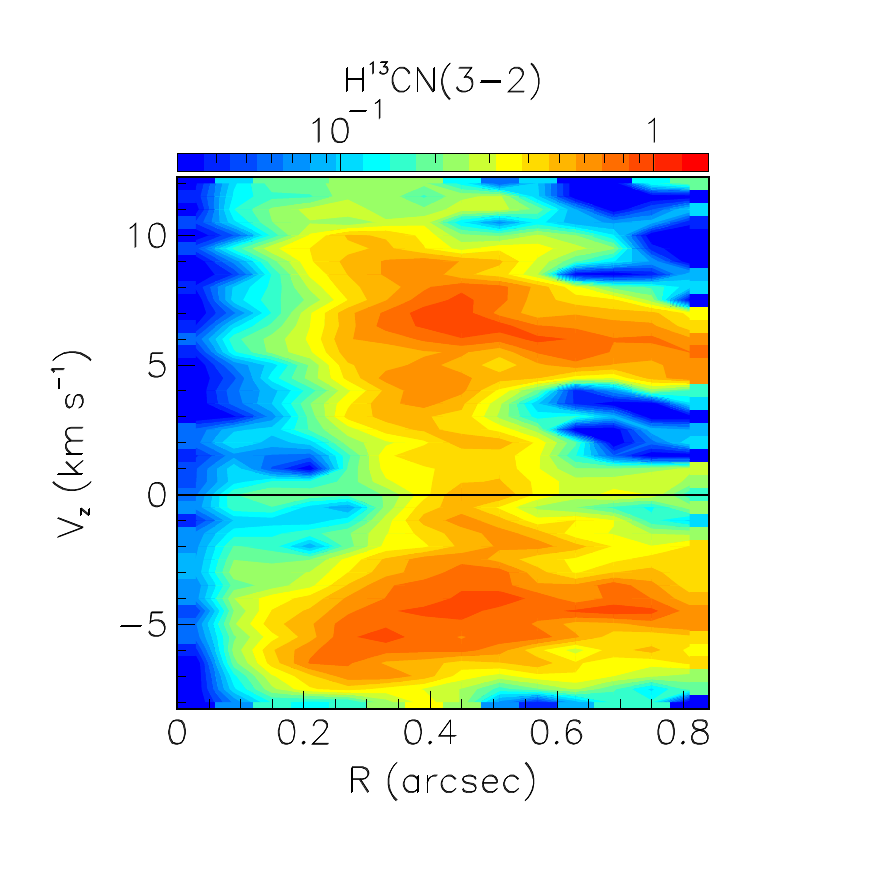}
  \includegraphics[width=0.23\linewidth, trim= 0.9cm 1.1cm 2.3cm 1.cm,clip]{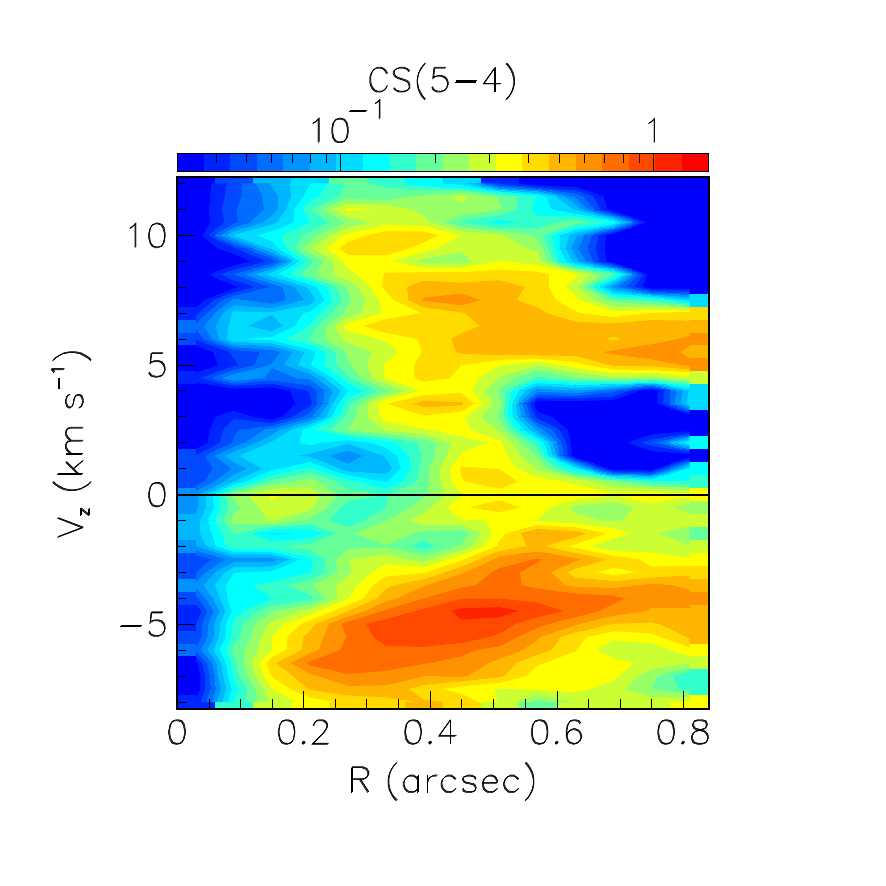}
  \includegraphics[width=0.23\linewidth, trim= 0.9cm 1.1cm 2.3cm 1.cm,clip]{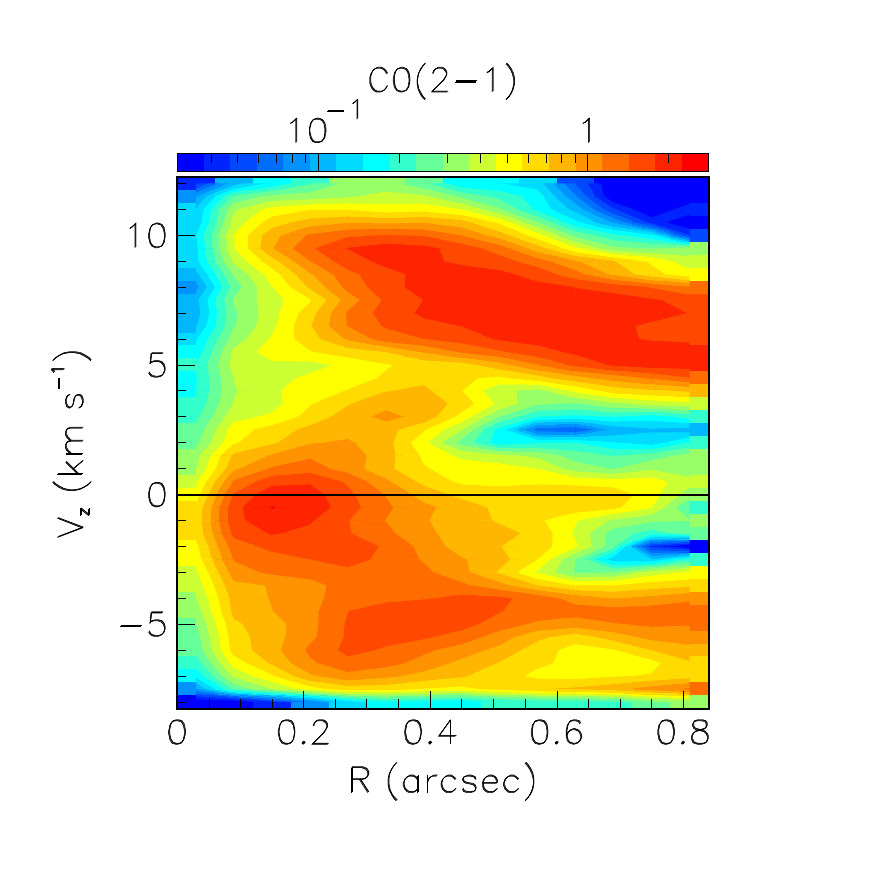}
  \includegraphics[width=0.23\linewidth, trim= 0.9cm 1.1cm 2.3cm 1.cm,clip]{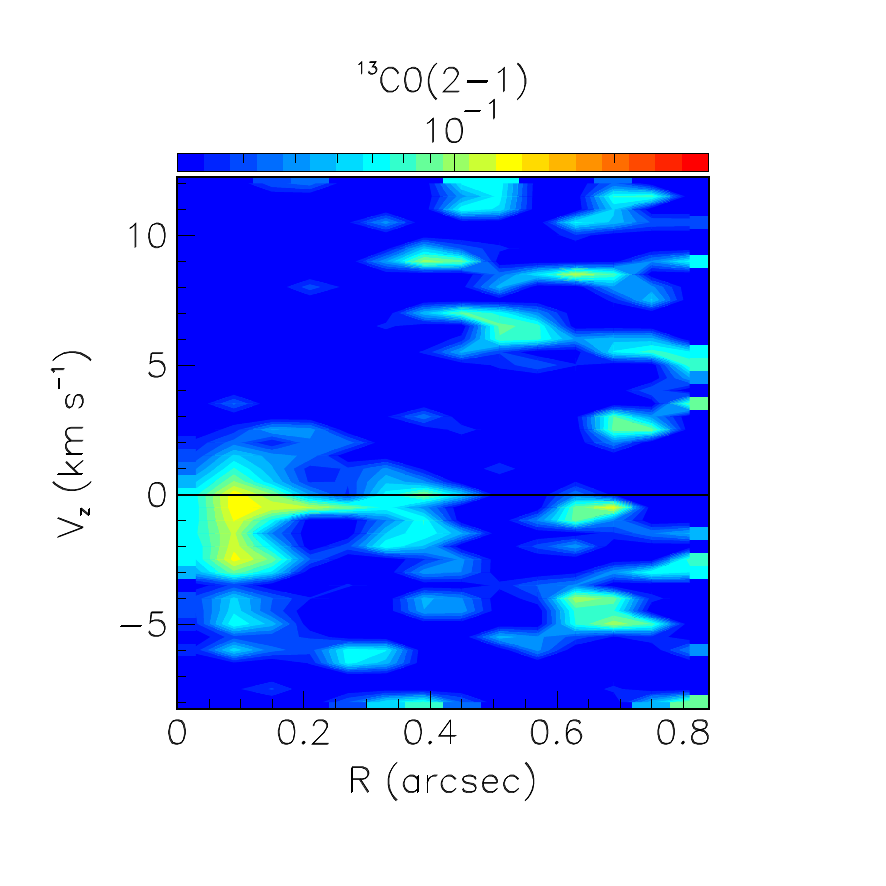}
  \caption{$V_z$ vs $R$ maps integrated over 270\dego$<$$\omega$$<$360\dego (A-configuration 2023 observations). Lines are indicated on top of each panel. The colour scale is in units of Jy\,arcsec$^{-1}$. The mean noise level is 0.14 Jy\,arcsec$^{-1}$.}
  \label{fig9}
\end{figure*}

Intensity maps integrated over the high Doppler velocity interval, 5$<$$V_z$$<$12 \kms, are displayed in Figure~\ref{fig8}. With the exception of the $^{13}$CO(2-1) line emission, which is in the noise, all other line emissions are enhanced in the north-western quadrant, including the H$^{12}$CN(3-2) line emission which also hosts the high Doppler velocity tail of the narrow component discussed in the preceding sub-section. In order to better illustrate the morpho-kinematics at stake in the north-western quadrant, 270\dego$<$$\omega$$<$360\dego, we display in Figure~\ref{fig9}, for each line emission separately, $V_z$ vs $R$ maps. They show patterns reminiscent of that observed in EP Aquarii by \citet{Nhung2024}, suggesting an interpretation in terms of a localised and episodic mass ejection, expected to be associated with shock waves near maximal pulsation amplitude, and having enhanced intensity and radial velocity over a particular convective cell. In such a case, the shock waves push away the gas to larger distances from the star with radial velocities larger than the normal expansion velocity: a depression of emission develops in their wake, the size of which increases with time. The enhancement of emission seen in the high Doppler velocity interval, 5$<$$V_z$$<$12 \kms, is associated with the gas having been pushed outward and a depression of emission is seen at lower values of $V_z$. However, in the present case, the situation is not as clear as it was in the case of EP Aqr because the data are of lesser sensitivity and lower angular resolution. The present data, by themselves, while being consistent with such a scenario, cannot exclude other possible interpretations. Higher values of the angular resolution, of the sensitivity and of the maximal recoverable scale would allow for more accuracy and better confidence in the reliability of this interpretation. If its validity were confirmed, it would suggest that the resulting mass ejection reaches between 1 and 2 arcsec from the centre of the star, implying that it occurred 5 to 10 decades ago over the red-shifted north-western quadrant with a radial velocity having reached $\sim$12 \kms. The $V_z$ vs $R$ maps displayed in Figure~\ref{fig9} show that the enhancement of mass ejection is not limited to the red-shifted hemisphere but also extends over the blue-shifted hemisphere where, however, the impact of absorption is stronger. This is consistent with the suggested interpretation, the pulsation shock waves covering the whole solid-angle and being simply enhanced over particular convective cells. Radiation pressure on dust formed in the gas originally compressed by the shock wave would accelerate the gas and thereby further increase the density in the affected region \citep[e.g.][]{Freytag2023}.

Exploring the light curve over the past 50 years reveals important variations of the pulsation amplitude but does not reveal obvious singular episodes of exceptionally large ones. The magnitude reached at maximal light varies between 3 and 6, spanning a factor $\sim$16 in intensity, and has a mean value of $\sim$4.5 and an rms value of $\sim$0.5 (a factor $\pm$1.6) with an approximately Gaussian distribution. The highest maximum occurred in 2006. Similar remarks can be drawn when exploring the light curve over the past century \citep{McAdam2000}. Probably worth mentioning is the presence of a small wiggle on the light curve at the time of the 2023 observations (Figure~\ref{fig1}). Such a wiggle is visible at mid-rise of each pulsation, however taking very variable forms; such is the case, in particular, after 1975, when \citet{Hinkle1982} claim evidence for the sudden formation and subsequent progressive decrease of an 800 K layer within some 200-300 mas away from the centre of the star. Similar features were first reported by \citet{LeBertre1992}, particularly clearly on C-star AFGL 1085, and commented upon by \citet{Winters1994}. They have now been observed on other S- and C-stars, but are not well understood \citep{MerchanBenitez2023}.

\subsection{Si$^{16}$O(5-4)/Si$^{16}$O(6-5) emission ratio: comparison with Local Thermal Equilibrium (LTE)}

In the LTE approximation, the unabsorbed emissivity $\varepsilon$ is proportional to $A_{ji}(2J+1)$e$^{-Eup/T}/T$, where $A_{ji}$ is the Einstein coefficient, $J$ the angular momentum of the upper state, $E_{up}$ its energy and $T$ the temperature. The ratio between Si$^{16}$O(5-4) and Si$^{16}$O(6-5) emissions, when neglecting opacity (more precisely when neglecting the difference between the effects of opacity on each of the two line emissions) and assuming level populations predicted by LTE, is therefore expected to be 0.48e$^{12.5/T}$ where 0.48 is the ratio of the Einstein coefficients multiplied by (2$J$+1) and 12.5 K the difference between the values of $E_{up}$ (see Table \ref{tab2}). The measured values listed in Table \ref{tab3}, using 2023 A-configuration data, give mean emission ratios of 0.86 in S$_1$ and 0.92 in S$_2$, corresponding to temperatures of 19 and 21 K. Using instead A+B-configuration data, which however mixes 2023 and 2024 observations, these values become 0.76 in S$_1$ and 0.68 in S$_2$, corresponding to temperatures of 27 and 36 K. Such temperatures are lower than expected: according to \citet{Justtanont2010}, they are only reached at distances exceeding 700 to 1000 au from the centre of the star. Assuming that the line emissions probe distances at arcsec scale, the temperature should exceed 100 K, implying that the emission ratio should be about half the measured value. In the same approximation, the ratio of $\varepsilon$ to the optical depth $\tau$ is proportional to $f^3$/(e$^{\Delta{E}/T}-1$), where $f$ is the frequency of the transition and $\Delta{E}$ its energy, proportional to $f$. For $T$$>>$$\Delta{E}$, as the Einstein coefficients are approximately proportional to $f^3$, we expect the optical depth to be proportional to $f$e$^{-Eu/T}/T^2$. Both $f$ and $T$ (the temperature probed by the transition) are larger for Si$^{16}$O(6-5) than for Si$^{16}$O(5-4) and their effects partly compensate each other; but, if absorption were much stronger for Si$^{16}$O(6-5) than for Si$^{16}$O(5-4), the observed discrepancy could be the simple result of a difference of opacity between the two lines.

\begin{figure*}
  \centering
  \includegraphics[width=0.3\linewidth, trim= 0.8cm 1.1cm 2.5cm 1.cm,clip]{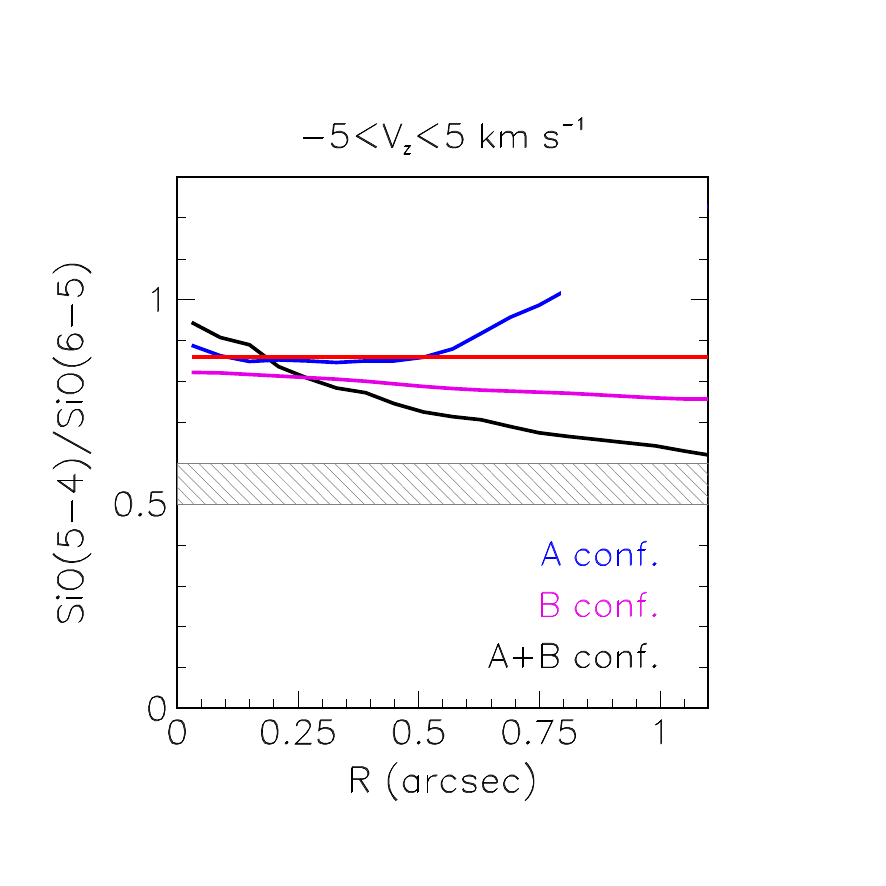}
  \includegraphics[width=0.3\linewidth, trim= 0.8cm 1.1cm 2.5cm 1.cm,clip]{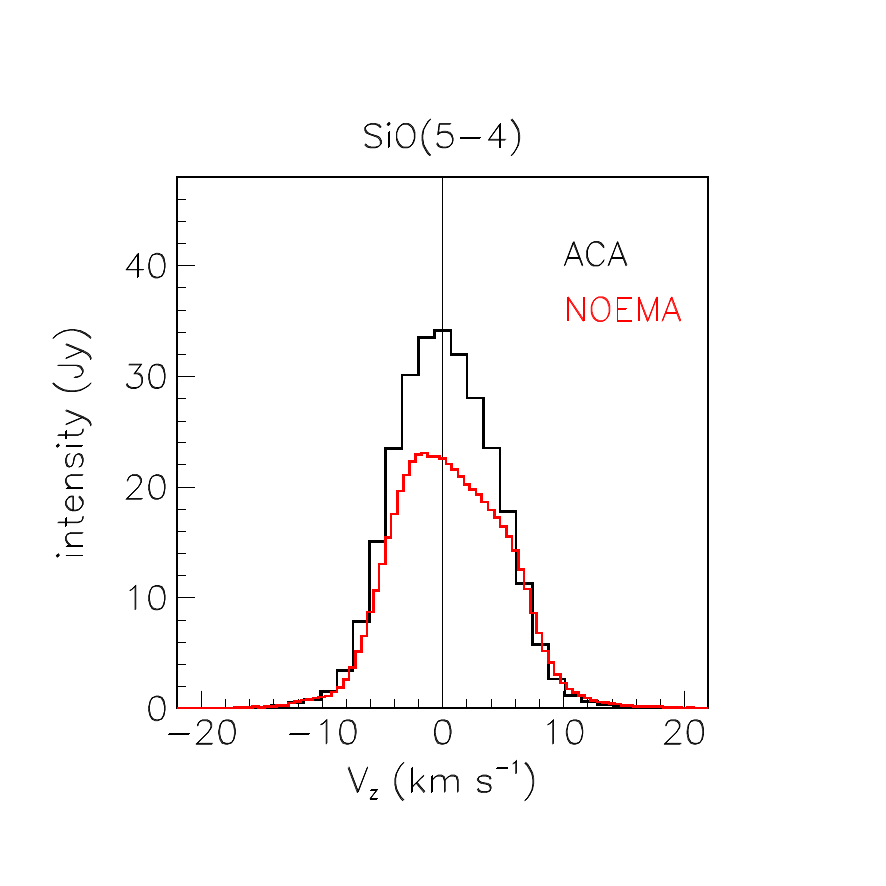}
  \includegraphics[width=0.3\linewidth, trim= 0.8cm 1.1cm 2.5cm 1.cm,clip]{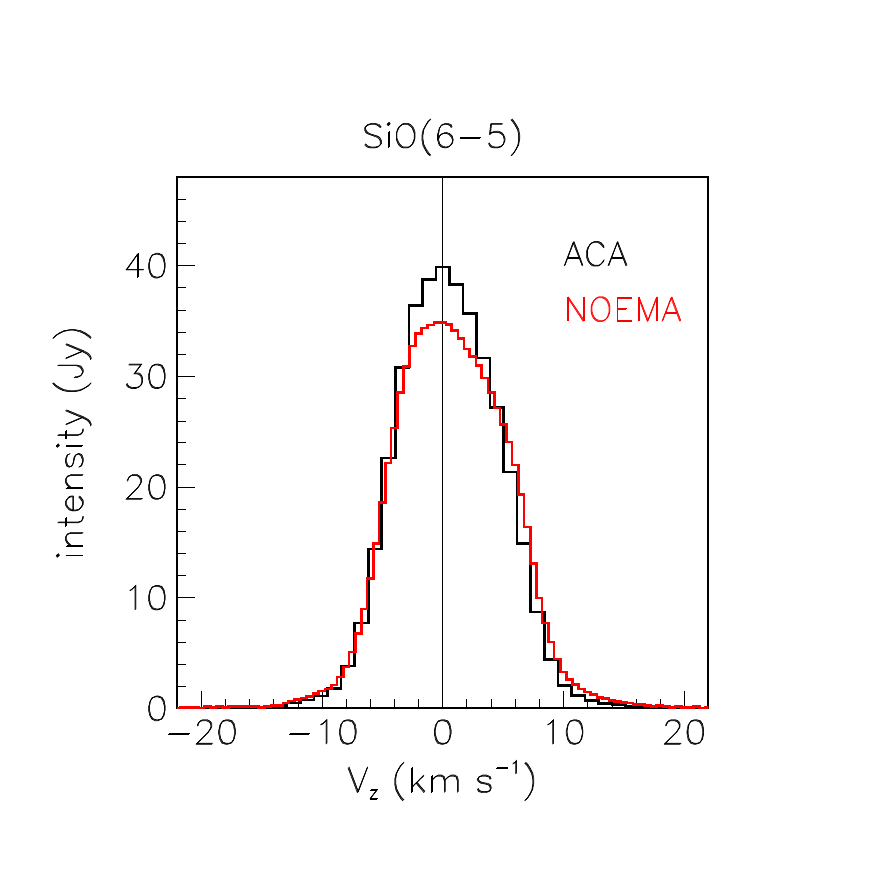}
  \caption{Left: radial distribution of the observed SiO(5-4)/SiO(6-5) intensity ratio. The band between 0.5 and 0.6 shows the prediction of a simple LTE model. The red line shows the value measured by ALMA/ACA. The blue curve uses the 2023 A-configuration observations, which suffer from a small maximal recoverable scale. The B- and A+B-configuration histograms, shown in magenta and black respectively, combine 2023 and 2024 observations and therefore suffer from a possible variability between the two years. Centre and right: Doppler velocity spectra measured using ALMA/ACA compared with the present NOEMA observations integrated over $R$$<$1.5 arcsec.}
  \label{fig10}
\end{figure*}

In order to get an independent evaluation of the emission ratio, we analysed archived ALMA/ACA observations of the SiO(5-4) and SiO(6-5) line emissions of $\chi$ Cygni from Project 2021.2.00147.S (PI M. Saberi). The Doppler velocity spectra are displayed in Figure~\ref{fig10}. The ACA measurement covers the whole emission while the NOEMA measurement is over $R$<1.5 arcsec only. As the extension of the SiO(5-4) emission is larger than that of the SiO(6-5) emission, the SiO(5-4)/SiO(6-5) emission ratio is expected to be smaller when measured by NOEMA than when measured by ACA. Indeed, it is $\sim$0.7 instead of 0.86, probably consistent within uncertainties.

Such large emission ratios have been measured earlier for other stars, such as, for example, 0.77 for R Dor \citep{DeBeck2018}. The importance of accounting properly for radiative transfer when interpreting line emission ratios has been underscored by several authors, such as \citet{Bieging2000} or \citet{Ramstedt2014}. Over 30 years ago, \citet{Bujarrabal1991} had already remarked, from their IRAM 30 m single dish observations of Si$^{16}$O lines from $\chi$ Cygni, that emission ratios such as Si$^{16}$O(2-1)/Si$^{16}$O(3-2) deviated significantly from LTE expectation and they were blaming it on the large opacity of the Si$^{16}$O lines. However, \citet{Lucas1992}, observing that the half-power radius of $\chi$ Cygni stays between 0.6 and 0.7 arcsec across the whole SiO(2-1) line profile, argued that such behaviour could not be accounted for by standard dynamics, even in the presence of large opacities. A situation very similar to the present case of $\chi$ Cygni was studied by \citet{Winters2022} in the case of RS Cnc, an MS-type, post-third-dredge-up semi-regular Long Period Variable. In that case, the Si$^{16}$O(5-4)/Si$^{16}$O(6-5) emission ratio, averaged over a projected distance from the centre of the star of $\sim$1 arcsec, is $\sim$0.65, closer to the LTE value than for $\chi$ Cygni, but still implying temperatures of $\sim$40-50 K, lower than expected. The authors attempted an LTE description, blaming the low temperature on a large optical depth, causing the line emission to explore preferentially the external layers of the CSE.

In such a context, it is instructive to obtain an estimate of the opacity of the SiO lines, which we do by producing a simple LTE model of the morpho-kinematics and integrating the radiative transfer equation over a spherical CSE. We find that standard radial distributions of the velocity, molecular abundance and temperature give good fits to the data. Precisely we use as reference parameters \citep{Justtanont2010, VandeSande2018} a mass-loss rate of 2.4 M$_\odot$\,yr$^{-1}$, a terminal velocity of 8 \kms, an SiO/H abundance ratio of 10$^{-5}$, meaning a density of 1.5 SiO/cm$^3$ at a distance of 165 au from the centre of the star, multiplied by a Gaussian having a $e$-folding radius of 215 au, a radial velocity increasing from 2 \kms\ at 17 au to 2+6(1$-$0.1/$r$)$^{0.8}$ at a distance $r$, a radial dependence of the temperature of the form $T$(K)=100(210/$r$[au])$^{0.78}$. However, when varying these within reasonable limits, we find that the emission ratio SiO(5-4)/SiO(6-5) remains between 0.5 and 0.6: absorption, even when strong, does not much affect the emission ratio and cannot explain the large values that are observed.  A systematic study of such emission ratios, including observations of other lines, would help, not only for $\chi$ Cygni but also for RS Cnc, R Dor and other AGB stars, to better tell apart the respective contributions of opacity and of level population and to understand what precisely is causing them to deviate from the LTE regime. In particular, non-LTE models have been mostly used to simulate single-dish observations and a detailed comparison with currently available interferometer data would help with telling apart different contributions; this is, however, beyond the scope of the present article.

\subsection{$^{12}$C/$^{13}$C, $^{16}$O/$^{17}$O and $^{28}$Si/$^{29}$Si abundance ratios from CO(2-1), HCN(3-2), SiO(5-4) and SiO(6-5) line emissions}
The availability of observations of CO(2-1) and HCN(3-2) line emissions for both $^{12}$C and $^{13}$C isotopologues should provide an estimate of the value taken by the $^{12}$C/$^{13}$C abundance ratio. Indeed, as both the Einstein coefficients and upper state energies take very similar values for both isotopes, we might expect the emission ratio to be a good approximation of the abundance ratio: the differences between the upper state energies are less than 1 K and the ratios of the Einstein coefficients are 1.14 for CO and 1.09 for HCN. Similarly, the Si$^{16}$O(6-5)/Si$^{17}$O(6-5) emission ratio, with a ratio of Einstein coefficients of 1.12 and a difference of upper states energies of only 1.7 K, should provide an estimate of the $^{16}$O/$^{17}$O abundance ratio; and the $^{28}$SiO(5-4)/$^{29}$SiO(5-4) emission ratio, with a ratio of Einstein coefficients of 1.04 and a difference of upper states energies of only 0.4 K, should provide an estimate of the $^{28}$Si/$^{29}$Si abundance ratio. However, opacity is expected to lower the measured emission of the $^{12}$C, $^{16}$O and $^{28}$Si isotopologues, namely to decrease the measured values of the $^{12}$C/$^{13}$C, $^{16}$O/$^{17}$O and $^{28}$Si/$^{29}$Si abundance ratios compared to their true values. In contrast with the case discussed in the preceding sub-section, we now expect a significant impact of the effect of opacity as the optical depth is larger for the $^{12}$C, $^{16}$O and $^{28}$Si than for the $^{13}$C, $^{17}$O and $^{29}$Si isotopologues. This has been discussed in the published literature in the case of the CO(1-0), CO(2-1) and CO(3-2) line emissions from $\chi$ Cygni observed using single-dish telescopes \citep{DeBeck2010, Wallerstein2011, Ramstedt2014} with a $^{12}$C/$^{13}$C abundance ratio between 33 and 40, about twice as large as the measured line emission ratio.

\begin{figure*}
  \centering
   \includegraphics[width=1\linewidth, trim= 0cm 0.5cm 0cm .5cm,clip]{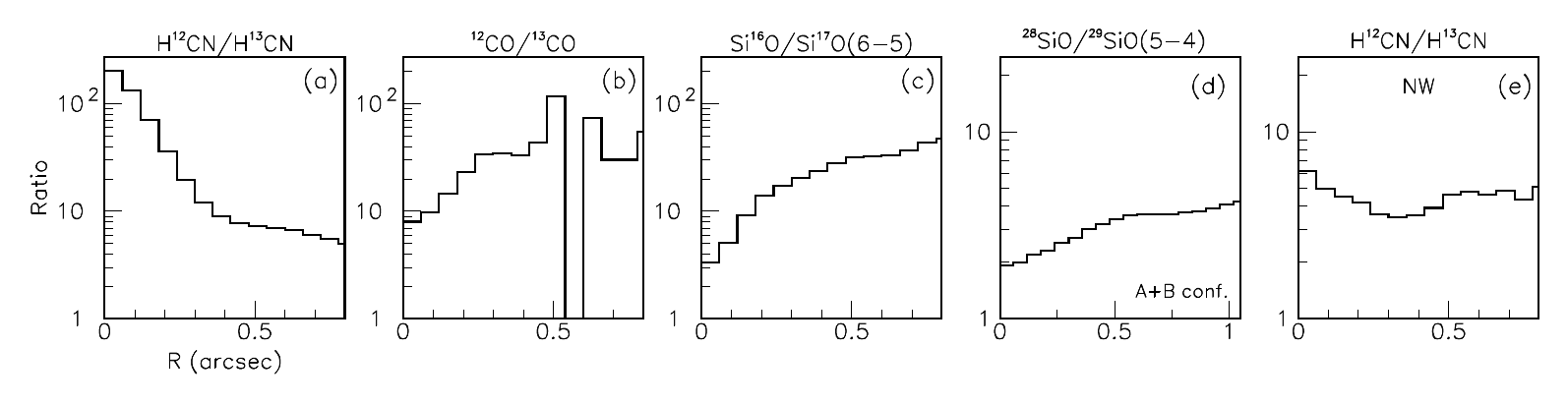}
  \caption{Dependence on $R$ of various emission ratios: a) H$^{12}$CN(3-2)/H$^{13}$CN(3-2), $|V_z|$$<$5 \kms; b) $^{12}$CO(2-1)/$^{13}$CO(2-1), $|V_z|$$<$5 \kms; c) Si$^{16}$O(6-5)/Si$^{17}$O(6-5), $|V_z|$$<$5 \kms; d) $^{28}$SiO(5-4)/$^{29}$SiO(5-4), $|V_z|$$<$5 \kms; e) H$^{12}$CN(3-2)/H$^{13}$CN(3-2), 5$<$$V_z$$<$12 \kms, 270\dego$<$$\omega$$<$360\dego.}
  \label{fig11}
\end{figure*}

For each of the HCN(3-2), CO(2-1), SiO(6-5) and SiO(5-4) line emissions, Figure~\ref{fig11} displays the dependence on $R$ of the relevant line emission ratios in the low Doppler velocity interval (in the case of the SiO(5-4) line emission, we show A+B merged data as the line was observed exclusively in 2023). They share a common feature: close to the star, in the low Doppler velocity interval, the emission ratios take different values than elsewhere. In the case of the HCN line, it is not surprising, it was already clear from the analysis presented in Sub-section 3.2 and in particular in the right panel of Figure~\ref{fig7}. However, in the case of the CO and SiO lines, it is unexpected. In particular, in both the CO(2-1) and SiO(6-5) cases the emission ratios are abruptly decreasing below $R$$\sim$0.3 arcsec and it is tempting to blame high opacities; but we see from Figure~\ref{fig5} that the decrease of the ratios close to the star is in part the result of an increase of the rare isotope emissions, rather than a decrease of the $^{12}$C and $^{16}$O isotopologue emissions as high opacities produces. In contrast with elsewhere, both $^{13}$CO(2-1) and Si$^{17}$O(6-5) emission ratios are observed with a reasonable signal to noise ratio in the low Doppler velocity intervals of region S$_1$ where $^{12}$CO(2-1)/$^{13}$CO(2-1)=12$\pm$2 and Si$^{16}$O(6-5)/Si$^{17}$O(6-5)=9$\pm$1 using data of the A+B-configuration, both lower than expected. Using data of the A-configuration only does not make much difference: the Si$^{16}$O(6-5)/Si$^{17}$O(6-5) ratio becomes 7$\pm$1. In the case of the $^{28}$SiO(5-4)/$^{29}$SiO(5-4) ratio, the decrease close to the star is more progressive: its value is $\sim$1.9$\pm$0.3 in region S$_1$ and increases to $\sim$3$\pm$0.5 elsewhere within $R$$<$1 arcsec.  In the case of HCN(3-2) emission it is instead in the high Doppler velocity interval that the rare isotopologue is observed with a reasonable signal-to-noise ratio, giving a surprisingly low emission ratio H$^{12}$CN(3-2)/H$^{13}$CN(3-2)=5.0$\pm$0.5 using 2023 data of the A-configuration.

As already mentioned, single-dish measurements of the $^{12}$C/$^{13}$C emission ratio are available for CO only and are at least half the associated abundance ratio of 33-40, significantly larger than the values reported here for the CO emission ratio in region S$_1$ at low Doppler velocities and of the HCN emission ratio in region S$_1$+S$_2$ at large Doppler velocities. Very low abundance ratios are typical of J-type carbon-stars, such as Y CVn or RY Dra with values of 2 and 2.5, respectively \citep{Ramstedt2014}. Known processes that may produce such low ratios are either of nuclear or chemical origin. Nuclear processes can be of the ``cool bottom processing'' type \citep{Boothroyd1999}, particularly efficient in low mass Red Giants; in regions where the CNO cycle is operative, $^{13}$C can be formed by the decay of the $\beta$-unstable isotope $^{13}$N produced in the reaction $^{12}$C(p,$\gamma$)$^{13}$N, and then dredged-up to the stellar surface \citep{Sengupta2013}. A typical example of a chemical process is the fractionation of CO molecules by interaction with carbon ions, $^{13}$C$^+$+$^{12}$CO$\rightarrow$$^{12}$C$^+$+$^{13}$CO \citep{Mamon1988}. 

In the case of the $^{16}$O/$^{17}$O abundance ratio, using observations of CO near-infrared overtone band transitions, \citet{Hinkle2016} measured $^{16}$O/$^{17}$O=328$\pm$100 for $\chi$ Cygni, \citet{Smith1990} $^{16}$O/$^{17}$O$\sim$710 for the MS-type star RS Cnc, \citet{Harris1985} 625$\leq$$^{16}$O/$^{17}$O$\leq$3000 for a sample of five MS- and three S-type stars; from the emission ratio of CO(1-0) and CO(2-1) lines, \citet{Kahane1992} measured 242$\leq$$^{16}$O/$^{17}$O$\leq$840 for a sample of five carbon-rich envelopes. However, \citet{DeBeck2018} measured an emission ratio Si$^{16}$O(6-5)/Si$^{17}$O(6-5) of only 66 for the CSE of R Dor and \citet{Winters2022} measured only $\sim$50 for the CSE of RS Cnc, remarking that the larger optical depth of the Si$^{16}$O(6-5) line causes this ratio to be smaller than the $^{16}$O/$^{17}$O abundance ratio.

Finally, in the case of the $^{28}$Si/$^{29}$Si abundance ratio \citet{Schoier2011} measure 8$\pm$2 using Herschel/HIFI .

In summary, we can state with confidence that in the regions where the rare isotopologue emissions are measured with a reasonable signal-to-noise ratio, we observe unexpectedly low emission ratios: Si$^{16}$O(6-5)/Si$^{17}$O(6-5)$\sim$7-9, $^{12}$CO(2-1)/$^{13}$CO(2-1)$\sim$13 and $^{28}$SiO(5-4)/$^{29}$SiO(5-4)$\sim$2  in the low Doppler velocity interval of region S$_1$ and H$^{12}$CN(3-2)/H$^{13}$CN(3-2)$\sim$5 in the high Doppler velocity interval within $R$$<$1 arcsec. Moreover we observe a decrease of the Si$^{16}$O(6-5)/Si$^{17}$O(6-5) and $^{12}$CO(2-1)/$^{13}$CO(2-1) emission ratios close to the star, caused in part by an increase of the rare isotopologue emissions. In other regions, the lower values of the signal-to-noise ratios prevent a reliable analysis but estimates using B-configuration or merged A+B data suggest that the isotopic emission ratios are consistent, within large uncertainties, with the ratios measured using single-dish telescopes.

\subsection{High Doppler velocity wings}
The presence of high Doppler velocity wings confined within a few stellar radii from the centre of the star is a common feature of millimetre observations of M- and S-types AGB stars \citep{Darriulat2024}, sometimes very clear, as for example in $o$ Cet \citep{Nhung2022}, EP Aqr \citep{Nhung2024} or RS Cnc \citep{Winters2022}, sometimes much less as in the cases of L$_2$ Pup \citep{Hoai2022a, Homan2017}. The observed effective line broadening is commonly understood as revealing the simultaneous presence of out-flowing and in-falling gas in the inner layers of the CSE where shocks associated with pulsation and convection are dominant. However, a detailed description is still lacking, in particular concerning the dependence on stellar phase and the different time scales of variability.

\begin{figure*}
  \centering
  \includegraphics[width=0.8\linewidth,trim= 0cm  1.4cm 0cm 0cm,clip]{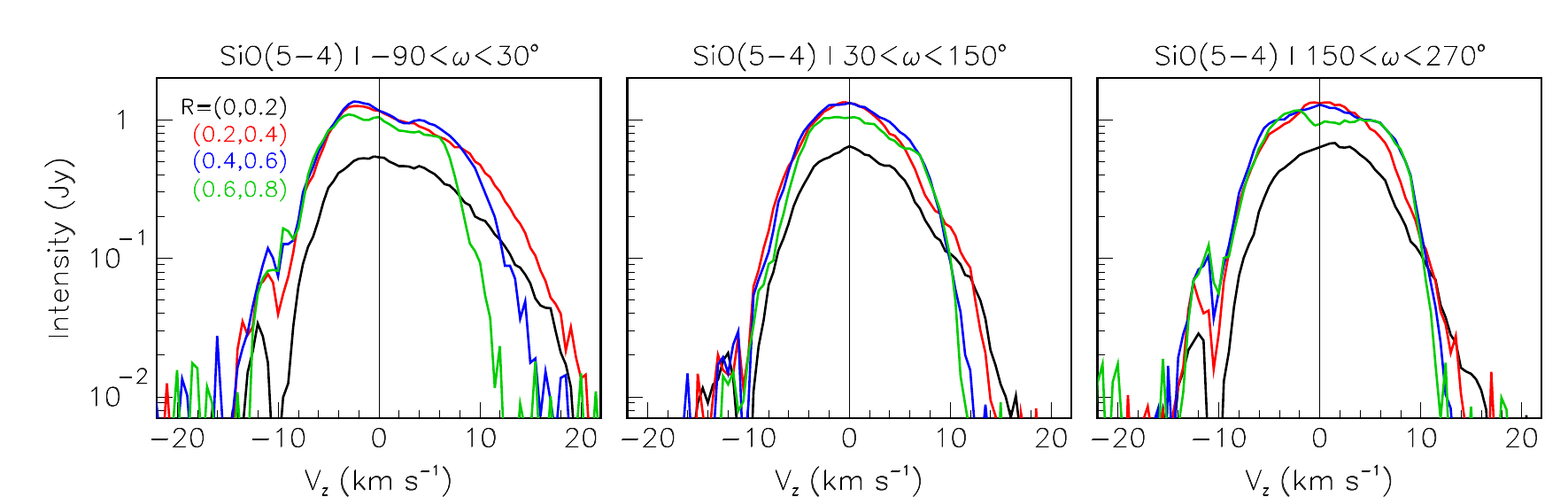}
  \includegraphics[width=0.8\linewidth,trim= 0cm  0cm 0cm 1.4cm,clip]{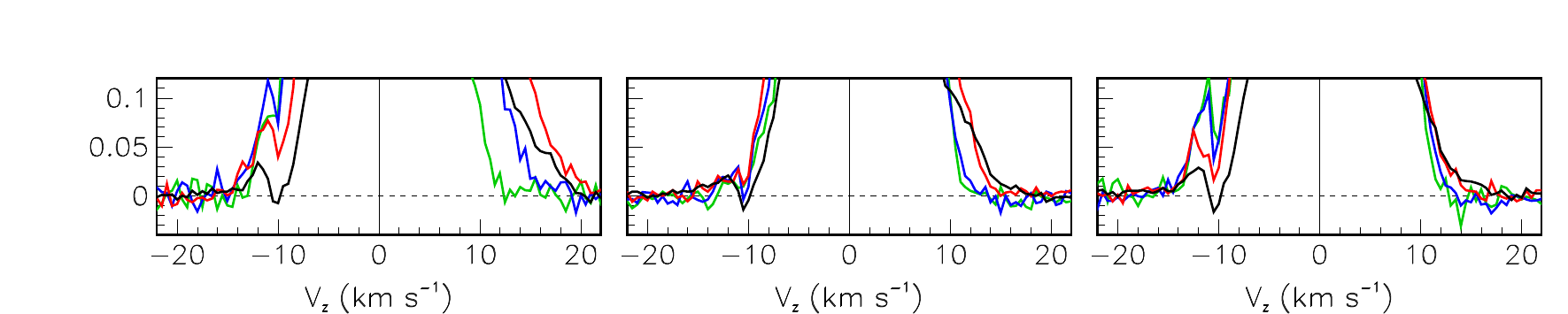}
  \includegraphics[width=0.8\linewidth,trim= 0cm  1.4cm 0cm 0cm,clip]{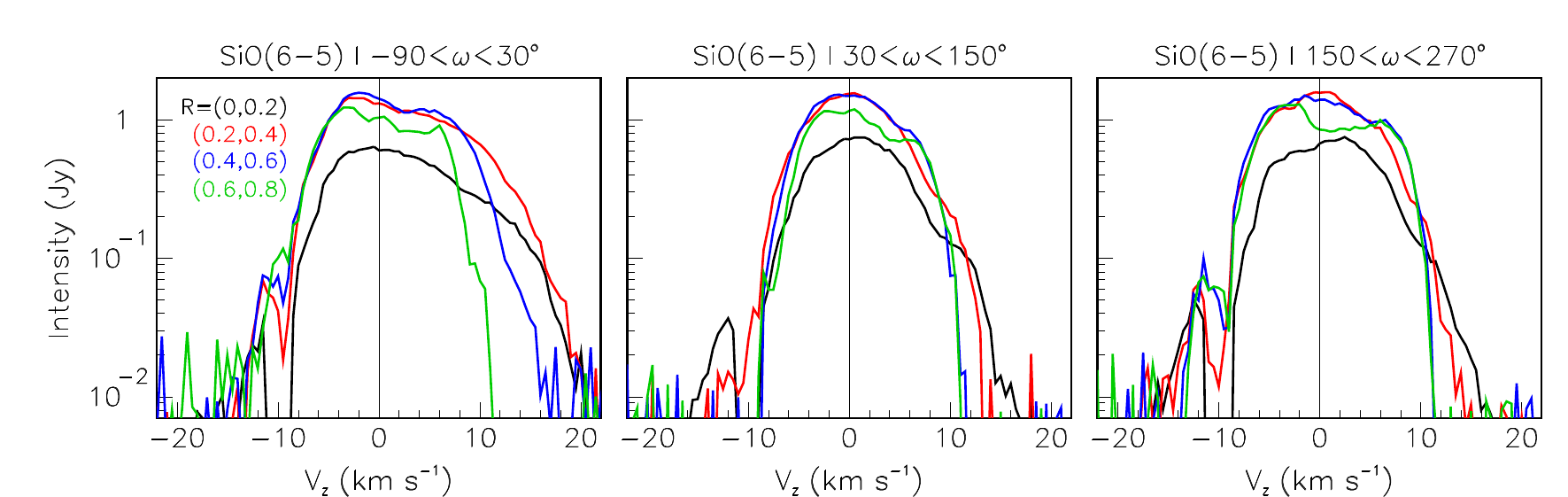}
  \includegraphics[width=0.8\linewidth,trim= 0cm  0cm 0cm 1.4cm,clip]{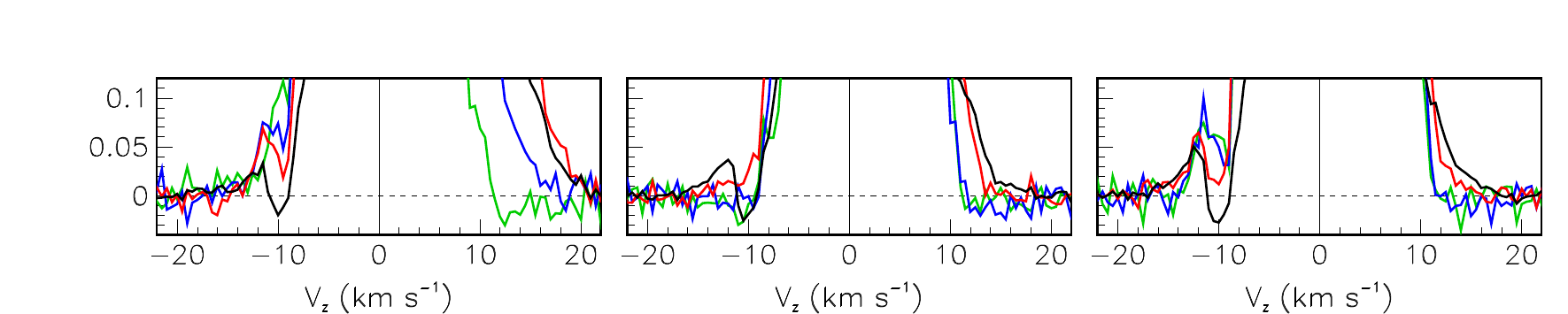}
  \includegraphics[width=0.8\linewidth,trim= 0cm  1.35cm 0cm 0cm,clip]{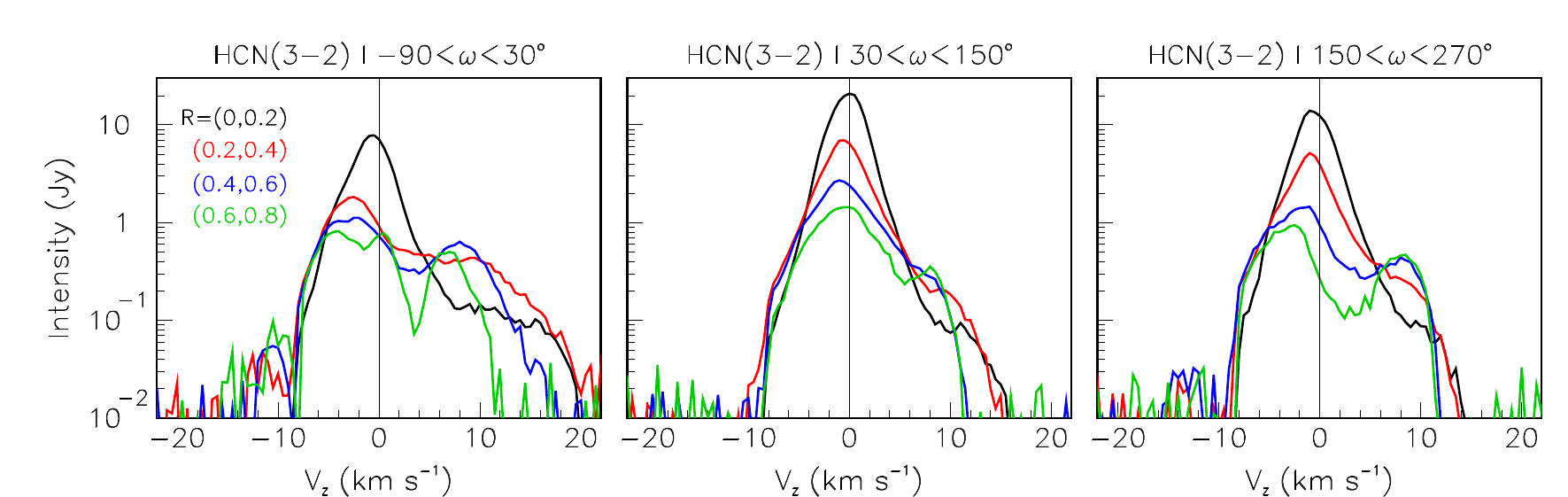}
  \includegraphics[width=0.8\linewidth,trim= 0cm  0cm 0cm 1.45cm,clip]{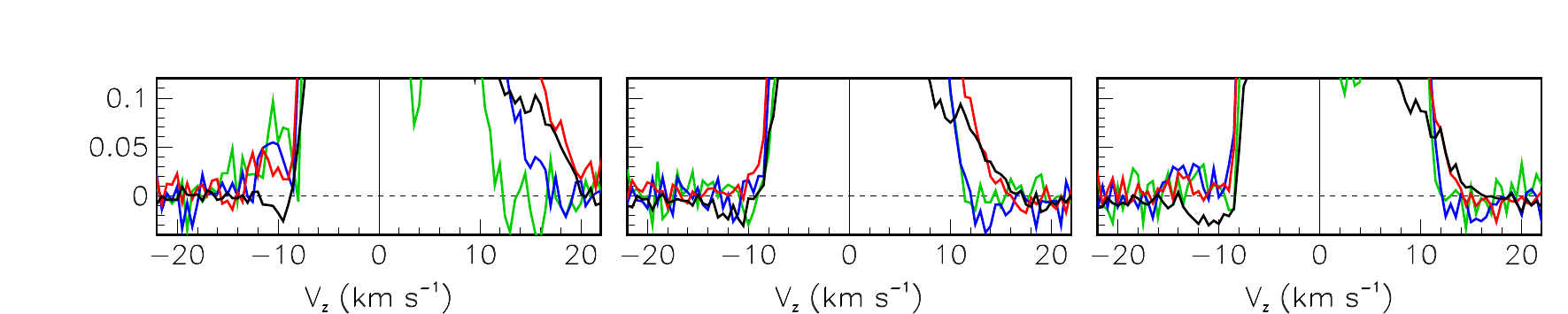}
  \caption{Doppler velocity spectra integrated over three intervals of position angles and four intervals of $R$. The lines and position angle intervals are indicated on top of each panel. $R$ intervals are 0 to 0.2 arcsec (black), 0.2 to 0.4 arcsec (red), 0.4 to 0.6 arcsec (blue) and 0.6 to 0.8 arcsec (green). In each panel, the upper part shows the line profiles in logarithmic scale and the lower part shows them in linear scale. }
  \label{fig12}
\end{figure*}

	In the present case of $\chi$ Cygni, the presence of a high Doppler velocity mass ejection in the north-western quadrant complicates the analysis. Figure~\ref{fig12} displays Doppler velocity spectra integrated over three intervals of position angles and four intervals of $R$ for three of the line emissions discussed in the preceding sections. As CO(2-1) emission shows no significant line broadening, consistent with the dominance of large radial distances explored by this line, it is not displayed in the figure. On the blue-shifted side, where absorption complicates the analysis, H$^{12}$CN(3-2) shows no clear line broadening while Si$^{16}$O(6-5) and Si$^{16}$O(5-4) show some with $|V_z|$ increasing from 12-13 to 18-20 \kms, confined within 0.4 arcsec. On the red-shifted side, H$^{12}$CN(3-2) is completely dominated by the mass ejection, which is also strong in Si$^{16}$O(5-4) and Si$^{16}$O(6-5). In summary, absorption and mass ejection make it difficult to reveal high Doppler velocity wings in the spectrum of $\chi$ Cygni; if they are present, which one cannot assert with certitude, they are weak.

\subsection{Si$^{16}$O($\nu$=1,$J$=6-5) maser}
\begin{figure*}
  \centering
  \includegraphics[height=4.7cm,trim= .5cm 2.2cm 17.3cm 9.8cm,clip]{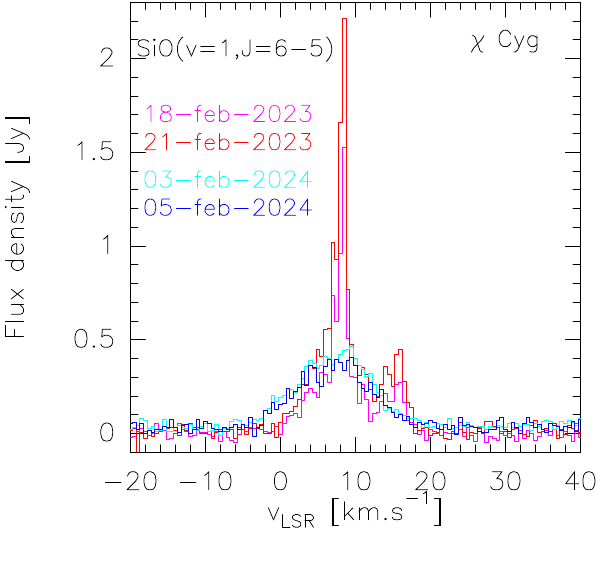}
  \includegraphics[height=4.3cm,trim= 0.5cm -0.2cm 0.2cm 0.2cm,clip]{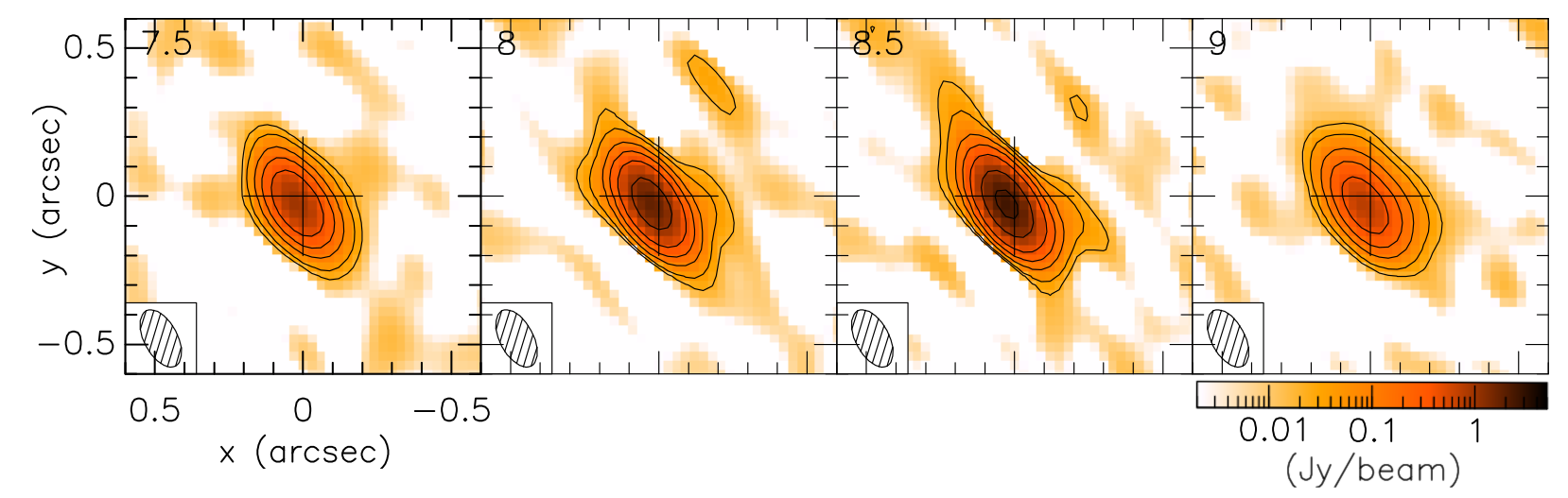}
  \caption{Si$^{16}$O($\nu$=1,$J$=6-5) line emission. Left: Doppler velocity spectra integrated over a square of 2$\times$2 arcsec$^2$ centred at the origin. Right: channel maps (Jy beam$^{-1}$) of the 2023 A-configuration observations between $V_z$=$-$2.5 and $-$1 \kms\ (values of V$_{lsr}$ are indicated in the upper-left corner of each panel).The contours are at 3, 6, 15, 30, 60, 150 and 300 times the noise level (see Table \ref{tab2}). }
  \label{fig13}
\end{figure*}

The emission of the vibrationally-excited line of Si$^{16}$O($\nu$=1,$J$=6-5) was observed in the extended A-configuration in February 2023. Figure~\ref{fig13} illustrates the result. The Doppler velocity spectrum displays a profile characteristic of maser emission, with a narrow peak at $\sim$$-$1.5 \kms. This is indeed the only remarkable feature visible on the channel maps, which are otherwise dominated by the convolution with the beam of the stellar disc emission, elongated in the NE-SW direction. Another enhancement of emission is observed in the 3$<$$V_z$$<$8 \kms\ interval of Doppler velocity. The high value of the energy of the $J$=6 upper level, 1812 K, implies that the line emission probes the very close neighbourhood of the photosphere. However, the angular resolution is insufficient to map reliably the maser emission and possibly reveal multiple emitting spots simultaneously present in the close neighbourhood of the star, as commonly observed using very long baseline observations on several other stars (e.g. Cotton et al. 2004) and as well established in the case of SiO maser lines observed from $\chi$ Cygni. Small differences between the spectra observed on February 18 and February 21 may reveal variability, as can be expected from such maser emission.

Observation of the same line emission using the B-configuration, also illustrated in Figure~\ref{fig13}, was made in February 2024. It shows that both the identified maser and the enhanced emission between 3 and 8 \kms\ have disappeared, suggesting that the latter feature was also caused by masers. While the broader beam of the B-configuration would smear the image of point-like maser enhancements, this has no impact on the Doppler velocity spectrum and cannot account for the disappearance.

As mentioned in the introduction, SiO masers in $\chi$ Cygni have been observed on many occasions, however usually in lower rotational transitions, mostly $J$=1-0 and 2-1. To our knowledge the observation of a maser in the $J$=6-5 transition is presented here for the first time. Observations reported in the available literature cover a broad range of vibrational excitations, from $\nu$=0 (ground state) in many cases and up to $\nu$=6 for $\chi$ Cygni \citep{Rizzo2021} and give evidence for strong day-to-day variability, more moderate for Miras than for semi-regular stars \citep{GomezGarrido2020} but, still, in excess of 10\% in the case of the $\nu$=1,2, $J$=2-1 transitions for $\chi$ Cygni \citep{Rizzo2021}.

\section{The $^{12}$CO circumstellar envelope}
\begin{figure*}
  \centering
  \includegraphics[height=4.3cm, trim= 0.5cm 0.7cm 0cm 1.8cm,clip]{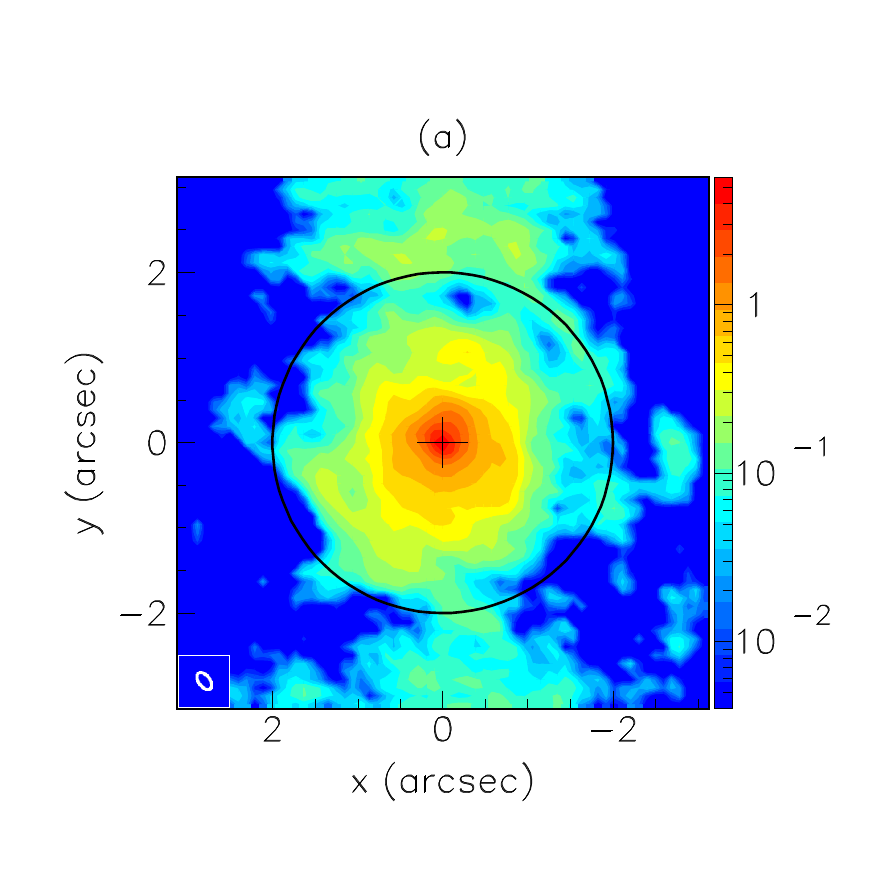}
  \includegraphics[height=4.3cm, trim= 0.5cm 0.7cm 2.5cm 1.8cm,clip]{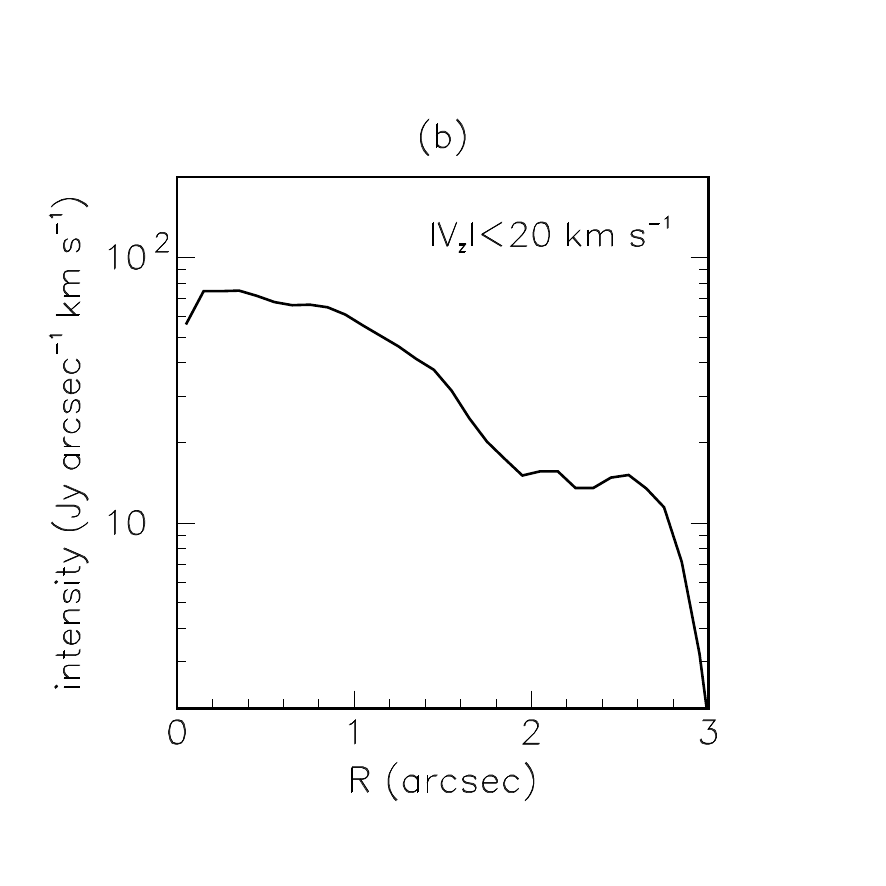}
  \includegraphics[height=4.3cm, trim= 0.5cm 0.7cm 2.5cm 1.8cm,clip]{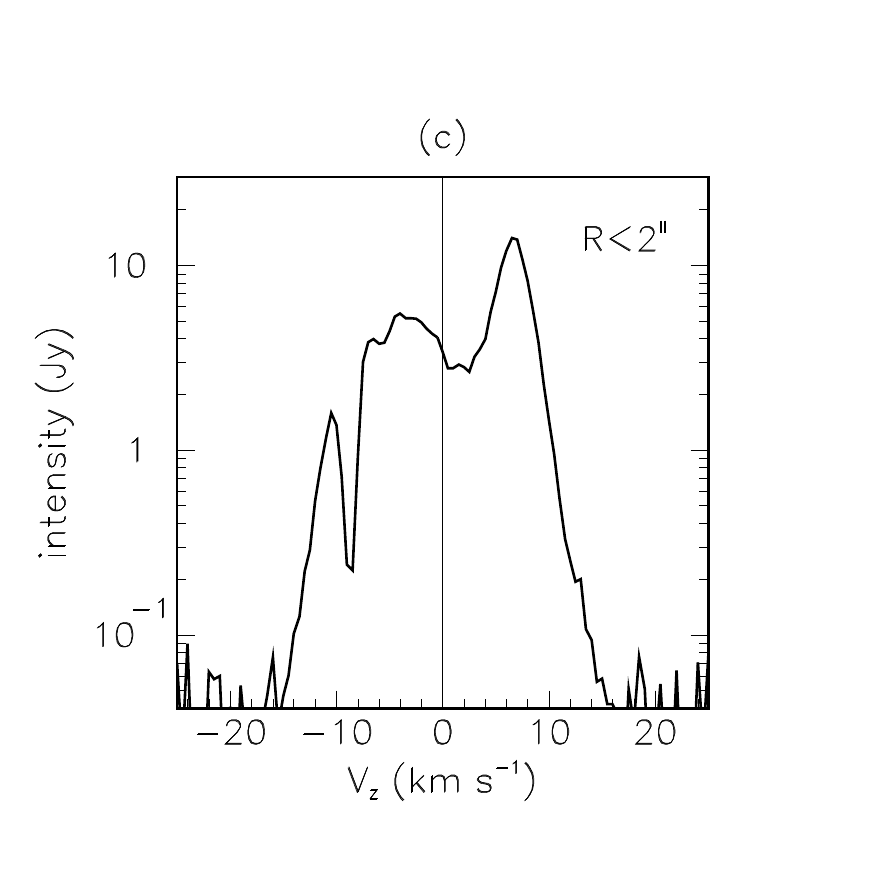}
  \includegraphics[height=4.3cm, trim= 0.5cm 0cm 0cm .5cm,clip]{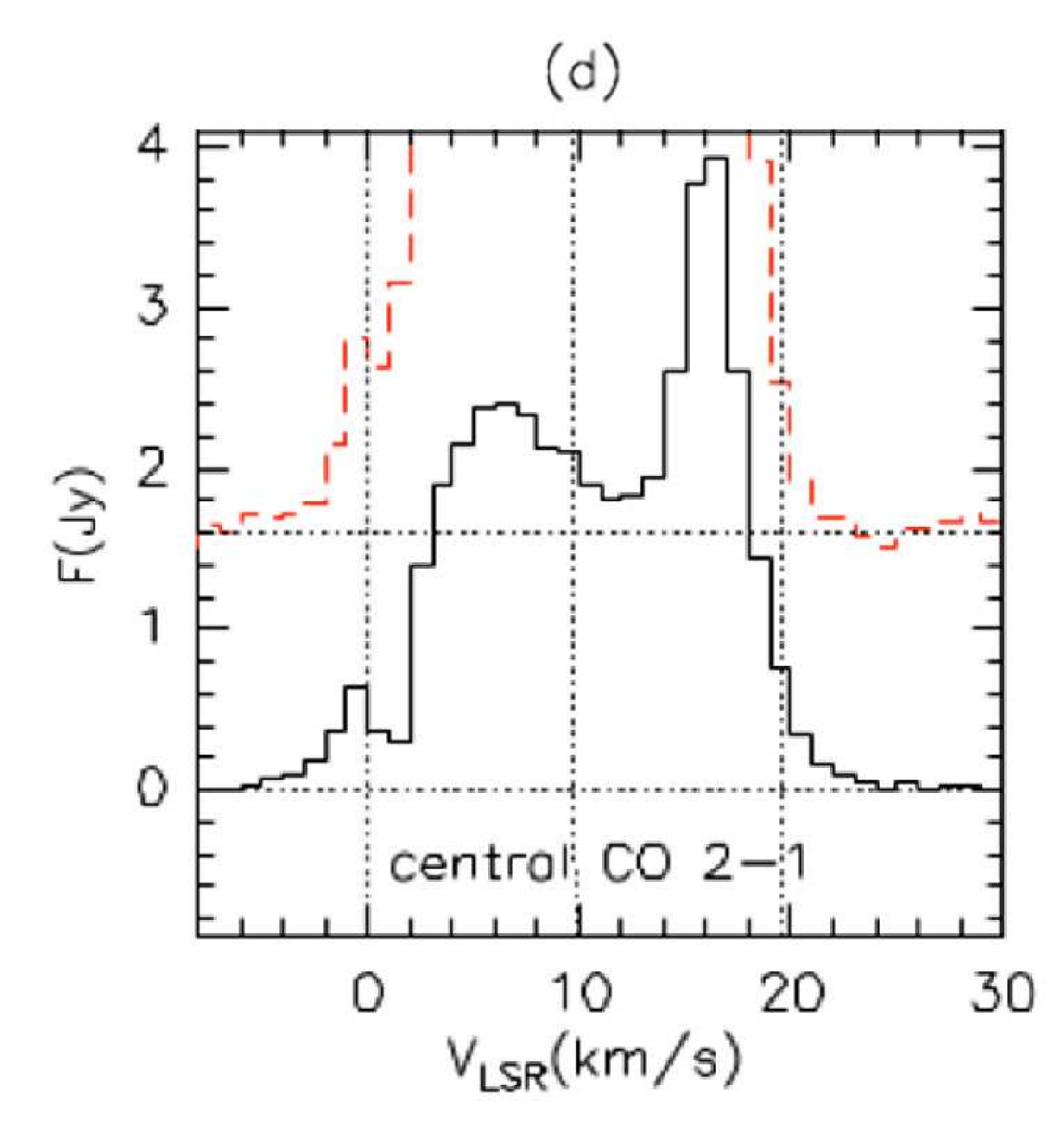}
  \caption{$^{12}$CO(2-1) line emission. From left to right: intensity map (Jy\,beam$^{-1}$\,\kms) integrated in the interval $|V_z|$$<$20 \kms\ (the circle has a radius of 2 arcsec), radial distribution integrated in the same interval of $V_z$; Doppler velocity spectrum integrated over $R$$<$2 arcsec. The rightmost panel (linear scale) is from \citet{CastroCarrizo2010}. The noise level is 24 mJy\,beam$^{-1}$\,\kms. }
  \label{fig14}
\end{figure*}

\begin{figure*}
  \centering
  \includegraphics[width=0.7\linewidth]{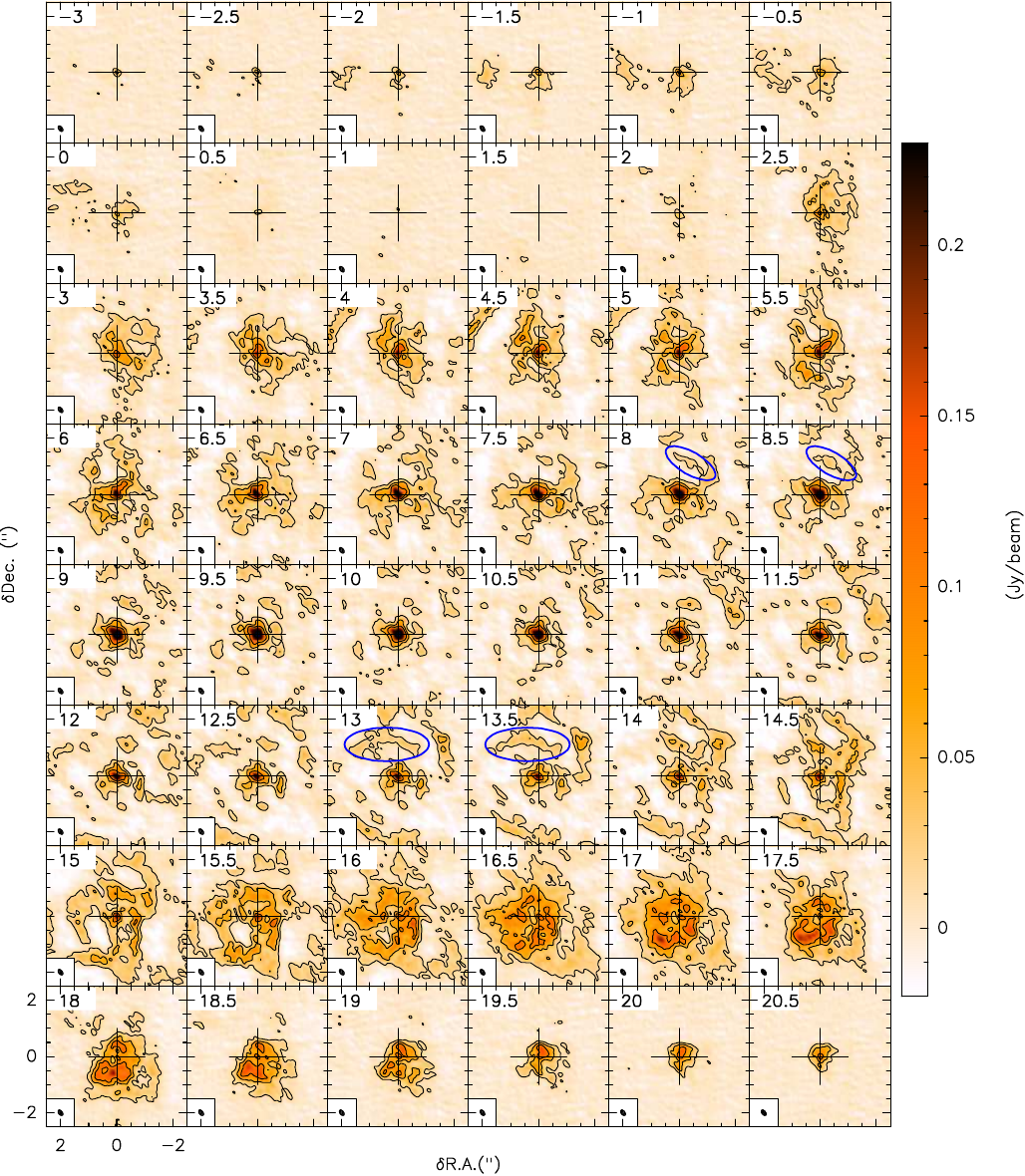}
  \caption{ $^{12}$CO(2-1) line emission, channel maps (Jy\,beam$^{-1}$) in the central $|x,y|$$<$2 arcsec square. The central LSR velocity is indicated in each panel (before subtraction of the 10 \kms\ systemic velocity of the star). Two examples of possible arcs seen in successive channel maps are circled in blue (at 8-8.5 \kms\ and 13-13.5 \kms, respectively). Close to the star, within $\sim$0.5 arcsec from its centre (Figures~\ref{fig16} and \ref{fig17}), there is evidence for the presence of a reservoir of $^{12}$CO gas, possibly partly gravitationally bound, with an rms Doppler velocity of $\sim$2-3 \kms. There is also evidence for an enhancement of emission at large values of $V_z$, where it is seen to cover dominantly the north-western red-shifted octant. Both features have been discussed in Section 3. The contours are at 3, 10 and 20 times the noise level (see Table \ref{tab2}).}
  \label{fig15}
\end{figure*}

$^{12}$CO(2-1) emission observations using both A and B antenna configurations have been merged after having been processed without continuum subtraction. The projected radial intensity distribution is displayed in Figure~\ref{fig14} together with the intensity map and Doppler velocity spectrum. We studied the decrease of the detected flux with the angular distance to the centre of the star by imaging discs of different sizes and different brightness radial distributions, with the result that reliable imaging is obtained up to $\sim$2.0 arcsec. Figure~\ref{fig15} displays channel maps.

\citet{CastroCarrizo2010} measured a radial extension of the line emission in excess of 40 arcsec and quote a photo-dissociation radius of 38 arcsec. Accordingly, the present observations cover only a very small central fraction of the CSE and are dominated by the emission of its outer layers.

The bulk of the emission covers the Doppler velocity interval $|V_z|$$<$15 \kms\ with a clear absorption peak revealing a terminal velocity of $\sim$9$\pm$1 \kms\ that must have been reached at short distances from the star. Above a continuum level of $\sim$30 mJy, to a first approximation, the spectrum (Figure~\ref{fig14}c) is very similar to that observed by \citet{CastroCarrizo2010} (Figure \ref{fig14}d) using a broad angular coverage, which reveals the presence of a sphere expanding radially. However, the patchy nature of the channel maps reveals the presence of important fluctuations of the amplitude of the line emission, both in time and in direction.

\begin{figure*}
  \centering
  \includegraphics[height=5.3cm, trim= 0.7cm 0.8cm 2.5cm 1.3cm,clip]{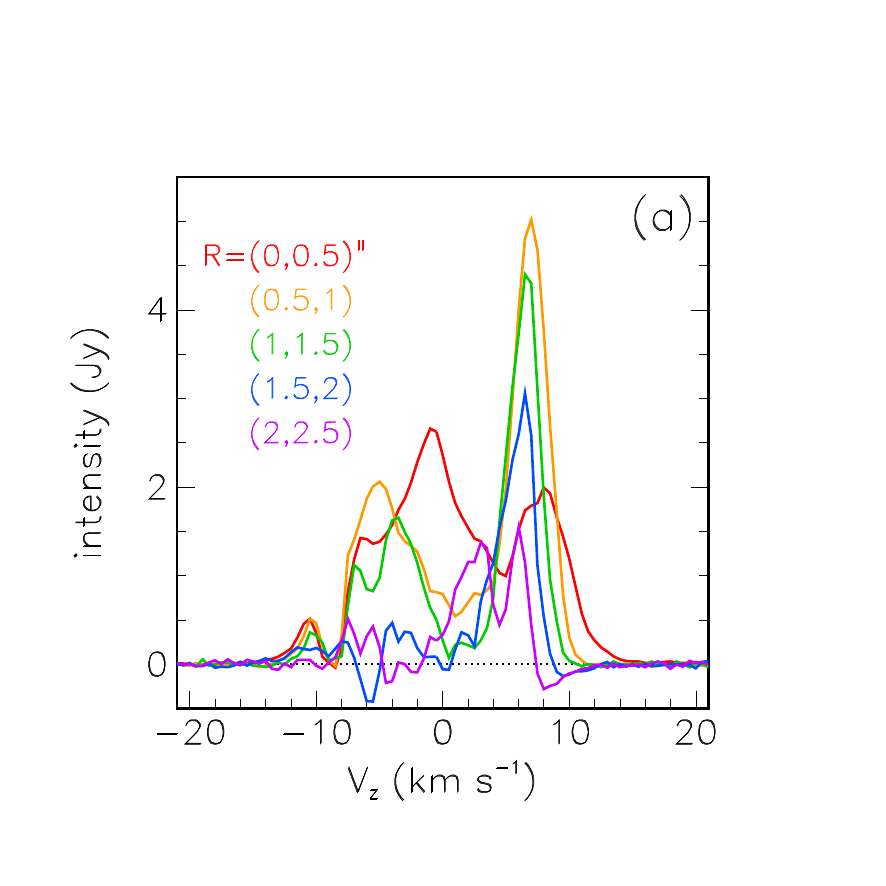}
  \includegraphics[height=5.3cm, trim= 0.5cm 0.7cm 2.4cm .8cm,clip]{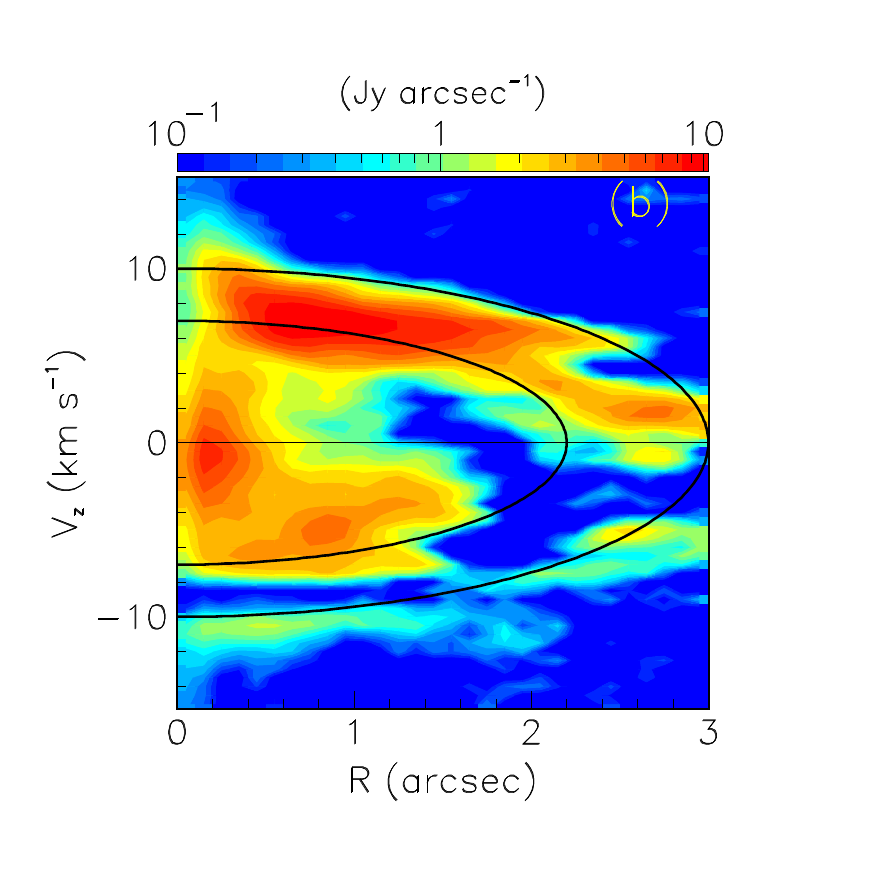}
  \includegraphics[height=5.3cm, trim= 1.7cm 0.7cm 2.4cm .8cm,clip]{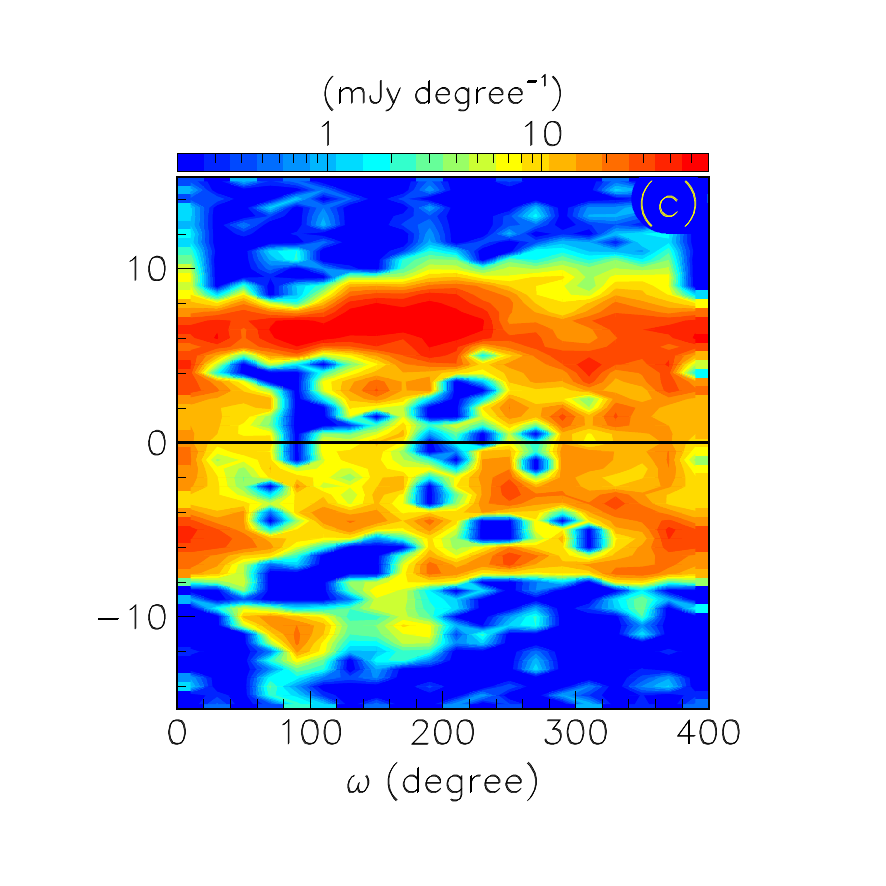}
  \includegraphics[height=5.3cm, trim= 1.7cm 0.7cm 2.4cm .8cm,clip]{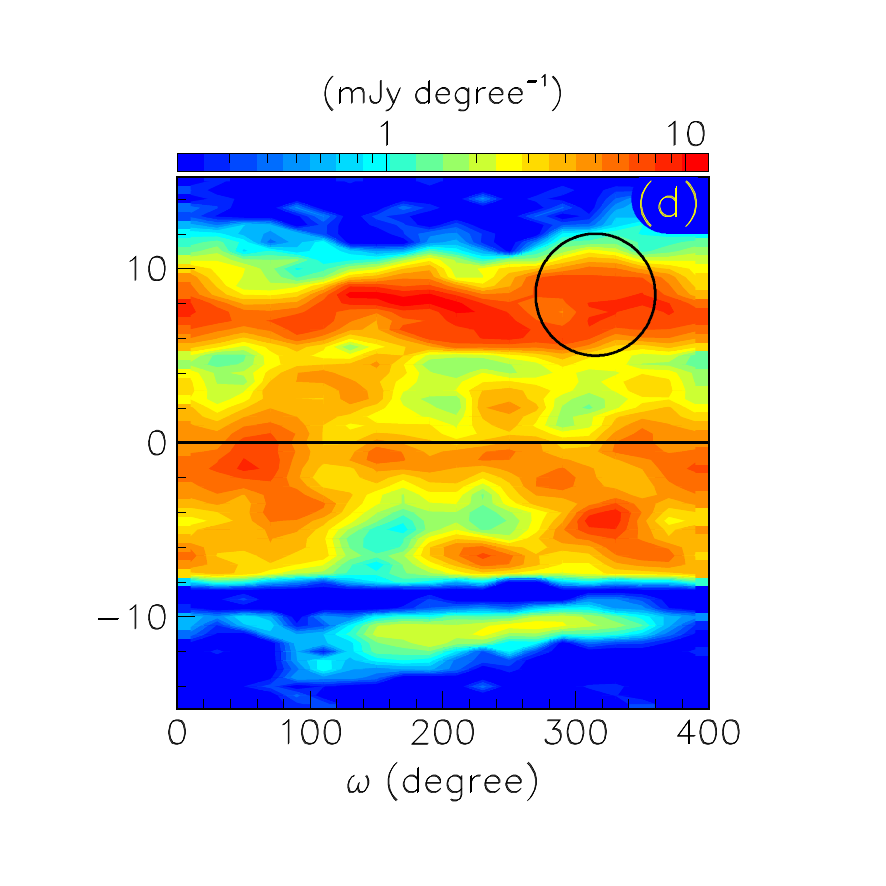}
  \caption{$^{12}$CO(2-1) line emission (merged data). a) Doppler velocity spectra in different intervals of $R$ as indicated in the insert. b) $V_z$ vs $R$ map integrated over position angles. The lines show thin spherical shells emitted 250 years ago with radial velocities of 10 and 7 \kms, respectively. c) $V_z$ vs $\omega$ map integrated in the interval 0.2$<$$R$$<$2.5 arcsec. d) $V_z$ vs $\omega$ map integrated in the interval 0.2$<$$R$$<$0.6 arcsec. The circle indicates the region covered by the red-shifted north-western octant discussed in Sub-section 3.3.}
  \label{fig16}
\end{figure*}

\begin{figure*}
  \centering
  \includegraphics[width=0.95\linewidth,trim= 0.5cm 1.4cm 0cm 0cm,clip]{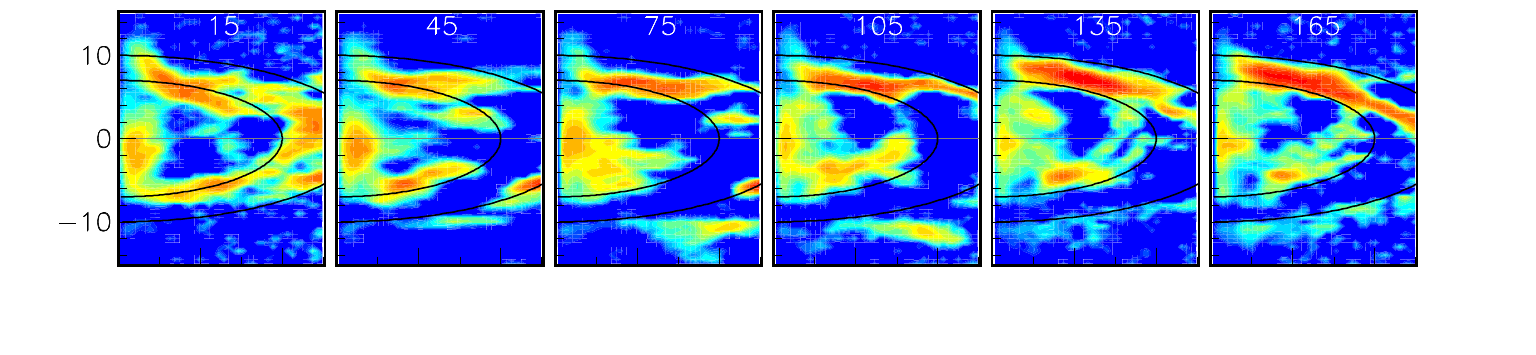}
  \includegraphics[width=0.95\linewidth,trim= 0.5cm 0cm 0cm 0cm,clip]{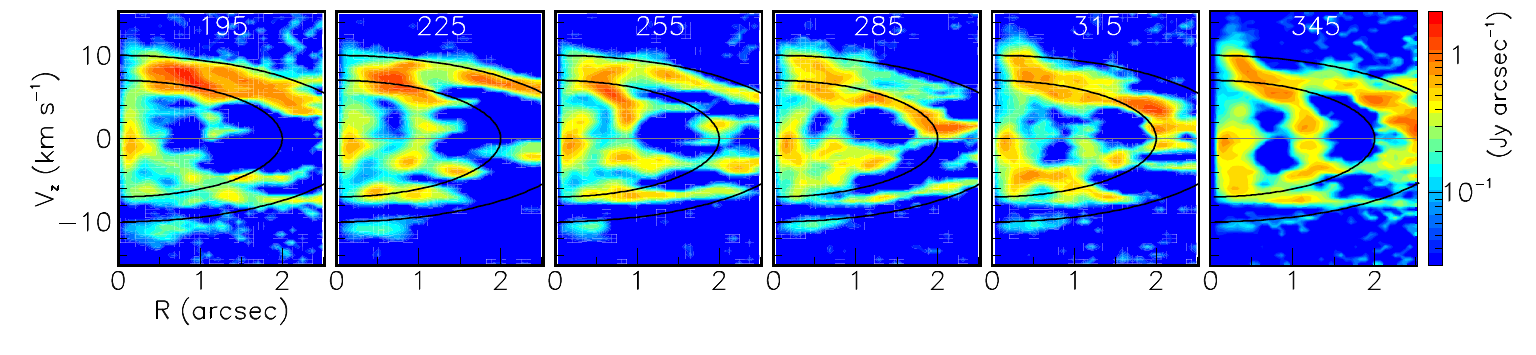}  
  \caption{$^{12}$CO(2-1) line emission (Jy\,arcsec$^{-1}$), combined data. $V_z$ vs $R$ maps in 12 sectors of $\omega$. The values of $\omega$ are given in the inserts. The mean noise level is 80 mJy\,arcsec$^{-1}$.}
  \label{fig17}
\end{figure*}

Figures~\ref{fig16} and \ref{fig17} display different projections of the data-cube. The $V_z$ vs $R$ maps are reminiscent of similar morpho-kinematics observed in the outflows of AGB stars EP Aqr \citep{Nhung2024} and R Leo \citep{Hoai2023}. Together with the observation of arcs of enhanced emission in the channel maps, they suggest that mass loss proceeds via the emission of shell caps covering solid angles at steradian scale and episodically enhanced with a time scale of a few decades. Indeed, in a $V_z$ vs $R$ map, a sphere ejected at time $t$ and expanding radially at velocity $V$ appears as an ellipse of equation $V_z^2+R^2/t^2$=$V^2$. At $R$=0, $V_z$=$V$ independently from the value of $t$. In contrast, for $V_z=$0, $R$=$Vt$ gives a measure, $t$=$R/V$, of the time of ejection. A succession of enhancements and depressions on the $V_z$=0 axis is therefore evidence for variability and episodes of enhanced mass loss. The $V_z$ vs $R$ map of Figure~\ref{fig16} displays as reference thin spherical shells emitted 250 years ago with radial velocities of 10 and 7 \kms, respectively. Compared with observations, they suggest that the CSE is made of a succession of enhanced shell cap emissions, together forming a thick spherical envelope seen in emission in the red-shifted hemisphere and in absorption in the blue-shifted hemisphere (as also illustrated on the Doppler velocity spectrum displayed in Figure~\ref{fig14}). While appealing, this interpretation, supported by the presence of enhancements of emission in the form of arcs in the channel maps, remains speculative and cannot be claimed with strong confidence from the currently available observations. If it were confirmed, it would favour an interpretation in terms of emissions covering solid angles associated with the surface of convective cells, enhanced on a time scale of a few decades and obeying the dynamics described by standard 3-D hydro-dynamical models such as described by \citet{Hofner2019}. The fine time-structure expected to be associated with pulsations having a period of 408 days, at the scale of $\sim$10 mas, cannot be resolved by the present observations. 

\section{ Possible interpretations}
\subsection{The singularity of H$^{12}$CN line emission}
The presence in 2023 of a strong H$^{12}$CN(3-2) line emission confined within some 200 mas from the centre of the star challenges interpretation. As we already remarked, its expansion to larger radii in 2024, at the scale of $\sim$100 mas in a year, meaning $\sim$80 \kms, contrasts with the narrow width of the line profile, $\sim$$\pm$2 \kms. An explanation in terms of maser emission would account for both the confinement of the emission and the exclusivity of the H$^{12}$CN(3-2) line but would suggest variability in intensity at a same location, not variability in location at a same intensity as is observed. Moreover, as remarked by \citet{Bieging2001}, ground vibrational state HCN masers have only been observed in optically bright carbon stars with relatively low mass-loss rates. These include, in particular, a H$^{13}$CN(1-0) maser from Y CVn \citep{Izumiura1987} and a H$^{12}$CN(1-0) maser from six low mass loss rate carbon stars \citep{Izumiura1995}. \citet{Olofsson1998}, from an extensive study of a large sample of AGB stars, in particular W Ori and X TrA, suggest that the H$^{12}$CN(1-0) maser emission observed in low mass-loss-rate ($<$$\sim$5$\times$10$^{-7}$ M$_\odot$ yr$^{-1}$) carbon stars is due to radiative excitation with radial amplification. However, \citet{Bieging2001} remarks that an alternative interpretation may be the heating effects of shock waves that are driven periodically through the extended stellar atmosphere by the large-amplitude pulsations of the stars, causing HCN molecules to be successively destroyed in the shock and reformed in the post-shock gas: shocks might then produce a ``chemical pump'' similar to that discussed by \citet{Schilke2000} for IRC+10216, the basic idea being that the HCN molecules are formed in random vibrational states, population inversion being caused by wide differences in the radiative decay rates of the different vibrational levels. The idea that a shock wave might produce masers travelling outward is of course attractive, but too speculative.

\begin{figure}
  \centering
   \includegraphics[width=0.9\linewidth]{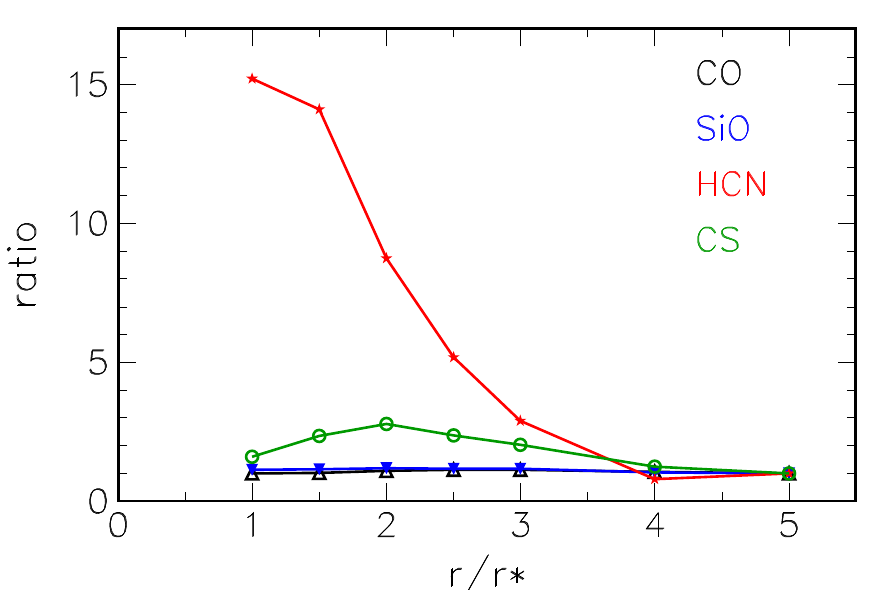}
  \caption{Radial dependence of the molecular abundances normalized to their values at $r$=5 stellar radii ($r^*$) of H$^{12}$CN (red), SiO and CO (black and blue) and CS (green) molecules as predicted by a model of a C/O=0.95 star hit by a shock of 25 \kms\ formed at one stellar radius \citep{Cherchneff2006}.}
  \label{fig18}
\end{figure}

Yet, the observation of unexpectedly low $^{12}$CO/$^{13}$CO, $^{28}$Si/$^{29}$Si and $^{16}$O/$^{17}$O emission ratios in the close neighbourhood of the star at small Doppler velocities, reported in Sub-section 3.5, adds to the difficulty to propose a sensible explanation of what happens in this region. In particular, the contrast observed at low Doppler velocities between the low value of the $^{12}$CO/$^{13}$CO and the high value of the H$^{12}$CN/H$^{13}$CN emission ratios challenges interpretation,  suggesting that two very different mechanisms may have successively dominated the dynamics in this region. It is useful, in this context, to recall the basic assumptions, explicitly spelled out by \citet{Willacy1998}, underlying shock chemistry models. Stellar pulsations cause large amplitude compressional waves to propagate outward and to steepen into shocks outside the stellar photosphere. These periodic shocks travel through the inner layers of the CSE and alter dramatically the temperature and the density of the gas. Layers of gas are then accelerated upwards but then decelerate and eventually fall back to their initial position under the inﬂuence of stellar gravity, leading to the formation of a stationary layer close to the stellar photosphere. The shock is assumed to form at stellar radius and its strength to be damped as it travels outwards. Using such a model, \citet{Duari2000} and \citet{Cherchneff2006} have studied a C/O=0.95 star hit by shocks formed at stellar radius, with velocities of 32 and 25 \kms, respectively. Figure~\ref{fig18} compares the radial dependence of the molecular abundances obtained by \citet{Cherchneff2006}, normalised at a distance of 5 stellar radii, for CO, SiO, CS and HCN molecules separately: the latter decreases by a factor $\sim$15 between 1.5 and 3.5 stellar radii while the others are nearly constant. Such a dramatic effect is however not commented by the author. According to the description of the relevant shock chemistry described in the paper, it is probably related to HCN formation via the chain of reactions H+C$_2$$\rightarrow$CH+C, CH+N$\rightarrow$CN+H, CN+H$_2$$\rightarrow$HCN+H, which relies on the formation of CH molecules in the shock front: the CH formation reaction has a very high activation barrier of over 3$\times$10$^4$ K, which prevents its occurrence in other regions and CH molecules do not contribute to the formation of CO and SiO molecules, and contribute only little to the formation of CS via CN+S$\rightarrow$CS+N, dominated instead by C+SO$\rightarrow$CS+O.
 
While offering an interesting illustration of the mechanisms at stake, such a model cannot be expected to provide a reliable description of reality. Indeed, \citet{Gobrecht2016} and \citet{Marigo2016}, using a similar model with different parameters find instead a deep depression of HCN abundance in the first few stellar radii. We simply retain from these arguments that the HCN abundance in the first few stellar radii, at variance with the abundance of the other molecules, seems to be extremely sensitive to the physical properties of the shock and to its amount of damping when moving outwards: within the limits of our current understanding of the shock dynamics at stake, an order of magnitude enhancement of HCN abundance within a few stellar radii, as predicted by \citet{Cherchneff2006} and \citet{Duari2000}, is quite plausible even if the very speculative nature of the above scenario cannot be understated. Moreover, the complexity of the dynamics at stake may also be expected to eventually offer favourable conditions for chemical pumping schemes causing significant maser emission. 

\subsection{Shocks and third-dredge-ups}
While the scenario developed in the previous sub-section rests on the important role played by strong shocks in the CSE of $\chi$ Cygni, the very large value taken by the H$^{12}$CN(3-2)/H$^{13}$CN(3-2) emission ratio in the inner layer may be related to the recent occurrence of third-dredge-up episodes. Indeed, both shocks and third-dredge-ups have been shown to play a dominant role, but the relation between the two, if any, is unclear. Indeed, for now several decades, as described in the introduction, both shocks and third-dredge-ups have been commonly accepted to play a dominant role, but the present observations raise questions about their precise respective contributions.

An important argument in proposing the observations reported in the present article was the S-type nature of $\chi$ Cygni, with a large C/O ratio of 0.95-0.98 \citep{Duari2000,Justtanont2010}. Indeed, we know from \citet{Uttenthaler2013} and \citep{Uttenthaler2019} that Tc-rich and Tc-poor stars populate different regions of the period-luminosity diagram, suggesting that third-dredge-ups, bringing new elements to the star surface, play a significant role in the evolution of the star. The large value of the $^{12}$C/$^{13}$C ratio implies that $\chi$ Cygni is indeed in an advanced stage of evolution, and the presence of technetium in its spectrum is evidence for a third dredge-up event to have occurred recently. However, the observed confinement of the H$^{12}$CN(3-2) emission, contrasting with the absence of a similar feature in $^{12}$CS and H$^{13}$CN production, is not in itself evidence for approaching the carbon-rich regime. We find it difficult to identify precisely the respective contributions of shock waves and third dredge-ups to the dynamics of the CSE. There seems to be no clear relation between the two. Indeed, the simulations performed by \citet{Duari2000} and \citet{Cherchneff2006} show that the production of such molecules by shock chemistry is essentially independent from the C/O ratio, and very much so over the M- to S-type transition. The pulsation periods are not much larger than the values which they take for M-stars and the mass-loss rates have similar distributions for M-, S- and C-stars \citep{Ramstedt2009}. In general, the episodic and inhomogeneous nature of the wind morpho-kinematics, and particularly the occurrence of brief episodes of mass-loss, are also observed in Tc-poor stars such as R Leo \citep{Hoai2023}, R Dor \citep{Nhung2019} or W Hya \citep{Hoai2022b} and cannot be interpreted as signalling the approach of the carbon-star regime. 

\section{Summary and conclusion}
The results obtained in the present article, taking advantage of the good sensitivity and high spectral and angular resolutions provided by the upgraded NOEMA array, have been discussed in the context of earlier studies of the generation of the nascent wind of $\chi$ Cygni, giving evidence for the important role played by shock waves and for the recent occurrence of a third dredge up.

We have identified three different components in the CSE:
\begin{itemize}
\item An extended spherical envelope of cool CO gas, previously observed by \citet{CastroCarrizo2010} to extend up to a radius of some 600 au, where it is photo-dissociated. The present work probes it along the line of sight, over which SiO, HCN and CS molecules are confined to smaller radii. It displays radial expansion with a mean velocity of 9$\pm$1 \kms. It gives evidence, within the available angular resolution of 100-200 mas FWHM, for both episodic and inhomogeneous enhancements of emission, with respective scales of a few decades and of a steradian, reminiscent of those observed in the CSE of other AGB stars, such as EP Aqr \citep{Nhung2024}, and suggesting the impact of shock waves.\\
  
\item A volume of gas containing CO, SiO, HCN and CS molecules surrounding the star and expanding radially with average velocities of only a few \kms, but reaching up to $\sim$8 \kms\ in the case of SiO. Both spectral and projected radial distributions are much narrower for H$^{12}$CN, particularly in 2023, and, but with less confidence, for Si$^{17}$O and possibly $^{13}$CO than for the other molecules. Evidence has been obtained for the unresolved enhancement of HCN(3-2) emission confined very close to the photosphere to be almost exclusively from the H$^{12}$CN isotopologue.\\
  
\item A recent mass ejection that occurred some decades ago and is enhanced over the red-shifted north-western octant at radii of $\sim$100-200 au and radial velocities of $\sim$10-15 \kms, suggesting that it was produced by a series of shock waves associated with large amplitude stellar pulsations and enhanced over a particular convective cell.\\
\end{itemize}

The present study has revealed unexpected features that challenge interpretation and provide new important information on the physics and chemistry at stake inside $\chi$ Cygni and its CSE, in particular on two of its major specificities: the presence of strong shocks and the past occurrence of third-dredge-up episodes. These include, in particular, evidence for strong and nearly unresolved H$^{12}$CN(3-2) emission in 2023, significantly more extended in 2024 but still confined to the close neighbourhood of the star, and for H$^{12}$CN(3-2)/H$^{13}$CN(3-2), $^{12}$CO(2-1)/$^{13}$CO(2-1) and Si$^{16}$O(6-5)/Si$^{17}$O(6-5) emission ratios displaying important variations across the CSE. These add to a series of apparent inconsistencies related to the measurement of isotopic abundance ratios and reported in the literature, several of which are referred to in the text. While some possible interpretations have been speculated, in particular referring to shock chemistry and maser emission, none has been found convincing enough and a clear understanding of the mechanisms at stake calls for new measurements of high resolution and sensitivity, in particular at different stellar phases.

\section*{Data availability}
The channel maps of the Si$^{16}$O(5-4), Si$^{16}$O(6-5), H$^{12}$CN(3-2), H$^{13}$CN(3-2), Si$^{16}$O($\nu$=1,$J$=6-5), CS(5-4) and Si$^{17}$O(6-5) line emissions are available at https://doi.org/10.5281/zenodo.13841303

\begin{acknowledgements}
   We are deeply grateful to Professors Stefan Uttenthaler and Iain McDonald for clarifications concerning the evolution of S stars, the third dredge up and the C/O ratio and to Professors Wouter Vlemmings and Hans Olofsson for sharing with us their views on the problems related to the evaluation of isotopic abundance ratios. We thank Dr Ka Tat Wong for a careful reading of the manuscript and for useful comments.We also thank the anonymous referee for a careful reading of the manuscript. We thank the staff at the NOEMA observatory for their support in performing these observations. This work is based on observations carried out under project number W22BL with the IRAM NOEMA interferometer. IRAM is supported by INSU/CNRS (France), MPG (Germany) and IGN (Spain). This research is funded by Vietnam Academy of Science and Technology under grand number THTETN.03/24-25. Financial support from the World Laboratory, the Vingroup Fellowship Programme, the Odon Vallet Foundation and the Vietnam National Space Center is gratefully acknowledged. 
\end{acknowledgements}

%
%

\begin{appendix}
\section{Other dectected lines}
\begin{figure*}[h!]
  \centering
  \includegraphics[width=0.65\linewidth]{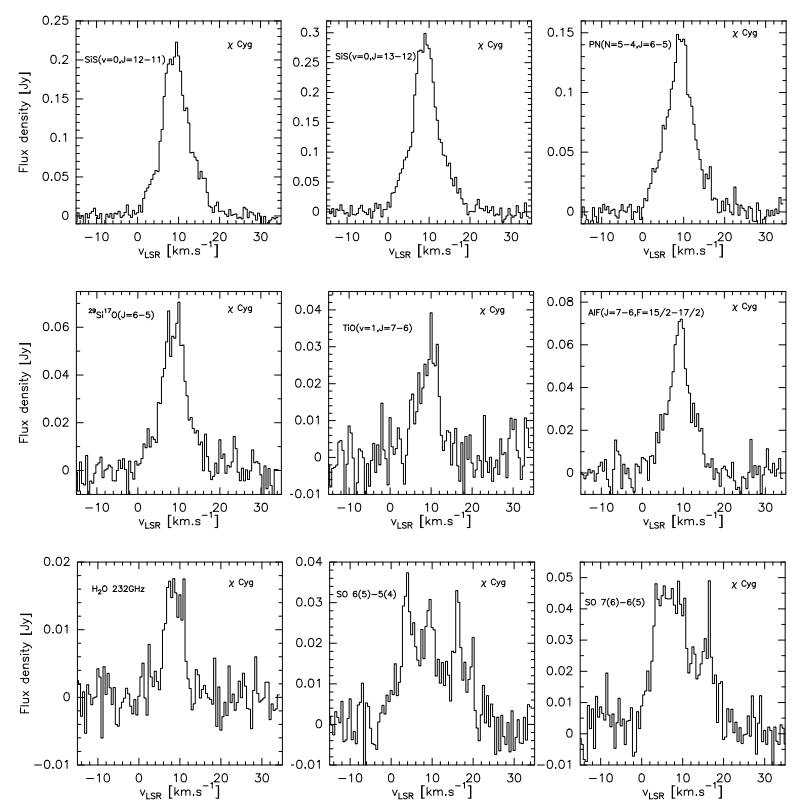}
  \caption{From upper-left to lower-right: Line profiles of SiS, PN, $^{29}$Si$^{17}$O, TiO, AlF, H$_2$O and SO emissions integrated over the central 1 arcsec$\times$1 arcsec square aperture.}
  \label{figb1}
\end{figure*}

Appendix A present the line profiles (Figure~\ref{figb1}) and properties (Table~\ref{tabb1}) of the other detected molecular line emissions not discussed in the present article.
  
\begin{table*}[h!]
  \centering
  \caption{Properties of other lines detected in this work. Line parameters are from the CDMS (Muller et al. 2005), except for H$_2$O which are from Belov et al. 1987. All cubes were imaged with natural weighting and the spectral resolution is 0.5 \kms\ in all cases. Fluxes are determined from the central 1 arcsec$\times$1 arcsec square aperture.}
  \label{tabb1}
  \begin{tabular}{ccccccccc}
    \hline
  Line&Frequency&$E_u/k$&$A_{ji}$&Integrated flux&Beam size&PA&1$\sigma$ noise\\
  &GHz&K&s$^{-1}$&Jy\,\kms&arcsec$^2$&deg&mJy\,beam$^{-1}$\\
  \hline
$^{29}$Si$^{17}$O($\nu$=0,$J$=6-5)&247.48153&41.6&7.82$\times$10$^{-4}$&0.48&0.44$\times$0.26&23&2.7\\
TiO($\nu$=1,$J$=7-6)&220.33268&1479.7&6.35$\times$10$^{-4}$&0.19&0.45$\times$0.23&26&2.3\\
AlF($J$=7-6,$F$=15/2-17/2)&
230.79389&
44.3&
1.38$\times$10$^{-4}$&
0.47&
0.44$\times$0.23&
26&
2.5\\
H$_2$O($\nu_2$=1,5(5,0)-6(4,3))&
232.68670&
3462&
4.76$\times$10$^{-6}$&
0.13&
0.44$\times$0.22&
26&
2.7\\
PN($N$=5-4,$J$=6-5)&
234.93569&
33.8&
5.18$\times$10$^{-4}$&
1.18&
0.44$\times$0.22&
26&
2.9\\
SO(6(5)-5(4))&
219.94944&
35&
1.34$\times$10$^{-4}$&
0.38&
0.46$\times$0.23&
26&
2.3\\
SO(7(6)-6(5))&
261.84372&
47.6&
2.28$\times$10$^{-4}$&
0.68&
0.42$\times$0.25&
19&
3\\
SiS($\nu$=0,$J$=12-11)&
217.81766&
68&
1.74$\times$10$^{-4}$&
1.75&
0.47$\times$0.24&
25&
3.1\\
SiS($\nu$=0,$J$=13-12)&
235.96136&
79.3&
2.22$\times$10$^{-4}$&
2.33&
0.43$\times$0.21&
25&
3\\
\hline
  \end{tabular}
  \end{table*}

\end{appendix}

\end{document}